\documentclass[a4paper,oneside,11pt]{article}
\pdfoutput=1
\usepackage{cprform}
\usepackage{amsmath,amstext,amsfonts,amsbsy,amssymb,amscd,slashed}
\usepackage{lscape,color,svg,epsfig,cite,bbm,multirow}
\usepackage{booktabs}
\usepackage{rotating}
\usepackage{tikz}
\usetikzlibrary{arrows,shapes}
\usetikzlibrary{trees}
\usetikzlibrary{matrix,arrows}                          
\usetikzlibrary{positioning}                            
\usetikzlibrary{calc,through}                           
\usetikzlibrary{decorations.pathreplacing}  
\usetikzlibrary{decorations.pathmorphing}       
\usetikzlibrary{decorations.markings}
\tikzset{
    vector/.style={decorate, decoration={snake}, draw},
        provector/.style={decorate, decoration={snake,amplitude=2.5pt}, draw},
        antivector/.style={decorate, decoration={snake,amplitude=-2.5pt}, draw},
    fermion/.style={draw=black, postaction={decorate},
        decoration={markings,mark=at position .55 with {\arrow[draw=black]{>}}}},
    fermionbar/.style={draw=black, postaction={decorate},
        decoration={markings,mark=at position .55 with {\arrow[draw=black]{<}}}},
    fermionnoarrow/.style={draw=black},
    gluon/.style={decorate, draw=black,
        decoration={coil,amplitude=4pt, segment length=5pt}},                           
    scalar/.style={dashed,draw=black, postaction={decorate},
        decoration={markings,mark=at position .55 with {\arrow[draw=black]{>}}}},
    scalarbar/.style={dashed,draw=black, postaction={decorate},
        decoration={markings,mark=at position .55 with {\arrow[draw=black]{<}}}},
    scalarnoarrow/.style={dashed,draw=black},
    electron/.style={draw=black, postaction={decorate},
        decoration={markings,mark=at position .55 with {\arrow[draw=black]{>}}}},
        bigvector/.style={decorate, decoration={snake,amplitude=4pt}, draw},
}

\newcommand{\ba}{\begin{array}}
\newcommand{\ea}{\end{array}}

\newcommand{\req}[1]{Eq.~(\ref{#1})}

\newcommand{\rep}[1]{\cite{#1}}

\newcommand{\dif}{{\rm d}}

\newcommand{\ci}[1]{\boldsymbol{#1}}

\newcommand{\Dslash}{\relax{\kern+.25em / \kern-.70em D}}

\newcommand{\alphas}{\alpha_{\rm\scriptscriptstyle s}}

\newcommand{\fm}{{\rm fm}}
\newcommand{\MeV}{{\rm MeV}}
\newcommand{\GeV}{{\rm GeV}}

\newcommand{\Real}{\relax{\mathsf{\Gamma\kern-.35em R}}}
\newcommand{\Int}{\relax{\mathsf{Z\kern-.40em Z}}}

\newcommand{\CF}{C_{\rm F}}

\newcommand{\ihalf}{{\scriptstyle{{i\over 2}}}}

\newcommand{\NC}{N}
\newcommand{\NF}{N_\mathrm{\scriptstyle f}}

\newcommand{\MSbar}{{\overline{\rm MS}}}

\newcommand{\gbar}{\kern1pt\overline{\kern-1pt g\kern-0pt}\kern1pt}
\newcommand{\mbar}{\kern2pt\overline{\kern-1pt m\kern-1pt}\kern1pt}
\newcommand{\obar}[1]{\kern3pt\overline{\kern-2pt #1\kern-0pt}\kern1pt}

\newcommand{\lQCD}{\Lambda_{\rm\scriptscriptstyle QCD}}
\newcommand{\mrgi}{M}
\newcommand{\orgi}[1]{\hat #1}

\newcommand{\mcrit}{m_{\rm cr}}
\newcommand{\hopc}{\kappa_{\rm c}}

\newcommand{\fP}{f_{\rm\scriptscriptstyle P}}

\newcommand{\fA}{f_{\rm\scriptscriptstyle A}}

\newcommand{\kV}{k_{\rm\scriptscriptstyle V}}
\newcommand{\kT}{k_{\rm\scriptscriptstyle T}}
\newcommand{\hT}{h_{\rm\scriptscriptstyle T}}

\newcommand{\ZP}{Z_{\rm\scriptscriptstyle P}}

\newcommand{\ZT}{Z_{\rm\scriptscriptstyle T}}

\newcommand{\sigmaT}{\sigma_{\rm\scriptscriptstyle T}}

\newcommand{\SigmaT}{\Sigma_{\rm\scriptscriptstyle T}}

\newcommand{\Oa}{\mbox{O}(a)}
\newcommand{\Oasq}{\mbox{O}(a^2)}

\newcommand{\icsw}{c_{\rm sw}}
\newcommand{\ict}{c_{\rm t}}
\newcommand{\icttil}{\tilde c_{\rm t}}

\newcommand{\icA}{c_{\rm\scriptscriptstyle A}}
\newcommand{\icT}{c_{\rm\scriptscriptstyle T}}

\newcommand{\abar}{\kern1pt\overline{\kern-1pt a\kern-0.5pt}\kern1pt}

\newcommand{\cO}{{\cal O}}

\newcommand{\cZ}{{\cal Z}}

\newcommand{\vx}{\mathbf{x}}
\newcommand{\vy}{\mathbf{y}}

\makeatletter
\def\hlinewd#1{%
\noalign{\ifnum0=`}\fi\hrule \@height #1 %
\futurelet\reserved@a\@xhline}
\makeatother
\renewcommand{\Hline}{\hlinewd{0.8pt}}
\begin{document}
\begin{titlepage}
\vspace*{-30truemm}
\begin{flushright}
IFT-UAM/CSIC-17-050\\
FTUAM-17-10\\[10pt]
\end{flushright}
\vspace{15truemm}
\centerline{\Bigrm Non-perturbative renormalization of tensor currents:}
\centerline{\Bigrm strategy and results for $N_f=0$ and $N_f=2$ QCD}
\vskip 10 true mm
\begin{center}
\epsfig{figure=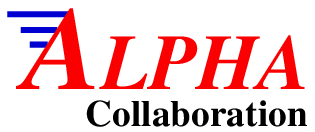, width=25 true mm}\\
\end{center}
\vskip -2 true mm
\centerline{\bigrm C.~Pena$^{a,b}$, D.~Preti$^{b}$}
\vskip 4 true mm
\centerline{\it $^a$ Departamento de F\'{\i}sica Te\'orica, Universidad Aut\'onoma de Madrid}
\centerline{\it Cantoblanco E-28049 Madrid, Spain}
\vskip 3 true mm
\centerline{\it $^b$ Instituto de F\'{\i}sica Te\'orica UAM-CSIC}
\centerline{\it c/~Nicol\'as Cabrera 13-15, Universidad Aut\'onoma de Madrid}
\centerline{\it Cantoblanco E-28049 Madrid, Spain}
\vskip 20 true mm
\noindent{\tenbf Abstract: }
{\tenrm
Tensor currents are the only quark bilinear operators lacking a non-perturbative determination
of their renormalisation group (RG) running between hadronic and electroweak scales. We develop
the setup to carry out the computation in lattice QCD via standard recursive finite-size
scaling techniques, and provide results for the RG running of tensor currents
in $N_f=0$ and $N_f=2$ QCD in the continuum for various Schr\"odinger Functional schemes.
The matching factors between bare and renormalisation group invariant currents 
are also determined for a range of values of the lattice spacing relevant for
large-volume simulations, thus enabling a fully non-perturbative renormalization
of physical amplitudes mediated by tensor currents.
}
\vspace{10truemm}
\eject
\end{titlepage}
\section{Introduction}

Hadronic matrix elements of tensor currents play an important r\^ole in
several relevant problems in particle physics. Some prominent examples are
rare heavy meson decays that allow to probe the consistency
of the Standard Model (SM) flavour sector
(see, e.g.,~\cite{Buchalla:2008jp,Antonelli:2009ws,Blake:2016olu} for an overview),
or precision measurements of $\beta$-decay
and limits on the neutron electric dipole moment
(see, e.g.,~\cite{Bhattacharya:2015esa,Bhattacharya:2016zcn,Abramczyk:2017oxr}
for an up-to-date lattice-QCD perspective).  

One of the key ingredients in these computations is the renormalization of
the current. Indeed, partial current conservation ensures that non-singlet
vector and axial currents require at worst finite normalizations, and fixes
the anomalous dimension of scalar and pseudoscalar densities to be minus the
quark mass anomalous dimension. They however do not constrain the tensor current,
which runs with the only other independent anomalous dimension among quark
bilinears. Controlling the current renormalization and running at the non-perturbative
level, in the same fashion achieved for quark masses~\cite{Capitani:1998mq,DellaMorte:2005kg,Campos:2016eef,Campos:2016vxh},
is therefore necessary in order to control systematic uncertainties,
and allow for solid conclusions in new physics searches.

The anomalous dimension of tensor currents is known to three-loop order
in continuum schemes~\cite{Gracey:2000am,Almeida:2010ns}, while on the lattice perturbative studies
have been carried out to two-loop order~\cite{Skouroupathis:2008mf}.
Non-perturbative determinations of
renormalization constants in RI/MOM schemes, for the typical few-$\GeV$
values of the renormalization scale accessible to the latter,
have been obtained for various numbers of dynamical flavours and
lattice actions~\cite{Gockeler:1998ye,Becirevic:2004ny,Aoki:2007xm,Sturm:2009kb,Constantinou:2010gr,Alexandrou:2012mt,Constantinou:2014fka}.
The purpose of this work is to set up the strategy for the application
of finite-size scaling techniques based on the Schr\"odinger Functional (SF)~\cite{Luscher:1992an},
in order to obtain a fully non-perturbative determination of both current
renormalization constants at hadronic energy scales, and the running of
renormalized currents to the electroweak scale. This completes the
ALPHA Collaboration non-perturbative renormalization programme for non-singlet
quark field bilinears~\cite{Capitani:1998mq,DellaMorte:2005kg,Campos:2016eef,Campos:2016vxh,massnf3,Blossier:2010jk,Bernardoni:2014fva} and four-quark operators~\cite{Guagnelli:2005zc,Palombi:2005zd,Dimopoulos:2006ma,Dimopoulos:2007ht,Palombi:2007dr,Papinutto:2014xna,Papinutto:2016xpq}.

As part of the strategy, we will set up a family of SF renormalization schemes,
and perform a perturbative study with the main purpose of computing the
perturbative anomalous dimension up to two loops, in order to make safe
contact with perturbative physics at the electroweak scale.
Preliminary results of this work have already appeared as
proceedings contributions~\cite{Fritzsch:2015lvo}.\footnote{During the development of
this work, Dalla Brida, Sint and Vilaseca have performed a related
perturbative study as part of the setup of the chirally rotated Schr\"odinger
Functional~\cite{Brida:2016rmy}. Their results for the one-loop matching factor required to compute
the NLO tensor anomalous dimensions in SF schemes coincide with ours 
(cf.~section~\ref{sec:sfpt}), previously published in~\cite{Fritzsch:2015lvo}. This constitutes
a strong crosscheck of the computation.}
We will then apply our formalism to the fully non-perturbative renormalization
of non-singlet tensor currents in $\NF=0$ and $\NF=2$ QCD. Our results for
$\NF=3$ QCD, that build on the non-perturbative determination of the running
coupling~\cite{Brida:2016flw,DallaBrida:2016kgh,Bruno:2017gxd} and the renormalization of quark
masses~\cite{Campos:2016eef,Campos:2016vxh,massnf3}, will be provided in a separate publication~\cite{tensornf3}.

The layout of this paper is as follows.
In section~2 we will introduce our notation and discuss the relevant
renormalization group equations.
In section~3 we will introduce our SF schemes, generalizing the ones
employed for quark mass renormalization.
In section~4 we will study these schemes in one-loop perturbation theory,
and compute the matching factors that allow to determine the NLO values
of anomalous dimensions.
In section~5 we will discuss our non-perturbative computations, and
provide results for the running of the currents between hadronic and
high-energy scales and for the renormalization constants needed to
match bare hadronic observables at low energies. Section~6 contains
our conclusions. Some technical material, as well as several tables
and figures, are gathered in appendices.

\section{Renormalization Group}
Theory parameters and operators are renormalized at the renormalization scale $\mu$. The scale dependence of these quantities is given by their Renormalization Group (RG) evolution. The Callan-Symanzik equations satisfied by the gauge coupling and quark masses are of the form  
\begin{align}
  \label{coupling_RGE}
  \mu\frac{\partial\gbar}{\partial\mu} &\,=\, \beta(\gbar(\mu)) \,,\\
  \label{mass_RGE}
  \mu\frac{\partial\mbar_i}{\partial\mu} &\,=\, \tau(\gbar(\mu))\mbar_i(\mu) \, ,
\end{align}
respectively, with renormalized coupling $\gbar$ and masses $\mbar_i$; the index $i$ runs over flavour.  
The renormalization group equations (RGEs) for composite operators have the same form as~\req{mass_RGE}, with the
anomalous dimensions of the operators $\gamma$ in the place of $\tau$. Starting from
the RGE for correlation functions, we can write the renormalization group
equation for the insertion of a multiplicatively renormalizable local composite
operator $O$ in an on-shell correlation function as:
\begin{gather}
  \label{oper_RGE}
  \mu\frac{\partial\obar{O}(\mu)}{\partial\mu} \,=\, \gamma(\gbar(\mu))\obar{O}(\mu) \,.
\end{gather}
where $\obar{O}(\mu)$ is the renormalized operator. The latter is connected to the bare
operator insertion $O(g_0^2)$ through
\begin{gather}
\obar{O}(\mu)=\lim_{a \to 0} Z_O(g_0^2,a\mu) O(g_0^2)\,.
\label{renormalizedO}
\end{gather}
where $g_0$ is the bare coupling, $Z_O$ is a renormalization constant, and $a$ is some inverse ultraviolet cutoff ---
the lattice spacing in this work.
We assume a mass-independent scheme, such that both the $\beta$-function and the anomalous dimensions $\tau$ and $\gamma$ depend only on the coupling and the number of flavours (other than on
the number of colours $\NC$); examples of such schemes are the $\MSbar$ scheme
of dimensional regularization~\cite{tHooft:1973mfk,Bardeen:1978yd}, RI schemes~\cite{Martinelli:1994ty}, or the SF schemes
we will use to determine the running non-perturbatively~\cite{Luscher:1992an,Jansen:1995ck}.
The RG functions then admit asymptotic expansions of the form:
\begin{align}
  \beta(g) \ &\underset{g \sim 0}{\approx} \
  -g^3\big(b_0 + b_1 g^2 + b_2 g^4 + \ldots \big) \,, \\
  \tau(g) \ &\underset{g \sim 0}{\approx} \
  -g^2\big(d_0 + d_1 g^2 + d_2 g^4 + \ldots \big) \,, \\
  \gamma(g) \ &\underset{g \sim 0}{\approx} \
  -g^2\big(\gamma_0 + \gamma_1 g^2 + \gamma_2 g^4 + \ldots \big) \,.
\end{align}
The coefficients $b_0$, $b_1$ and $d_0$, $\gamma_0$ are independent of the
renormalization scheme chosen. In particular~\cite{vanyaterent,Khriplovich:1969aa,thooft,Gross:1973id,Politzer:1973fx,Caswell:1974gg,Jones:1974mm}
\begin{align}
  b_0 &= \frac{1}{(4\pi)^2}\left ( \frac{11}{3}\NC-\frac{2}{3}\NF \right ) \,, \\
  b_1 &= \frac{1}{(4\pi)^4}\left [ \frac{34}{3}\NC^2-\left ( \frac{13}{3}\NC-\frac{1}{\NC} \right ) \NF \right ]  \,,
\end{align}
and
\begin{gather}
  d_0 = \frac{6\CF}{(4\pi)^2}\,,
\end{gather}
with $\CF=\frac{\NC^2-1}{2\NC}$. 

The RGEs in
Eqs.~(\ref{coupling_RGE}--\ref{oper_RGE})
can be formally solved in terms of the renormalization group invariants (RGIs)
$\lQCD$, $\mrgi_i$ and $\orgi{O}$, respectively,
as:\footnote{Overall normalizations are a matter of convention, apart from
  that of $\lQCD$, which is universal. Our choice for $\mrgi_i$ follows Gasser and
  Leutwyler~\rep{Gasser:1982ap,Gasser:1983yg,Gasser:1984gg}, whereas for~\req{rgi_operator} we have chosen
  the most usual normalization with a power of $\alphas$.}
\begin{align}
  \label{rgi_coupling}
  \lQCD &\,=\, \mu \,
  \frac{[b_0\gbar^2(\mu)]^{-b_1/2b_0^2}}{e^{1/2b_0\gbar^2(\mu)}} \,
    \exp\left\{-\int_0^{\gbar(\mu)}\dif g
      \left[\frac{1}{\beta(g)}+\frac{1}{b_0g^3}-\frac{b_1}{b_0^2g}\right]\right\}
    \,, \\
  \label{rgi_mass}
  \mrgi_i &\,=\, \mbar_i(\mu) \, [2b_0\gbar^2(\mu)]^{-d_0/2b_0} \,
    \exp\left\{-\int_0^{\gbar(\mu)}\dif g
      \left[\frac{\tau(g)}{\beta(g)}-\frac{d_0}{b_0g}\right]\right\}
    \,, \\
  \nonumber
  \orgi{O} &\,=\, \obar{O}(\mu) \, \left[\frac{\gbar^2(\mu)}{4\pi}\right]^{-\gamma_0/2b_0} \,
    \exp\left\{-\int_0^{\gbar(\mu)}\dif g \,
      \left[\frac{\gamma(g)}{\beta(g)}-\frac{\gamma_0}{b_0g}\right]\right\}\\
  \label{rgi_operator}
    &\,\equiv\,\orgi{c}(\mu)\obar{O}(\mu)\,.
\end{align}
While the value of the $\lQCD$ parameter depends on the renormalization
scheme chosen, $\mrgi_i$ and $\orgi{O}$ are the same for all schemes. In
this sense, they can be regarded as meaningful physical quantities, as opposed
to their scale-dependent counterparts. The aim of the non-perturbative determination
of the RG running of parameters and operators is to connect the RGIs --- or,
equivalently, the quantity renormalized at a very high energy scale, where
perturbation theory applies --- to the bare parameters or operator insertions,
computed in the hadronic energy regime. In this way the three-orders-of-magnitude
leap between the hadronic and weak scales can be bridged without significant
uncertainties related to the use of perturbation theory.

In order to describe non-perturbatively the scale dependence of the gauge coupling and composite operators,
we will use the step-scaling functions (SSFs) $\sigma$ and $\sigma_O$, respectively,
defined as
\begin{gather}
-\log(s)=\int_{\sqrt{u}}^{\sqrt{\sigma(u)}} \, \frac{dg'}{\beta(g')} \,, \label{eqsigma}\\
\sigma_O(s,u)=\exp \left \{ \int_{\sqrt{u}}^{\sqrt{\sigma(u)}}\, \frac{\gamma(g')}{\beta(g')} dg' \right \}\,, \label{eqsigmaT}
\end{gather}
or, equivalently,
\begin{align}
\sigma(s,u) &= \left.\gbar^2(\mu/s)\right|_{u=\gbar^2(\mu)} \,, \\
\sigma_O(s,u) &= U(\mu/s,\mu)\,,
\end{align}
where
\begin{gather}
U(\mu_2,\mu_1)=\exp \left \{ \int_{\sqrt{\gbar^2(\mu_1)}}^{\sqrt{\gbar^2(\mu_2)}}\, \frac{\gamma(g')}{\beta(g')} dg' \right \}
\label{eq:rgevol}
\end{gather}
is the RG evolution operator for the operator at hand, which connects renormalized
operators at different scales as $\obar{O}(\mu_2)=U(\mu_2,\mu_1)\obar{O}(\mu_1)$.
The SSFs are thus completely determined by, and contain the same information as,
the RG functions $\gamma$ and $\beta$. In particular, $\sigma_O(s,u)$ corresponds
to the RG evolution operator of $\obar{O}$ between the scales $\mu/s$ and $\mu$;
from now on, we will set $s=2$, and drop the parameter $s$ in the dependence. 
The SSF can be related to renormalization constants via the identity
\begin{gather}
\label{Sigma}
\sigma_O(u) = \lim_{a \to 0}\Sigma_O(u,a\mu)\,, \qquad
\Sigma_O(u,a\mu) = \left . \frac{Z_O(g_0^2, a\mu/2)}{Z_O(g_0^2, a\mu)} \right |_{u=\gbar^2(\mu)}\,.
\end{gather} 
This will be the expression we will employ in practice to determine
$\sigma_O$, and hence operator anomalous dimensions, for a broad range of
values of the renormalized coupling $u$.

In this work we will focus on the renormalization of tensor currents.
The (flavour non-singlet) tensor bilinear is defined as  
\begin{gather}
  T_{\mu\nu}(x) = i\,\bar\psi_{s_1}(x)\sigma_{\mu\nu}\psi_{s_2}(x) \,,
  \label{deftensor}
\end{gather}
where $\sigma_{\mu\nu}=\ihalf[\gamma_\mu,\gamma_\nu]$, and $s_1 \ne s_2$
are flavour indices. Since all the Lorentz components have the same anomalous dimension, as far as renormalization is concerned it is enough to consider the ``electric'' operator $T_{0k}$. 
As already done in the introduction, it is important to observe that the tensor current is the 
only bilinear operator that evolves under RG transformation in a different way respect to the 
quark mass ---
partial conservation of the vector and axial currents protect them from
renormalization, and fixes the anomalous dimension of both scalar and
pseudoscalar densities to be $-\tau$.
The one-loop (universal) coefficient of the tensor anomalous dimension is
\begin{gather}
  \gamma_{\rm\scriptscriptstyle T}^{(0)}=\frac{2\CF}{(4\pi)^2} \,.
  \end{gather}

\section{Schr\"odinger Functional renormalization schemes}
The renormalization schemes we will consider are based on the Schr\"odinger
Functional~\cite{Luscher:1992an}, i.e. on the QCD partition function
$\cZ=\int{\rm D}[A,\bar\psi,\psi]e^{-S[A,\bar\psi,\psi]}$ on a finite Euclidean
spacetime of dimensions $L^3 \times T$ with lattice spacing $a$, where periodic boundary conditions
on space (in the case of fermion fields, up a to a global phase $\theta$)
and Dirichlet boundary conditions at times $x_0=0,T$ are imposed.
A detailed discussion of the implementation and notation that we will follow
can be found in~\cite{Luscher:1996sc}. We will always consider
$L=T$ and trivial gauge boundary fields (i.e. there is no background field induced by the latter).
The main advantage of SF schemes is that they allow to compute
the scale evolution via finite-size scaling, based on the identification
of the renormalization scale with the inverse box size, i.e. $\mu=1/L$.

\begin{figure}[t!]
\begin{center}
\includegraphics[width=60mm]{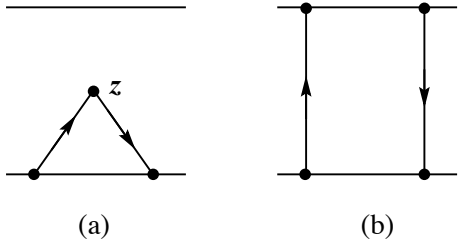}
\vspace{-2mm}
\captionof{figure}{Sketch of correlation function in the SF: bilinear insertion on the left and boundary-to-boundary on the right.}
\label{fig:fpf1}
\end{center}
\end{figure}

To define suitable SF renormalization conditions we
can follow the same strategy as in~\cite{Capitani:1997mw,Sint:1998iq,DellaMorte:2005kg,Campos:2015fka}, which has been
applied successfully also to several other composite
operators both in QCD~\cite{Guagnelli:1998ve,Guagnelli:1999wp,Shindler:2003vx,Guagnelli:2005zc,Palombi:2005zd,Dimopoulos:2006ma,Dimopoulos:2007ht,Palombi:2007dr,Blossier:2010jk,Bernardoni:2014fva} and other theories.\footnote{See, e.g.,~\cite{Pica:2017gcb} for a recent review.}
We first introduce the two-point functions
\begin{gather}
  \label{kT}
  \kT(x_0) = - \frac{1}{6}\sum_{k=1}^3
  \langle T_{0k}(x)\,\cO[\gamma_k]\rangle \,,
\end{gather}
\begin{gather}
  \label{f1}
  f_1 = -\frac{1}{2L^6}
  \langle\cO'_{s_2s_1}[\gamma_5]\cO_{s_1s_2}[\gamma_5]\rangle \,,
\end{gather}
and
\begin{gather}
  \label{k1}
  k_1 = -\frac{1}{6L^6}
  \langle\cO'_{s_2s_1}[\gamma_k]\cO_{s_1s_2}[\gamma_k]\rangle \,.
\end{gather}
where
\begin{gather}
  \label{source}
  \cO[\Gamma] =
  a^6\sum_{\vx,\vy}\bar{\zeta}_{s_2}(\vx)\,\Gamma\,\zeta_{s_1}(\vy)
\end{gather}
is a source operator built with the $x_0=0$ boundary fields $\zeta,\bar\zeta$.
A sketch of the correlation function in the SF is provided in Fig.\ref{fig:fpf1}.
The renormalization constant $\ZT$ is then defined by
\begin{gather}
  \label{renorm_condition}
  \ZT(g_0,a/L)\frac{\kT(L/2)}{f_1^{1/2-\alpha}k_1^{\alpha}} =
  \left.\frac{\kT(L/2)}{f_1^{1/2-\alpha}k_1^{\alpha}}\right|_{m_0=\mcrit, \, g_0^2=0} \, ,
\end{gather}
where we have already fixed $\mu=1/L$, $m_0$ is the bare quark mass, and
$\mcrit$ is the critical mass, needed if Wilson fermions are
used in the computation --- as will be our case.
The factor $f_1^{1/2-\alpha}k_1^{\alpha}$ cancels the renormalization
of the boundary fields contained in $\cO[\Gamma]$, which holds for any
value of the parameter $\alpha$; we will restrict ourselves to the choices
$\alpha=0,1/2$. The only remaining parameter in Eq.~(\ref{renorm_condition})
is the kinematical variable $\theta$ entering spatial boundary conditions;
once its value is specified alongside the one of $\alpha$, the scheme is
completely fixed.
We will consider the values $\theta=0,0.5,1.0$ in the perturbative study
discussed in the next section, and in the non-perturbative computation
we will set $\theta=0.5$.

The condition in Eq.~(\ref{renorm_condition}) involves the correlation function
$\kT$, which is not $\Oa$ improved. Therefore, the scaling of
the renormalized current towards the continuum limit, given by Eq.~(\ref{renormalizedO}),
will be affected by $\Oa$ effects. The latter can be removed
by subtracting suitable counterterms, following the standard on-shell
$\Oa$ improvement strategy for SF correlation functions~\cite{Luscher:1996sc}.
On the lattice, and in the chiral limit, the $\Oa$
improvement pattern of the tensor current reads
\begin{gather}
T_{\mu \nu}^{\rm I}=T_{\mu \nu} + a \icT(g_0^2)(\tilde{\partial}_{\mu}V_{\nu} - \tilde{\partial}_{\nu}V_{\mu})\,,
\end{gather} 
where $\tilde\partial$ is the symmetrized lattice derivative and
$V_\mu=\bar\psi_{s_1}\gamma_\mu\psi_{s_2}$ is the vector current.
Focusing again only on the electric part, the above formula reduces to
\begin{gather}
\label{improvement}
T_{0 k}^{\rm I}=T_{0 k} + a \icT(g_0^2)(\tilde{\partial}_{0}V_{k} - \tilde{\partial}_{k}V_{0}) \, ,
\end{gather}
which results in an $\Oa$ improved version of the two-point function
$\kT$ of the form
\begin{gather}
\kT^{\rm I}(x_0)=\kT(x_0) + \icT(g_0^2)\tilde{\partial}_{0}\kV(x_0)   \, ,
\end{gather}
with
\begin{gather}
  \label{kV}
  \kV(x_0) = - \frac{1}{6}\sum_{k=1}^3
  \langle V_{k}(x)\,\cO[\gamma_k]\rangle \,.
\end{gather}
Note that the contribution involving the spatial derivative vanishes.
Inserting $\kT^{\rm I}$ in Eq.~(\ref{renorm_condition}), and the resulting
$\ZT$ in Eq.~(\ref{renormalizedO}) alongside the $\Oa$ improved current,
will result in $\Oasq$ residual cutoff effects in the value of the SSF $\SigmaT$
defined in~\req{Sigma},
provided the action and $\mcrit$ are also $\Oa$ improved.

\section{Perturbative study}
\label{sec:sfpt}

We will now study our renormalization conditions in one-loop perturbation theory.
The aim is to obtain the next-to-leading (NLO) anomalous dimension of the
tensor current in our SF schemes, necessary for a precise connection to RGI currents,
or continuum schemes, at high energies; and compute the leading perturbative contribution
to cutoff effects, useful to better control continuum limit extrapolations.

We will expand the relevant quantities in powers of the bare coupling $g_0^2$ as
\begin{gather}
X=\sum_{n=0}^{\infty} g_0^2 X^{(n)}
\end{gather}
where $X$ can be any of $\ZT$, $\kT$, $\kV$, $f_1$, or $k_1$.
To $\mathcal{O}(g_0^2)$, Eq.~(\ref{improvement}) can be written as
\begin{gather}
\kT^{\rm I}(x_0)=\kT^{(0)}(x_0) + g_0^2 \left [ \kT^{(1)}(x_0) + a\icT^{(1)}\tilde{\partial}_0\kV^{(0)}(x_0) \right ] + \mathcal{O}(ag_0^4) \,,
\end{gather}
with $\icT(g_0^2)=\icT^{(1)}g_0^2 + \mathcal{O}(g_0^4)$.
The renormalization constant for the improved tensor correlator $\kT^{\rm I}$ at one-loop is then given by
\begin{gather}
\begin{split}
&\ZT^{(1)}(a/L)  = \\ &-  \Bigg\{ \frac{1}{\kT^{(0)}(T/2)} \left[ \kT^{(1)}(T/2) + \icttil^{(1)}{\kT}_{\rm ;bi}^{(0)}(T/2) + \mcrit^{(1)}\frac{\partial \kT^{(0)}(T/2)}{\partial m_0} + \icT^{(1)} \tilde{\partial}_0 \kV^{(0)}(T/2) \right]  \\
&-\left(\frac{1}{2}-\alpha\right) \frac{1}{f_1^{(0)}} \left[ f_1^{(1)} + \icttil^{(1)}f_{1;{\rm bi}}^{(0)} + \mcrit^{(1)}\frac{\partial f_1^{(0)}}{\partial m_0} \right]\\ 
&-\alpha \frac{1}{k_1^{(0)}} \left[ k_1^{(1)} + \icttil^{(1)}k_{1;{\rm bi}}^{(0)} + \mcrit^{(1)}\frac{\partial k_1^{(0)}}{\partial m_0} \right]
  \Bigg\}
\end{split}
\end{gather}
where $\icttil$ is the coefficient of the counterterm that subtracts the $\Oa$ contribution coming from the fermionic action at the boundaries, and $a\mcrit^{(1)}$ is the one-loop value of the critical mass,
for which we employ the continuum values of $a\mcrit^{(1)}$ from~\cite{Palombi:2005zd,ss1996}.
The one-loop value of the improvement coefficient $\icT$ has been obtained using
SF techniques in~\cite{Sint:1997dj}.
We have repeated the computation of this latter quantity as a crosscheck of our perturbative
setup; a summary is provided in Appendix~\ref{app:cT}.

\begin{figure}[t!]
\begin{center}
\includegraphics[width=80mm]{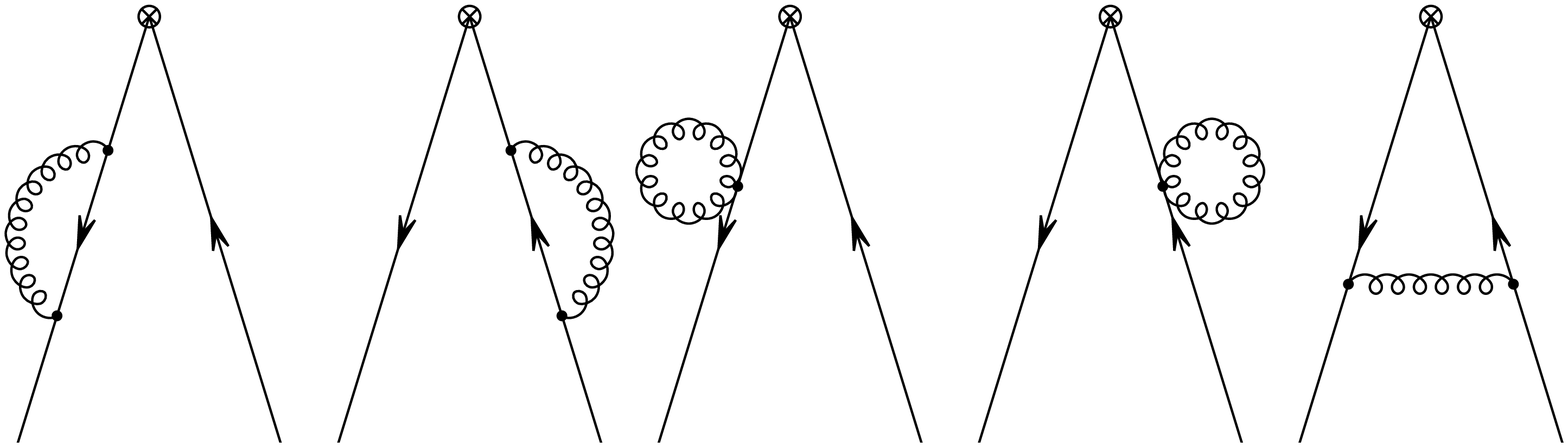}
\vspace{-2mm}
\label{fig:2f_mod1}
\caption{One-loop diagrams for boundary-to-bulk correlators.}
\end{center}
\begin{center}
\includegraphics[width=80mm]{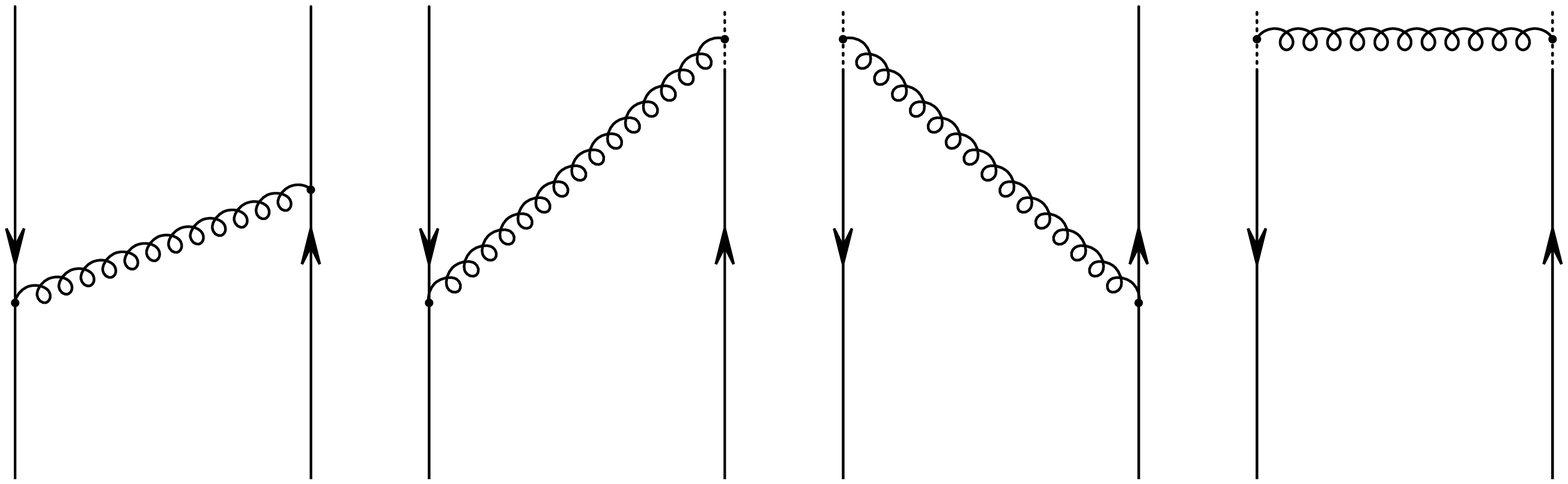}
\vspace{-2mm}
\end{center}
\begin{center}
\includegraphics[width=80mm]{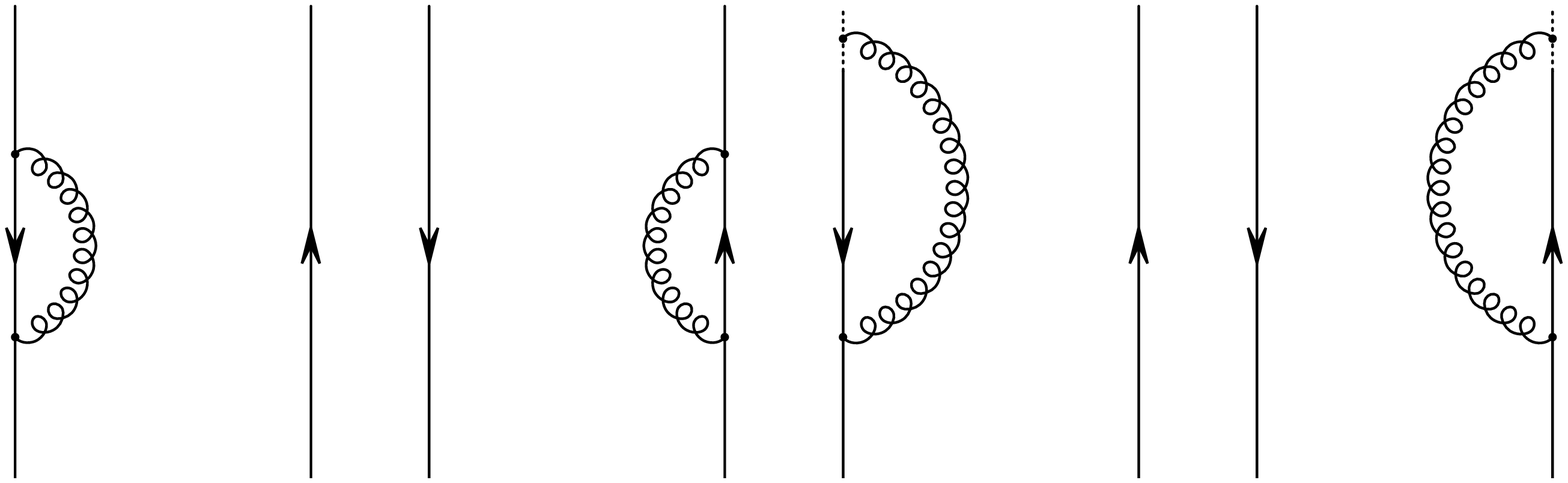}
\vspace{-2mm}
\end{center}
\begin{center}
\includegraphics[width=80mm]{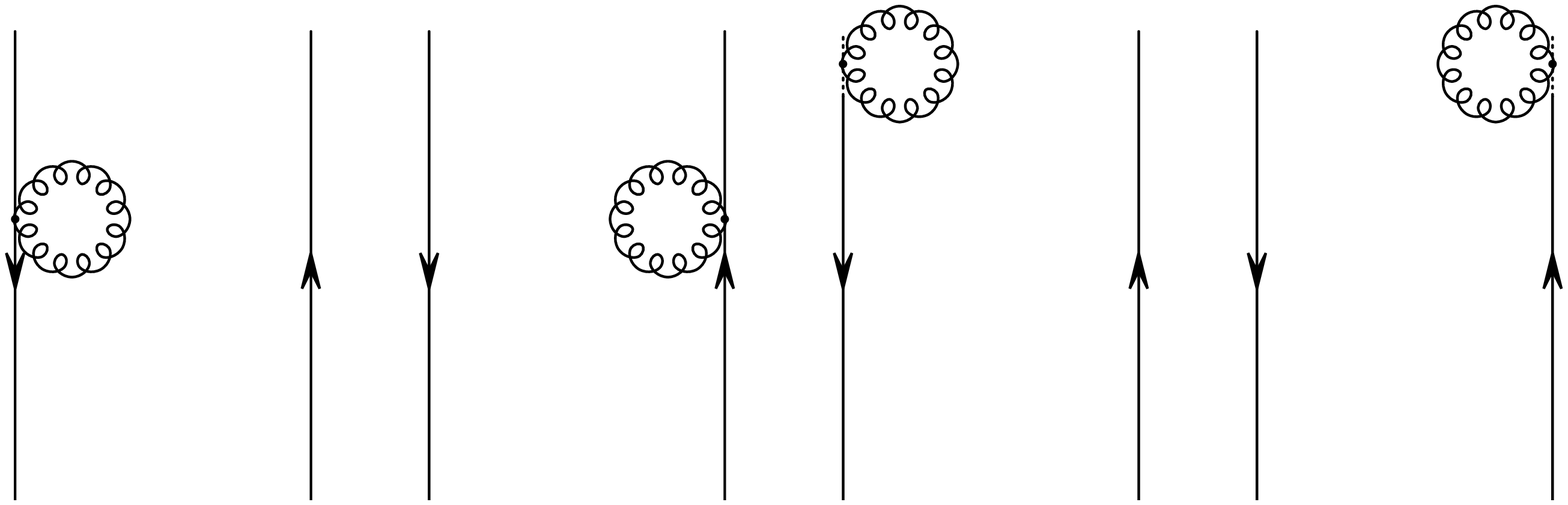}
\vspace{-2mm}
\label{fig:2f_mod2}
\caption{One-loop diagrams for boundary-to-boundary correlators.}
\end{center}
\end{figure}

\subsection{Perturbative scheme matching}

Any two mass-independent renormalization schemes (indicated by primed and unprimed
quantites, respectively) can be related by a finite parameter and operator
renormalization of the form 
\begin{align}
\gbar'^2 &= \chi_g(\gbar)\gbar^2, \label{gp} \\
\mbar'_i &= \chi_m(\gbar)\mbar_i , \quad i=1,\dots,\NF, \label{mp} \\
\obar{O}' &= \chi_O(\gbar)\obar{O}  \,, \label{op}
\end{align}
where we have assumed $O$ to be multiplicatively renormalizable.
The scheme change factors $\chi$ can be expanded perturbatively as 
\begin{gather}
\chi(g) \stackrel{g \sim 0}{\approx} 1 + \chi^{(1)}g^2 + \mathcal{O}(g^4)\,.
\end{gather}
Plugging Eqs.~(\ref{gp}, \ref{mp}, \ref{op}) into the Callan-Symanzik equations allows to relate a change in a renormalized quantity to the change in the corresponding RG function, viz.
\begin{align}
\beta'(g') &= \left [ \beta(g) \frac{\partial g'}{\partial g} \right ]_{g=g(g')} \,, \\
\tau'(g') &= \left [ \tau(g) + \beta(g) \frac{\partial }{\partial g} \log(\chi_m(g)) \right ]_{g=g(g')} \,,\\
\gamma'(g') &= \left [ \gamma(g) + \beta(g) \frac{\partial }{\partial g} \log(\chi_O(g)) \right ]_{g=g(g')}\,. \label{changegamma}
\end{align}
In particular, expanding Eq.~(\ref{changegamma}) to order $g^2$ provides a useful relation between the 2-loop coefficient of the anomalous dimension in the two schemes, viz.
\begin{gather}
\label{matching}
\gamma'_1=\gamma_1 + 2b_0\chi_O^{(1)}-\gamma_0\chi_g^{(1)}.
\end{gather}

The one-loop matching coefficient $\chi_g^{(1)}$ for the SF coupling
was computed in~\cite{Luscher:1993gh,Sint:1995ch},
\begin{gather}
 \chi_g^{(1)} =2b_0\log(L\mu) - \frac{1}{4\pi}(c_{1,0}+c_{1,1}\NF) \,,
\end{gather}
where the logarithm vanishes with our choice $\mu=1/L$,
and for the standard definition of
the SF coupling one has
\begin{gather}
c_{1,0}=1.25563(4)\, \quad \quad c_{1,1}=0.039863(2) \,.
\end{gather}
The other finite term $\chi_O$ in Eq.~(\ref{changegamma}) will provide
the operator matching between the lattice-regulated SF scheme and some
reference scheme where the NLO anomalous dimension is known, such as $\MSbar$
or RI, that we will label as ``cont''.
The latter usually are based on variants of the dimensional regularization
procedure; our SF schemes will be, on the other hand, regulated by a lattice.
The practical application of Eq.~(\ref{matching}) thus involves a two-step
procedure, in which the lattice-regulated SF scheme is first matched to a lattice-regulated
continuum scheme, that is in turned matched to the dimensionally-regulated continuum
scheme. This yields
\begin{gather}
[\chi_O^{(1)}]_{\rm SF;cont} = [\chi_O^{(1)}]_{\rm SF;lat} - [\chi_O^{(1)}]_{\rm cont;lat}\,. 
\end{gather} 
The one-loop matching coefficients $[\chi_O^{(1)}]_{\rm cont;lat}$ that we need can be extracted from the literature \cite{Capitani:2000xi,Skouroupathis:2008mf,Gracey:2003mr}, while the term $[\chi_O^{(1)}]_{\rm SF;lat}$ is obtained from our one-loop calculation of renormalization constants. Indeed, the asymptotic expansion for the one-loop coefficient of a renormalization constant in powers and logarithms of the
lattice spacing $a$ has the form
\begin{gather}
\label{Zasym}
Z^{(1)}(L/a) = \sum_{n \geq 0} \left ( \frac{a}{L} \right )^n \{ r_n + s_n \log (L/a) \}\,, 
\end{gather}
where $s_0=\gamma_{\rm\scriptscriptstyle T}^{(0)}$ and the finite part
surviving the continuum limit is the matching factor we need,
\begin{gather}
[\chi_0^{(1)}]_{\rm SF;lat} = r_0 \,.
\end{gather}

Our results for $[\chi_0^{(1)}]_{\rm SF;lat}$ have been obtained by computing
the one-loop renormalization constants on a series of lattices of sizes ranging
from $L/a=4$ to $L/a=48$, and fitting the results to Eq.~(\ref{Zasym})
to extract the expansion coefficients. The computation has been carried out with
$\Oa$ improved fermions for three values of $\theta$ for each scheme,
and without $\Oa$ improvement for $\theta=0.5$, which allows for a crosscheck of our
computation and of the robustness of the continuum limit (see below).
The results for the matching factors are provided in Table \ref{finite_parts};
details about the fitting procedure and the assignment of uncertainties are discussed in Appendix~\ref{app:fitpt}.

Inserting our results in Eq.~(\ref{matching}), we computed for the first time the NLO anomalous dimension in our family of SF schemes for the tensor currents, which are given in Table~\ref{gammaTNLO}.
We have crosschecked the computation by performing the matching with and without
$\Oa$ improvement, and proceeding through both $\MSbar$ and ${\rm RI}$ as reference
continuum schemes, obtaining the same results in all cases.
In this context we observe that the NLO correction to the running
is in general fairly large.
It is also worth mentioning that the choice of $\theta=0.5$, which leads to a
close-to-minimal value of the NLO mass anomalous dimension in SF schemes analogous
to the ones considered here~\cite{Sint:1998iq}, is not the optimal choice for the tensor current.
We will still use $\theta=0.5$ in the non-perturbative computation, since
our simulations were set up employing the optimal value for quark mass renormalization.

Finally, as already mentioned in the introduction, parallel to our work
Dalla Brida, Sint and Vilaseca have performed a related, fully independent
perturbative study as part of the setup of the chirally rotated Schr\"odinger
Functional~\cite{Brida:2016rmy}. Their results for the one-loop matching factors
$[\chi_O^{(1)}]_{\rm SF;lat}$ are perfectly consistent with ours, providing
a very strong crosscheck.

\begin{table}[t!]
\begin{center}
\noindent\begin{tabular}{cccc}
\Hline
$\theta$ & $\alpha$ & $r_{0;\rm SF}^{\alpha;\theta} \, \, {\rm (\icsw=0)}$ & $r_{0;\rm SF}^{\alpha;\theta} \, \, {\rm (\icsw=1)}$ \\
\Hline
\multirow{2}*{0.0} 
& 0 & n/a & $-0.0198519(3) \times \CF $\\ 
& 1/2 & n/a & $-0.0198519(3) \times \CF$\\ 
\hline
\multirow{2}*{0.5} 
& 0 &  $-0.096821(5) \times \CF $ &$-0.05963(4) \times \CF $\\ 
& 1/2 & $-0.099979(5) \times \CF $&$ -0.06279(4) \times \CF $\\
\hline
\multirow{2}*{1.0} 
& 0 & n/a & $-0.0827(2) \times \CF $\\ 
& 1/2 & n/a &  $-0.0866(2) \times \CF$ \\ 
\Hline
\end{tabular}
\end{center}
\caption{Finite parts of one-loop renormalization constants in the scheme specified by the parameters $\theta$ and $\alpha$ for the unimproved and $\Oa$-improved fermion actions}.
\label{finite_parts}
\end{table}

\begin{table}[t!]
\begin{center}
\noindent\begin{tabular}{cccc}
\Hline
$\theta$ & $\alpha$ & $\gamma_{\rm\scriptscriptstyle T;SF}^{(1)}$ & $\gamma_{\rm\scriptscriptstyle T;SF}^{(1)}/\gamma_{\rm\scriptscriptstyle T}^{(0)}$\\
\Hline
\multirow{2}*{0.0} 
& 0 & $0.0143209(6) - 0.00067106(3) \times \NF  $& $0.84805(3) - 0.0397383(2) \times \NF $\\ 
& 1/2 & $0.0143209 (6)- 0.00067106(3) \times \NF $& $0.84805(3) - 0.0397383(2) \times \NF $\\ 
\hline
\multirow{2}*{0.5} 
& 0 & $0.0069469(8) - 0.00022415(5) \times \NF $& $0.41138(5) - 0.013273(6) \times \NF $\\ 
& 1/2 & $0.0063609(8) - 0.00018863(5) \times \NF  $& $0.37668(5) - 0.011170(6) \times \NF$ \\
\hline
\multirow{2}*{1.0} 
& 0 & $0.00266(3) + 0.000036(2) \times \NF $& $0.157(2) + 0.0021(1) \times  \NF $\\ 
& 1/2 & $0.00192(3) + 0.000081(2) \times \NF $& $0.114(2) + 0.0048(1) \times  \NF $\\  
\Hline
\end{tabular}
\end{center}
\caption{NLO anomalous dimensions for various SF schemes, labeled by the parameters
$\theta$ and $\alpha$. 
The ratio to the LO anomalous dimension is also provided, as an indicator of the
behaviour of the perturbative expansion. 
For comparison, $\gamma_{{\rm\scriptscriptstyle T};\overline{\rm\scriptscriptstyle MS}}^{(1)}/\gamma_{\rm\scriptscriptstyle T}^{(0)}=0.1910-0.091 \times \NF$}.
\label{gammaTNLO}
\end{table}

\subsection{One-loop cutoff effects in the step scaling function}

\begin{figure}[t!]
\begin{center}
\begin{minipage}[h!]{1.0\textwidth}
\begin{center}
\includegraphics[width=110mm]{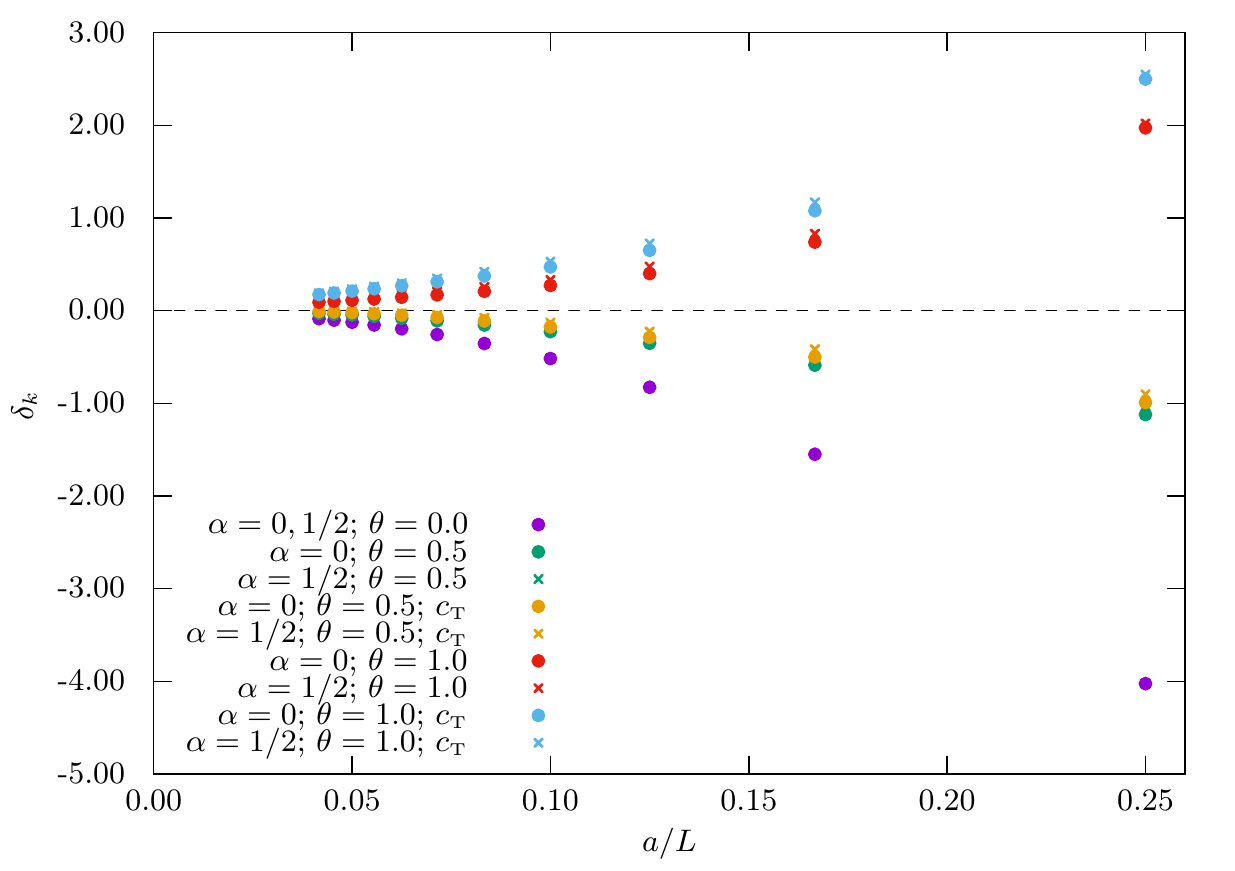}
\includegraphics[width=110mm]{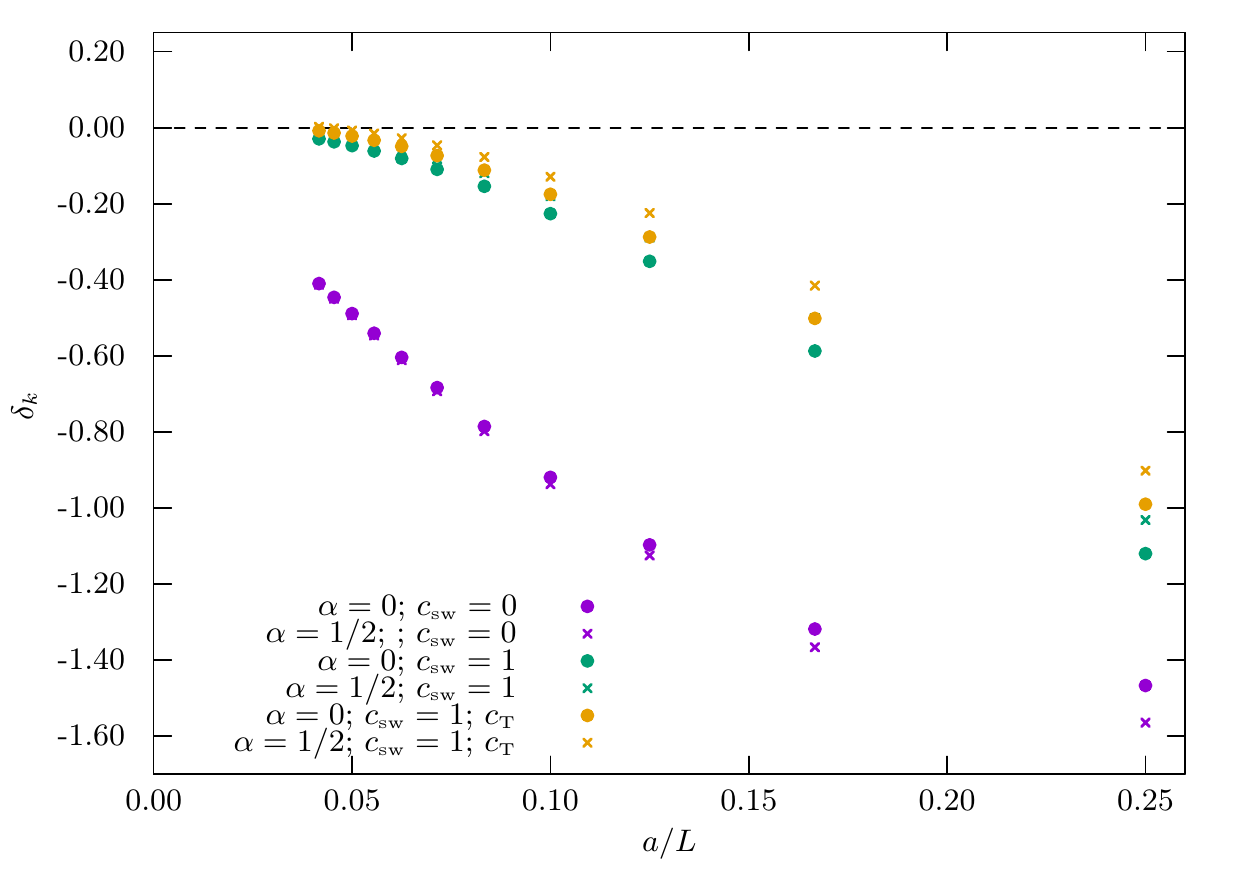}
\vspace{-5mm}
\end{center}
\end{minipage}
\end{center}
\caption{Upper panel: cutoff effects as a function of $a/L$ for the various schemes
considered and the $\Oa$ improved fermion action. Results with and without operator
improvement are provided.
Lower panel: zoom-in displaying only results for the schemes with $\theta=0.5$
(which will be the one employed in the non-perturbative computation),
also including those with an unimproved fermion action.
}
\label{fig:cutoff}
\end{figure}

As mentioned above, the RG running is accessed via SSFs, defined in Eq.~(\ref{Sigma}).
It is thus both interesting and useful to study the scaling of $\SigmaT$ within
perturbation theory.
Plugging the one-loop expansion of the renormalization constant in Eq.~(\ref{Sigma}), we
obtain an expression of the form
\begin{gather}
\SigmaT(u,L/a)=1 + k(L/a)\gbar^2 + \mathcal{O}(\gbar^4) \,,
\end{gather}
where 
\begin{gather}
k(L/a)=\ZT^{(1)}(2L/a) - \ZT^{(1)}(L/a) \,.
\end{gather}
In order to extract the cutoff effect which quantifies how fast the continuum limit
$\sigmaT$ is approached, we define 
\begin{gather}
k(\infty)=\gamma_{\rm\scriptscriptstyle T}^{(0)}\log(2) \,,
\end{gather}
and the relative cutoff effect $\delta_k$
\begin{gather}
\delta_k(L/a) = \frac{k(L/a)}{k(\infty)} - 1 \,.
\label{eq:deltak}
\end{gather}
The one-loop values of $\delta_k$ for both the improved and unimproved renormalization conditions are listed in
Table~\ref{tabcutoff}. The behaviour of $\delta_k$
as a function of the lattice size
is shown in 
Fig.~\ref{fig:cutoff}. The figure shows that the bulk of the linear cutoff effect is removed
by the improvement of the action, and that the improvement of the current has a comparatively small 
impact. Note also that $\theta=0.5$ leads to the smaller perturbative cutoff effects among
the values explored, cf. Table~\ref{tabcutoff}.

\section{Non-perturbative computations}

We will now present non-perturbative results for both $\NF=0$ and $\NF=2$ QCD.
The simulations underlying each of the two cases are those in~\cite{Guagnelli:2005zc} (which
in turn reproduced and extended the simulations in~\cite{Capitani:1998mq})
and~\cite{DellaMorte:2005kg}, respectively.
For $\NF=2$ simulations are performed with non-perturbatively $\Oa$ improved
Wilson fermions, whereas in the quenched case the computation was performed
both with and without $\Oa$ improvement, which, along with the finer lattices used,
allows for a better control of the continuum limit (cf. below).
A gauge plaquette action is always used.
In both cases, we rely on the computation
of the SF coupling and its non-perturbative running, given in \cite{Luscher:1993gh,Capitani:1998mq}
for $\NF=0$ and \cite{DellaMorte:2004bc} for $\NF=2$.

\subsection{$\NF=0$}
Simulation details for the quenched computation are given in~\cite{Guagnelli:2005zc}.
Simulation parameters have been determined by tuning $\beta$ such that the value of the
renormalized SF coupling is kept constant with changing $L/a$, and fixing the bare
quark mass to the corresponding non-perturbatively tuned value of $\hopc$.
A total of fourteen values of the renormalized coupling have been considered,
namely, $u=\{0.8873,$ $0.9944,$ $1.0989,$ $1.2430,$ $1.3293,$ $1.4300,$ $1.5553,$ $1.6950,$ $1.8811,$ $2.1000,$ $2.4484,$ $2.7700,$ $3.1110,$ $3.4800\}$,
corresponding to fourteen different physical lattice lengths $L$.
In all cases the renormalization constants $\ZT$ are determined, in the two schemes given by
$\alpha=0,1/2$, on lattices of sizes $L/a=\{6,8,12,16\}$ and $2L/a=\{12,16,24,32\}$,
which allows for the determination of $\SigmaT(u,a/L)$ at four values of the lattice spacing.

As mentioned above,
two separate computations have been performed, with and without an $\Oa$ improved
fermion action with a non-perturbatively determined $\icsw$ coefficient.\footnote{The SF boundary improvement counterterms proportional to $\ict$ and $\icttil$ are taken into account at two-
and one-loop order in perturbation theory, respectively.}
This allows to improve our control over the continuum limit extrapolation
for $\sigmaT$, by imposing a common result for both computations based on universality.
It is important to note that the gauge ensembles for the improved and unimproved
computations are different, and therefore the corresponding results are fully
uncorrelated.
Another important observation is that the $\icT$ coefficient
for the $\Oa$ improvement counterterm of the tensor
current is not known non-perturbatively, but only to leading order
in perturbation theory. In our computation of $\ZT$ for $\NF=0$ we have thus never
included the improvement counterterm in the renormalization condition,
even when the action is improved, and profit only from the above universality
constraint to control the continuum limit, as we will discuss in detail below.
The resulting numerical values of the renormalization constants and SSFs are reported in Tables \ref{renormalization_constantnf0A}~and~\ref{renormalization_constantnf0B}.

\subsubsection{Continuum extrapolation of SSFs}

As discussed above, the continuum limit for $\SigmaT$ is controlled by
studying the scaling of the results obtained with and without an $\Oa$ improved
actions. To that respect, we first check that universality holds within our
precision, by performing independent continuum extrapolations of both datasets.
Given the absence of the $\icT$ counterterm, we always
assume that the continuum limit is approached linearly in $a/L$, and
parametrize
\begin{gather}
\label{sigmaTextrap}
\SigmaT^{\icsw=0}(u,a/L)=\sigmaT^{\icsw=0}(u) + \rho_{\rm\scriptscriptstyle T}^{\icsw=0}(u)\frac{a}{L} \,,\\
\SigmaT^{\icsw={\rm NP}}(u,a/L)=\sigmaT^{\icsw={\rm NP}}(u) + \rho_{\rm\scriptscriptstyle T}^{\icsw={\rm NP}}(u)\frac{a}{L} \,.
\end{gather}
We observe that, in general, fits that drop the coarsest lattice,
corresponding to the step $L/a=6 \rightarrow 12$, are of better quality;
when the $\SigmaT(L/a=6)$ datum is dropped,
$\sigmaT^{\icsw=0}(u)$ and $\sigmaT^{\icsw={\rm NP}}(u)$
always agree within $\sim 1\sigma$. The slopes
$\rho_{\rm\scriptscriptstyle T}^{\icsw={\rm NP}}(u)$ are systematically
smaller than $\rho_{\rm\scriptscriptstyle T}^{\icsw=0}(u)$, showing that
the bulk of the leading cutoff effects in the tensor current is subtracted
by including the Sheikholeslami-Wohlert (SW) term in the action.

We thus proceed to obtain our best estimate for $\sigmaT(u)$ from a constrained
extrapolation, where we set $\sigmaT^{\icsw=0}(u)=\sigmaT^{\icsw={\rm NP}}(u)=\sigmaT(u)$
in Eq.~(\ref{sigmaTextrap}), and drop the $L/a=6 \rightarrow 12$ step from the fit.
The results for both schemes are provided in
Table~\ref{tabcontlimnf0}, and illustrated in Figs.~\ref{clnf0f1},~\ref{clnf0k1}.

\subsubsection{Fits to continuum step-scaling functions}

In order to compute the RG running of the operator in the continuum limit, we fit
the continuum-extrapolated SSFs to a functional form in $u$. The simplest choice,
motivated by the perturbative expression for $\gamma_{\rm\scriptscriptstyle T}$ and $\beta$,
and assuming that $\sigmaT$ is a smooth function of the renormalized coupling
within the covered range of values of the latter, is a polynomial of the form
\begin{gather}
\sigmaT(u)=1+p_1u + p_2u^2 + p_3u^3+p_4u^4 + \dots \,.
\label{eq:ssfpoly}
\end{gather}
The perturbative prediction for the first two coefficients of \req{eq:ssfpoly} reads 
\begin{align}
p_1^{\rm\scriptscriptstyle pert}= & \gamma_{\rm\scriptscriptstyle T}^{(0)}\log(2) \,, \label{eq:p1} \\
p_2^{\rm\scriptscriptstyle pert}= & \gamma_{\rm\scriptscriptstyle T}^{(1)}\log(2) + \left [ \frac{(\gamma_{\rm\scriptscriptstyle T}^{(0)})^2}{2} + b_0\gamma_{\rm\scriptscriptstyle T}^{(0)} \right ] (\log(2))^2 \,.\label{eq:p2}
\end{align}
Note, in particular, that perturbation theory predicts a dependence on $\NF$ only
at $\cO(u^2)$.

\begin{figure}[!t]
\begin{center}
\includegraphics[width=69mm]{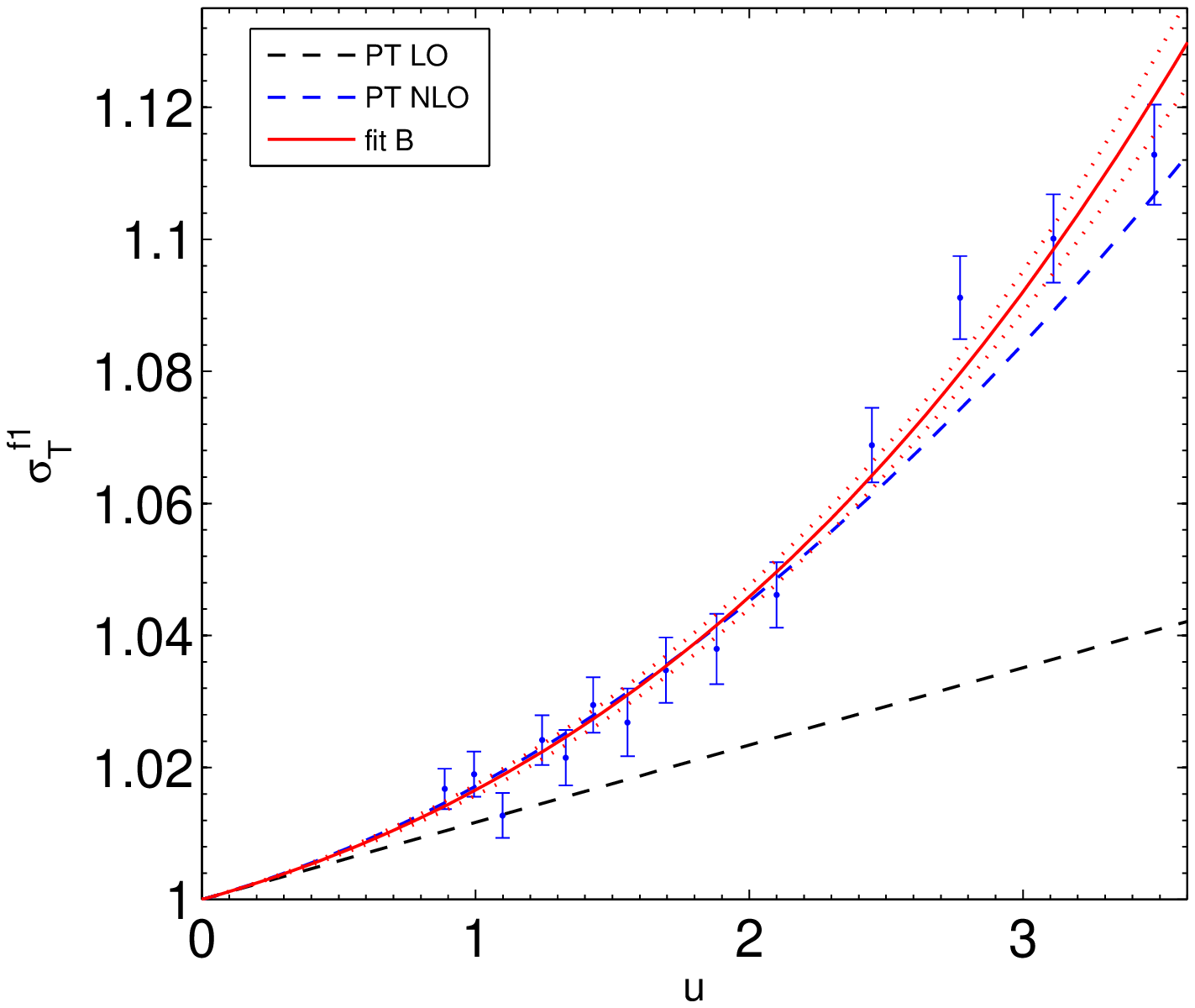}
\hspace{-1mm}
\includegraphics[width=69mm]{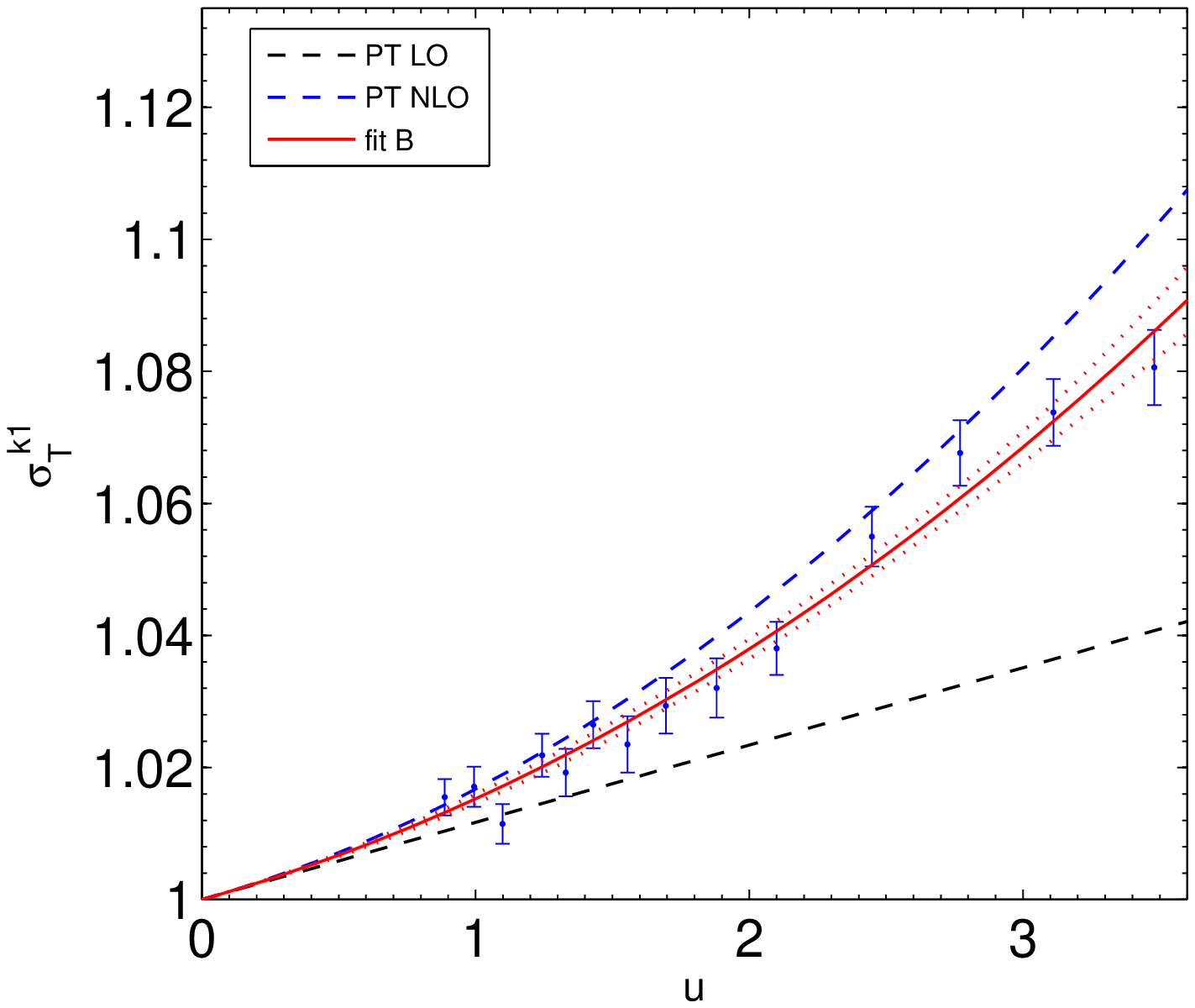}
\vspace{-10mm}
\end{center}
\caption{$\NF=0$ continuum-extrapolated SSFs in the schemes
$\alpha=0$ (left) and $\alpha=1/2$ (right), 
and their fitted functional forms following fit B in Table~\ref{fitssfnf0}.
The one- and two-loop perturbative predictions are also shown for comparison.}
\label{ssfnf0}
\end{figure}

We have considered various fit ans\"atze, exploring combinations of the order of the polynomial
and possible perturbative constraints, imposed by fixing either $p_1$ or both $p_1$ and $p_2$
to the values in Eqs.~(\ref{eq:p1},\ref{eq:p2}). We always take as input the results from the
joint $\icsw=0$ and $\icsw={\rm NP}$ extrapolation, discussed above.
The results for the various fits are shown in Table~\ref{fitssfnf0}.
All the fits result in a good description of the
non-perturbative data, with values of $\chi^2/{\rm d.o.f.}$ close to unity and
little dependence on the ansatz.
The coefficients of powers larger than $u^3$ are consistently compatible with zero
within one standard deviation.
We quote as our preferred fit the one that fixes $p_1$ to its perturbative
value, and reaches $\cO(u^3)$ (fit B in Table~\ref{fitssfnf0}). This provides
an adequate description of the non-perturbative data, without artificially decreasing
the goodness-of-fit by including several coefficients with large relative errors
(cf., e.g., fit E). The result for $\sigmaT$ from fit~B in our two schemes is
illustrated in Fig.~\ref{ssfnf0}.
It is also worth pointing out that the value for $p_2$ obtained
from fits A and B is compatible with the perturbative prediction within $1$ and $1.5$ standard
deviations, respectively, for the two schemes; this reflects the small observed 
departure of $\sigmaT$ from its two-loop value until the region $u \gtrsim 2$ is reached,
cf.~Fig.~\ref{ssfnf0}.

\subsubsection{Determination of the non-perturbative running factor}

Once a given fit for $\sigmaT$ is chosen, it is possible to compute the running between
two well-separated scales through a finite-size recursion. The latter is started
from the smallest value of the energy scale
$\mu_{\rm\scriptscriptstyle had}=L_{\rm\scriptscriptstyle had}^{-1}$, given by
the largest value of the coupling for which $\sigmaT$ has been computed, viz.
\begin{gather}
\gbar^2(2\mu_{\rm\scriptscriptstyle had})=3.48 \,.
\end{gather}
Using as input the coupling SSF $\sigma(u)$ determined in \cite{Capitani:1998mq},
we construct recursively the series of coupling values
\begin{gather}
u_{k+1} = \gbar^2(2^{k+2} \mu_{\rm had}) = \sigma^{-1}(u_k) \, , \quad u_0=3.48\,.
\end{gather}
This in turn allows to compute the product
\begin{gather}
U(\mu_{\rm\scriptscriptstyle had},\mu_{\rm\scriptscriptstyle pt})= \prod_{k=0}^n \sigmaT(u_k) \,, \quad \mu_{\rm\scriptscriptstyle pt}=2^{n+1}\mu_{\rm\scriptscriptstyle had}\,,
\end{gather}
where $U$ is the RG evolution operator in Eq.~(\ref{eq:rgevol}), here
connecting the renormalised operators at scales $\mu_{\rm\scriptscriptstyle had}$ and
$2^{n+1}\mu_{\rm\scriptscriptstyle had}$.
The number of iterations $n$ is dictated by the smallest value of $u$ at which
$\sigmaT$ is computed non-perturbatively, i.e. $u=0.8873$.
We find $u_7=0.950(11)$ and $u_8=0.865(10)$, corresponding respectively to $8$ and $9$ steps of recursion. The latter involves a short extrapolation
from the interval in $u$ covered by data, in a region where the SSF is strongly
constrained by its perturbative asymptotics. This point is used only to test the robustness of the recursion, 
but is not considered in the final analysis. The values of $u_k$ and the corresponding
running factors are given in Tables~\ref{tab:run_nf0_f1}~and~\ref{tab:run_nf0_k1}.

\begin{figure}
\begin{center}
\includegraphics[width=69mm]{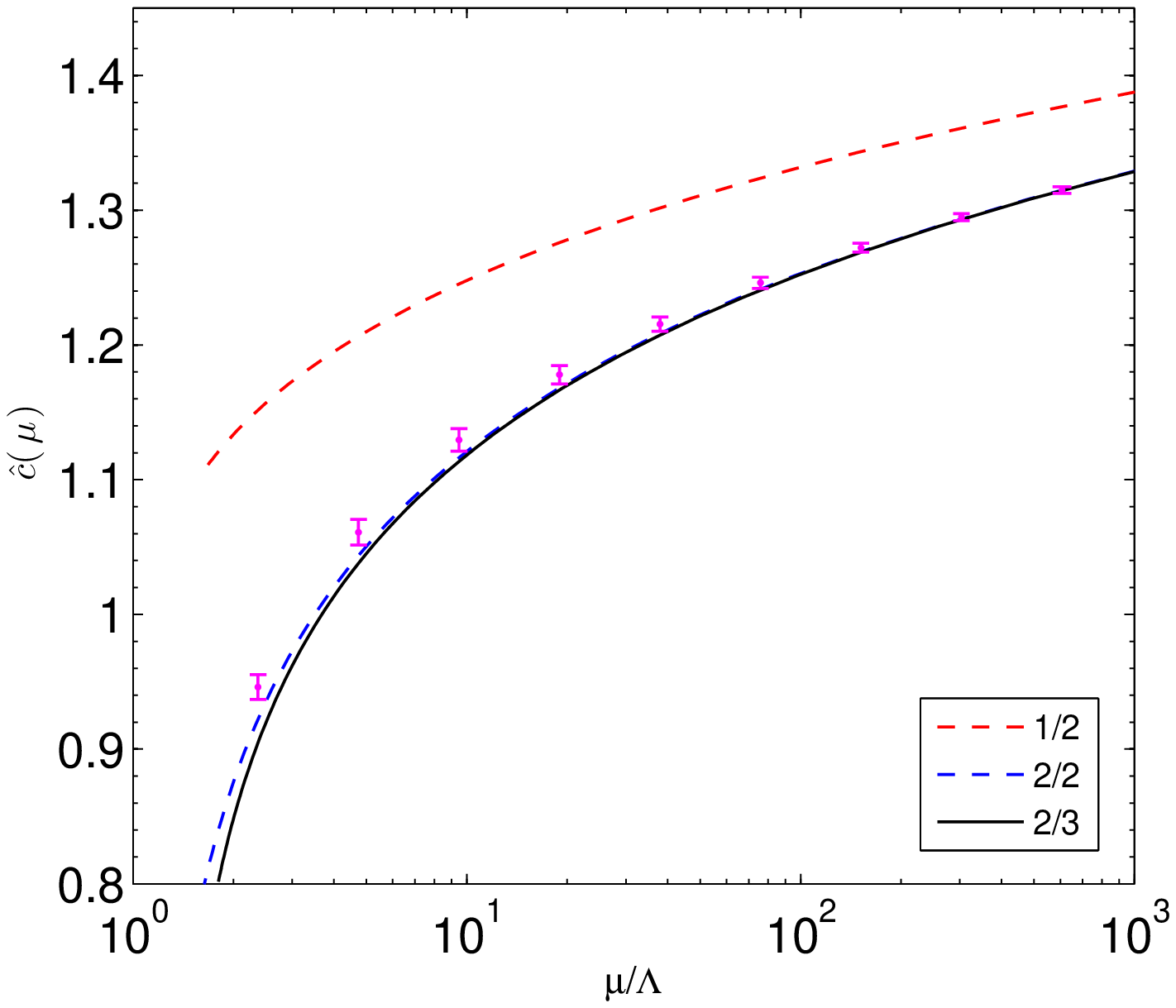}
\hspace{-2.0mm}
\includegraphics[width=69mm]{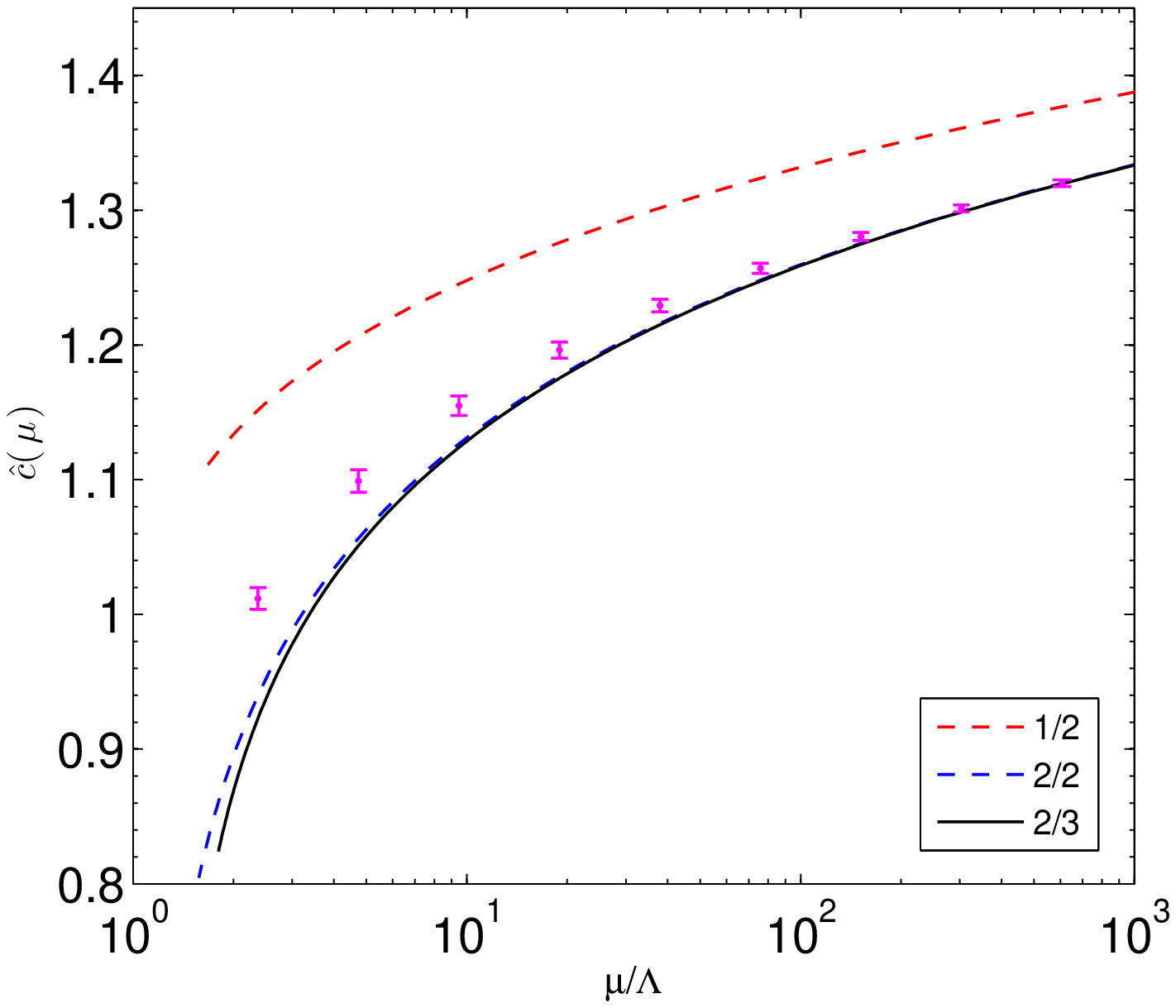}
\vspace{-5mm}
\end{center}
\caption{Running of the tensor current for $\NF=0$
in the schemes $\alpha=0$ (left) and $\alpha=1/2$ (right),
compared to perturbative predictions using the $1/2$-, $2/2$-, and $2/3$-loop
values for $\gamma_{\rm\scriptscriptstyle T}/\beta$.}
\label{figrunningnf0}
\end{figure}

Once $\mu_{\rm\scriptscriptstyle pt}=2^8\mu_{\rm\scriptscriptstyle had}$
has been reached, perturbation theory can be used to make contact with the RGI operator.
We thus compute the total running factor
$\orgi{c}(\mu)$ in \req{rgi_operator}
at $\mu=\mu_{\rm\scriptscriptstyle had}$ as
\begin{gather}
\orgi{c}(\mu_{\rm\scriptscriptstyle had}) =
\frac{\orgi{c}(\mu_{\rm\scriptscriptstyle pt})}{U(\mu_{\rm\scriptscriptstyle had},\mu_{\rm\scriptscriptstyle pt})}\,,
\end{gather}
where $\orgi{c}(\mu_{\rm\scriptscriptstyle pt})$ is computed using the highest available
orders for $\gamma$ and $\beta$ in our schemes (NLO and NNLO, respectively).
In order to assess the systematic uncertainty arising from the use of perturbation theory,
we have performed two crosschecks:
\begin{enumerate}
\item Perform the matching to perturbation theory at all the points in the recursion,
and check that the result changes within a small fraction of the error.
\item Match to perturbation theory using different combinations of perturbative
orders in $\gamma$ and $\beta$: other than our NLO/NNLO preferred choice, labeled
``$2/3$'' --- after the numbers of loops --- in Tables~\ref{tab:run_nf0_f1}~and~\ref{tab:run_nf0_k1},
we have used matchings at $1/2$-, $2/2$-, and $3/3$-loop order, where in the latter
case we have employed a mock value of the NNLO anomalous dimension given by
$\gamma^{(2)} \equiv (\gamma^{(1)})^2/\gamma^{(0)}$ as a means to have a guesstimate
of higher-order truncation uncertainties. 
\end{enumerate}
We thus quote as our final numbers
\begin{gather}
\begin{split}
\left.\orgi{c}(\mu_{\rm\scriptscriptstyle had})\right|_{\NF=0} & = 0.9461(95)\,, \quad \mbox{scheme~}\alpha=0\,;\\
\left.\orgi{c}(\mu_{\rm\scriptscriptstyle had})\right|_{\NF=0} & = 1.0119(83)\,, \quad \mbox{scheme~}\alpha=1/2\,.
\end{split}
\label{eq:rginf0}
\end{gather}
In Fig.~\ref{figrunningnf0} we plot the non-perturbative running of the operator
in our two schemes, obtained by running backwards from the perturbative matching point corresponding to the renormalized coupling
 $u_7=0.950(11)$. 
with our non-perturbative $\sigmaT$, and compare it with perturbation theory.
In order to set the physical scale corresponding to each value of the coupling,
we have used $\Lambda_{\rm\scriptscriptstyle SF}/\mu_{\rm\scriptscriptstyle had} = 0.422(32)$, from \cite{Capitani:1998mq}. The latter work also provides the
value of $\mu_{\rm\scriptscriptstyle had}$ in units of the Sommer scale $r_0$~\cite{Sommer:1993ce},
viz. $(2r_0\mu_{\rm\scriptscriptstyle had})^{-1}=0.718(16)$ --- which,
using $r_0=0.5~\fm$, translates into $\mu_{\rm\scriptscriptstyle had}=274(6)~\MeV$.
It is important to stress that the results in \req{eq:rginf0} are given
in the continuum, and therefore do not contain any dependence on the regularization
procedures employed to obtain them.

\subsubsection{Hadronic matching}

The final piece required for a full non-perturbative renormalization is to
compute renormalization constants at the hadronic scale $\mu_{\rm\scriptscriptstyle had}$
within the interval of values of the bare gauge coupling covered by non-perturbative
simulations in large, hadronic volumes. We have thus proceeded to obtain
$\ZT$ at four values of the bare coupling,
$\beta=\{6.0129$,$6.1628$,$6.2885$,$6.4956\}$, tuned to ensure that $L$ --- and hence the 
renormalized SF coupling --- stays constant when $L/a=\{8,10,12,16\}$, respectively.
The results, both with and without $\Oa$ improvement, are provided in 
Tables~\ref{renormalization_matchingnf0_f1}~and~\ref{renormalization_matchingnf0_k1}.
These numbers can be multiplied by the corresponding value of the running factor
in \req{eq:rginf0} to obtain the quantity
\begin{gather}
\hat{Z}_{\rm\scriptscriptstyle T}(g_0^2) = \orgi{c}(\mu_{\rm\scriptscriptstyle had})\ZT(g_0^2,a\mu_{\rm\scriptscriptstyle had})\,,
\end{gather}
which relates bare and RGI operators for a given value of $g_0^2$.
They are quoted in Table~\ref{RGInf0}; it is important to stress that
the results are independent of the scheme within the $\sim 1\%$ precision of our computation
--- as they should, since the scheme dependence is lost at the level of RGI operators,
save for the residual cutoff effects which in this case are not visible within errors.
A second-order polynomial fit to the dependence of the results in $\beta$
\begin{gather}
\hat{Z}_{\rm\scriptscriptstyle T}(g_0^2) = z_0 + z_1(\beta-6) + z_2(\beta-6)^2
\end{gather}
for the numbers obtained from the scheme $\alpha=1/2$, which turns out to
be slightly more precise, yields
\begin{gather}
\begin{split}
\icsw = {\rm NP}: & \quad
z_0 = 0.9814(9)\,, \quad
z_1 = 0.138(8)\,, \quad
z_2 = -0.06(2)\,; \\
\icsw = 0: & \quad
z_0 = 0.8943(4) \,, \quad
z_1 = 0.127(3) \,, \quad
z_2 = -0.024(6) \,,
\end{split}
\end{gather}
with correlation matrices among the fit coefficients
\begin{gather}
\begin{split}
C[\icsw={\rm NP}] & = \left(\ba{rrr}
  1.000 & -0.766 &  0.605 \\
 -0.766 &  1.000 & -0.955 \\
  0.605 & -0.955 &  1.000 \\
\ea\right)\,, \\[2.0ex]
C[\icsw=0] & = \left(\ba{rrr}
  1.000 & -0.768 &  0.615 \\
 -0.768 &  1.000 & -0.960 \\
  0.615 & -0.960 &  1.000 \\
\ea\right)\,.
\end{split}
\end{gather}
These continuous form can be obtained to renormalize bare matrix elements,
computed with the appropriate action, at any convenient value of $\beta$.


\subsection{$\NF=2$}

In this case all our simulations were performed using an $\Oa$ improved Wilson action, with the SW coefficient $\icsw$ determined in\cite{Jansen:1998mx}.
Renormalization constants have been computed at six different values of the SF renormalized coupling 
$u=\{0.9703$,$1.1814$,$1.5078$,$2.0142$,$2.4792$,$3.3340\}$, corresponding to six different physical lattice lengths $L$.
For each physical volume, three different values of the lattice spacing have been simulated, corresponding to lattices with $L/a=6,8,12$ and the double steps $2L/a=12,16,24$, for the computation of the renormalization constant $\ZT(g_0,a/L)$. 
All simulational details, including those referring to the tuning of $\beta$ and
$\kappa$, are provided in \cite{DellaMorte:2005kg}.

Concerning $\Oa$ improvement, the configurations at the three weaker values of the coupling were 
produced using the one-loop perturbative estimate of $\ict$~\cite{Luscher:1992an},
while for the three stronger couplings the two-loop value \cite{Bode:1999sm} was used.
In addition, for $L/a=6$, $\beta=7.5420$ and $L/a=8$, $\beta=7.7206$ separate simulations
were performed with the one- and two-loop value of $\ict$, which results in two
different, uncorrelated ensembles, with either value of $\ict$, being available for $u=1.5078$. 
For $\icttil$ the one-loop value is used throughout.
Finally, since, contrary to the quenched case, we do not have two separate (improved and unimproved)
sets of simulations to control the continuum limit, we have included in our analysis
the improvement counterterm to the tensor current, with the one-loop value of $\icT$ \cite{Sint:1997dj}.

The resulting values for the renormalization constants $\ZT$ and the SSF $\SigmaT$ are listed in Table \ref{tabZTnf2}. The estimate of autocorrelation times has been computed using the ``Gamma Method'' of~\cite{Wolff:2003sm}.

\subsubsection{Continuum extrapolation of SSFs}

In this case, our continuum limit extrapolations will assume an $\Oasq$ scaling of
$\SigmaT$. This is based on the fact that we implement $\Oa$ improvement of the action
(up to small $\cO(ag_0^4)$ effects in $\icttil$ and $\cO(ag_0^4)$ or $\cO(ag_0^6)$ in $\ict$,
cf. above); and that the residual $\cO(ag_0^4)$ effects associated to the use of the
one-loop perturbative value for $\icT$ can be expected to be small, based on the findings
discussed above for $\NF=0$. Our ansatz for a linear extrapolation in $a^2$ is thus of the form
\begin{gather}
\SigmaT(u,a/L) = \sigmaT(u) + \rho_{\rm\scriptscriptstyle T}(u)\left(\frac{a}{L}\right)^2 \,.
\end{gather}
Furthermore, in order to ameliorate the scaling we subtract the leading perturbative
cutoff effects that have been obtained in Sec.~\ref{sec:sfpt}, by rescaling our data
for $\SigmaT$ as
\begin{gather}
\SigmaT'(u,a/L)=\frac{\SigmaT(u,a/L)}{1+u\delta_k(a/L)\gamma_{\rm\scriptscriptstyle T}^{(0)}\log(2)} \,,
\end{gather}
where the values of the relative cutoff effects $\delta_k(a/L)$ are taken from Table \ref{tabcutoff}. Continuum extrapolations are performed both taking $\SigmaT$ and
the one-loop improved $\SigmaT'$ as input; the two resulting continuum limits
are provided in Tables~\ref{tabcontlimnf2NOimp}
and~\ref{tabcontlimnf2SIimp}, respectively. As showed in Fig.~\ref{contlimnf0k1}, the effect of 
including the perturbative improvement is in general non-negligible only for our coarsest
$L/a=6$ lattices. The slope of the continuum extrapolation is decreased by subtracting
the perturbative cutoff effects at weak coupling, but for $u \gtrsim 2$ the quality
of the extrapolation does not change significantly, and the slope actually flips sign.
The $u=1.5078$ case is treated separately, and a combined extrapolation to the continuum value is 
performed using the independent simulations carried out with the two different values of
$\ict$. We quote as our best results the extrapolations obtained from $\SigmaT'$.

\subsubsection{Fits to continuum step-scaling functions}

\begin{figure}
\begin{center}
\includegraphics[width=70mm]{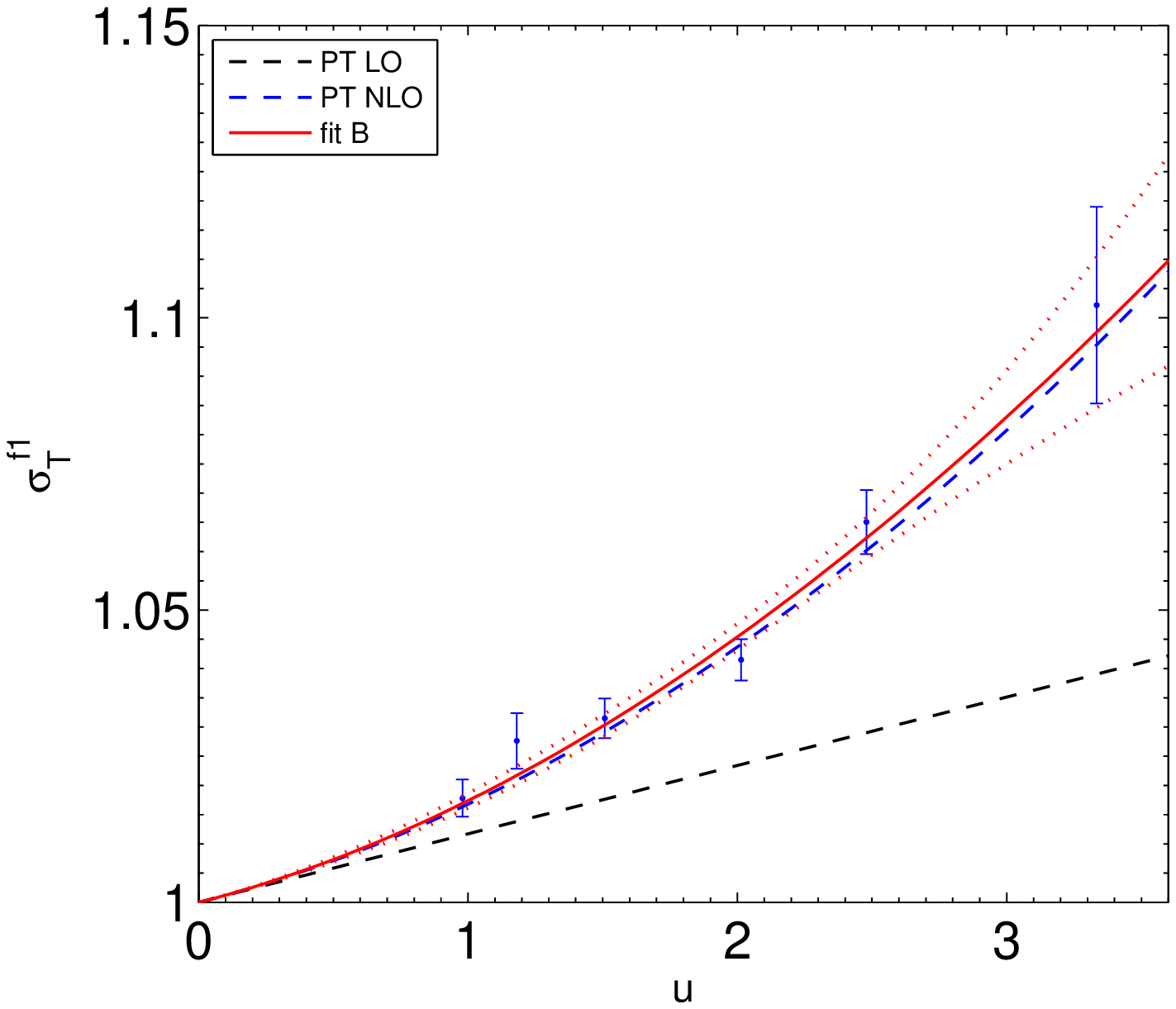}
\hspace{-3mm}
\includegraphics[width=70mm]{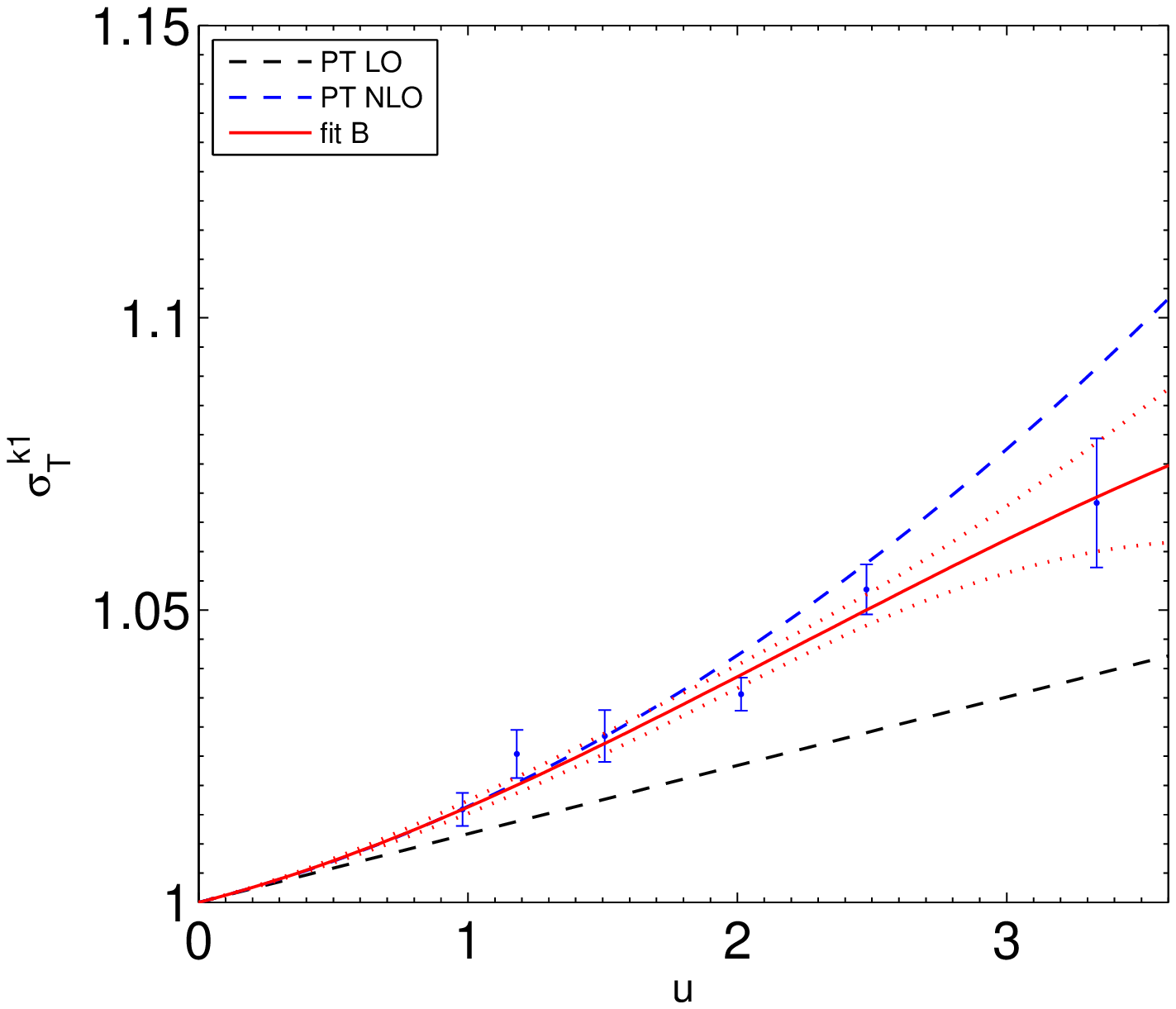}
\vspace{-10mm}
\end{center}
\caption{SSF for $\NF=2$ in the scheme $\alpha=0$ (left) and $\alpha=1/2$ (right),
compared with the LO and NLO perturbative predictions.}
\label{ssfnf2}
\end{figure}

Here we follow exactly the same strategy described above for $\NF=0$,
again considering several fit ans\"atze by varying the combination of the order of the polynomial 
and the number of coefficients fixed to their perturbative values. The results are listed in 
Table~\ref{fitssfnf2}. As in the quenched case, we quote as our preferred result
the fit obtained by fixing the first coefficient to its perturbative value and fitting
through $\mathcal{O}(u^3)$ (fit B). The resulting fit, as well as its comparison
to perturbative predictions, is illustrated in Fig.~\ref{ssfnf2}. 

\subsubsection{Non-perturbative running}

\begin{figure}
\begin{center}
\includegraphics[width=69mm]{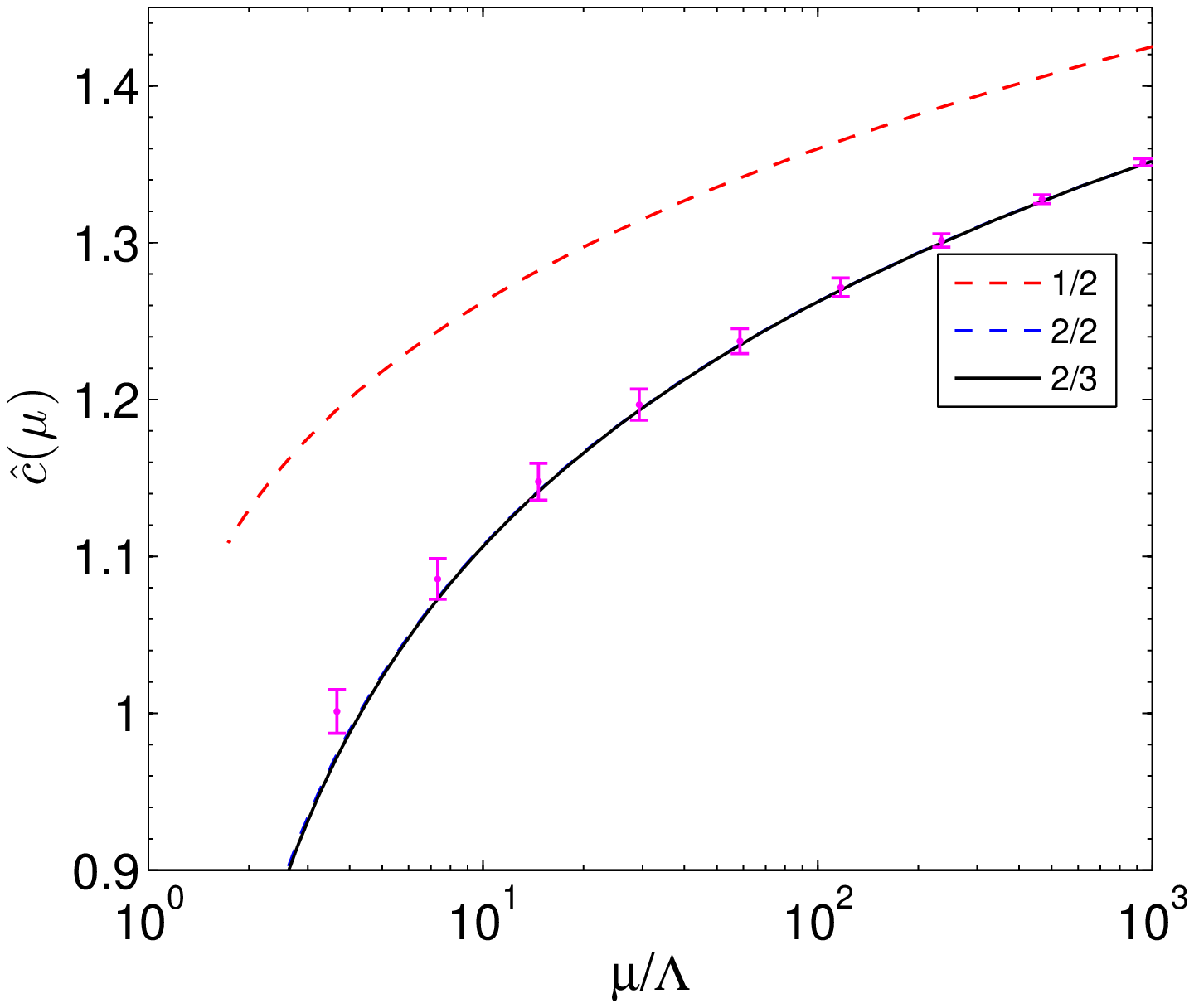}
\hspace{-2.0mm}
\includegraphics[width=69mm]{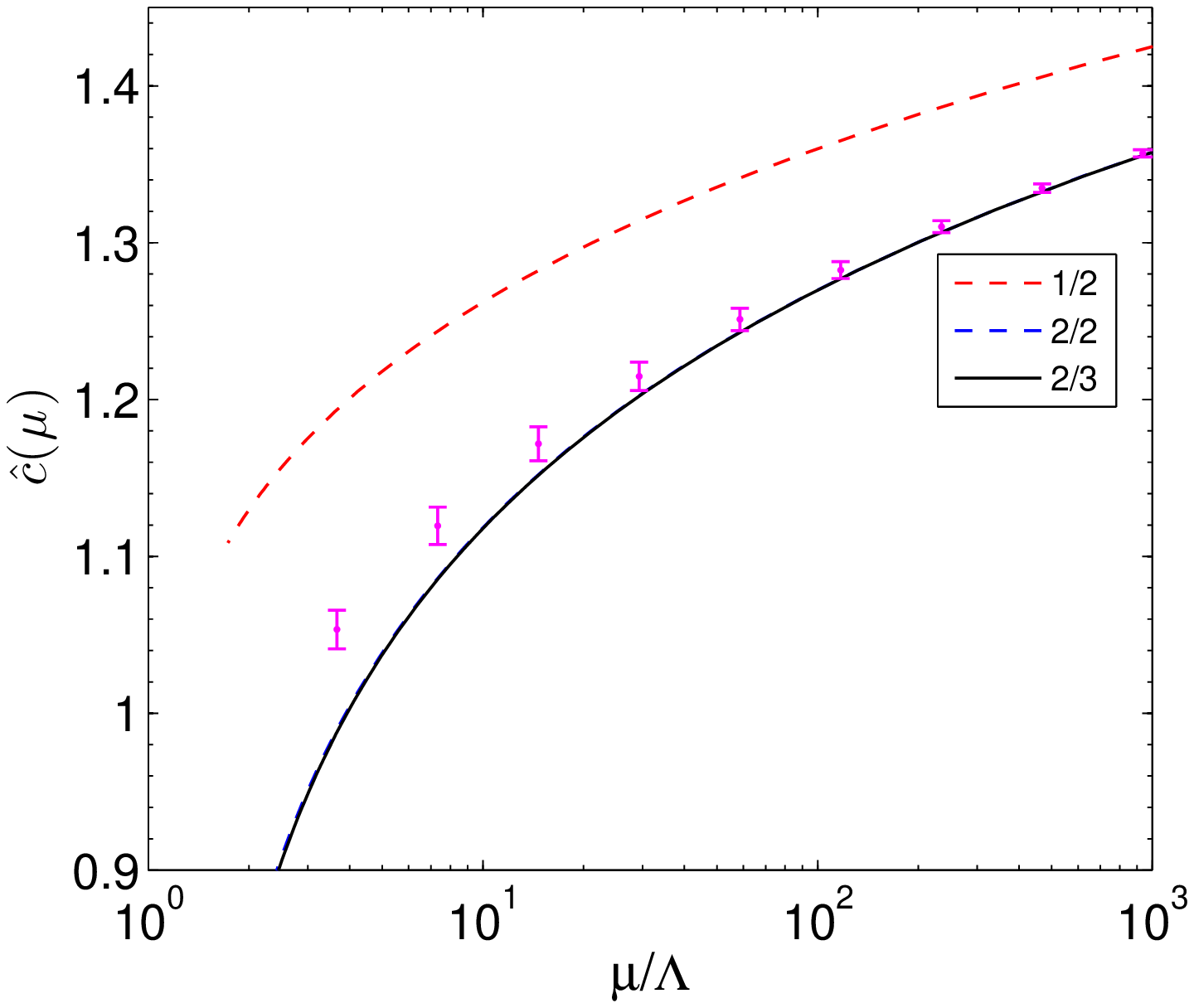}
\vspace{-5mm}
\end{center}
\caption{Running of the tensor current for $\NF=2$
in the schemes $\alpha=0$ (left) and $\alpha=1/2$ (right),
compared to perturbative predictions using the $1/2$-, $2/2$-, and $2/3$-loop
values for $\gamma_{\rm\scriptscriptstyle T}/\beta$.}
\label{figrunningnf2}
\end{figure}

Using as input the continuum SSFs, we follow the same strategy as in the quenched case
to recursively compute the running between low and high energy scales. In this case
the lowest scale reached in the recursion, following \cite{DellaMorte:2005kg}, is given by 
$\gbar^2_{\rm SF}(\mu_{\rm\scriptscriptstyle had})=4.61$.
Using the coupling SSF from~\cite{DellaMorte:2004bc}, the smallest value of the coupling that
can be reached via the recursion without leaving the interval covered by data
is $\gbar^2_{\rm SF}(\mu_{\rm\scriptscriptstyle pt})=1.017(10)$,
corresponding to $n=7$ (i.e. a total factor scale of $2^8$ in energy, like in the $\NF=0$ case). 
The matching to the RGI at $\mu_{\rm\scriptscriptstyle pt}$
is again performed using the 2/3-loop values of the $\gamma$/$\beta$ functions,
and the same checks to assess the systematics are carried out as in the quenched case.
Now the value obtained for
$\orgi{c}(\mu_{\rm\scriptscriptstyle had})$ remains within the quoted error
for all $n \geq 3$. Detailed results for the recursion in either scheme
are provided in Tables \ref{tab:run_nf2_f1}~and~\ref{tab:run_nf2_k1}.
We quote as our final results for the running factor
\begin{gather}
\begin{split}
\left.\orgi{c}(\mu_{\rm\scriptscriptstyle had})\right|_{\NF=2} & = 1.001(14)\,, \quad \mbox{scheme~}\alpha=0\,;\\
\left.\orgi{c}(\mu_{\rm\scriptscriptstyle had})\right|_{\NF=2} & = 1.053(12)\,, \quad \mbox{scheme~}\alpha=1/2\,.
\end{split}
\label{eq:rginf2}
\end{gather}
The running is illustrated, and compared with the perturbative prediction,
in Fig.~\ref{figrunningnf2}, where the value of
$\log(\Lambda_{\rm\scriptscriptstyle SF}/\mu_{\rm\scriptscriptstyle had})=-1.298(58)$
from~\cite{DellaMorte:2005kg} has been used. Using $r_0\Lambda_{\rm\scriptscriptstyle SF}=0.30(3)$ from~\cite{DellaMorte:2004bc} and $r_0=0.50~\fm$, this would correspond to a value of the
hadronic matching energy scale $\mu_{\rm\scriptscriptstyle had} \approx 432(50)~\MeV$.

\subsubsection{Hadronic Matching}

The computation of the renormalization constants at
$\mu_{\rm\scriptscriptstyle had}$ needed to match bare hadronic quantities
proceeds in a somewhat different way to the quenched case. The value of $\ZT$
in either scheme has been computed at three values of $\beta$, namely
$\beta=\{5.20,5.29,5.40\}$, again within the typical interval covered
by large-volume simulations with non-perturbatively $\Oa$ improved fermions
and a plaquette gauge action. For each of the values of $\beta$ two or three
values of the lattice size $L/a$ have been simulated, corresponding to different
values of $L$ and therefore to different values of the renormalized coupling.
The resulting values of $\ZT$ are given in Table~\ref{renormalization_matchingnf2}.

The lattice size $L/a=6$ used at $\beta=5.20$
corresponds within errors to $L=1/\mu_{\rm\scriptscriptstyle had}$;
for the other two values of $\beta$ linear interpolations can be performed
to obtain $\ZT$ at the correct value $u=4.610$; examples of such interpolations
are illustrated in Fig.~\ref{interpolation_hadronic}. The resulting values
of $\ZT$ can then be multiplied times the running factors in \req{eq:rginf2}
to obtain the RGI renormalization factors for each $\beta$. The result is
provided in Table~\ref{RGInf2}. In this case the $g_0^2$ dependence is barely visible
within the quoted errors, and the expected scheme independence holds only
up to $\sim 3\sigma$.

\section{Conclusions}

In this work we have set up the strategy for a non-perturbative determination
of the renormalization constants and anomalous dimension of tensor currents
in QCD using SF techniques, and obtained results for $\NF=0$ and $\NF=2$.
In the former case we employed both $\Oa$ improved and unimproved Wilson
fermions, and simulations were performed at four values of the lattice spacing
for each of the fourteen
different values of the renormalization scale, resulting in an excellent
control of the continuum limit. For $\NF=2$ our simulations were carried
out with $\Oa$ improved fermions, at only three values of the lattice
for each of the six renormalization scales.
The precision of the running factors up to the electroweak scale in the
schemes that allow for higher precision is $0.9\%$ and $1.1\%$, respectively.
The somewhat limited quality of our $\NF=2$ dataset, however, could result in
the quoted uncertainty for that case not being fully free of unquantified systematics.
We have also provided values of renormalization constants at the lowest
energy scales reached by the non-perturbative running, which allows to match
bare matrix elements computed with non-perturbatively $\Oa$ improved Wilson
fermions and the Wilson plaquette gauge action.

As part of the ALPHA programme, we are currently completing a similar study
in $\NF=3$ QCD~\cite{tensornf3}, that builds upon a high-precision determination
of the strong coupling~\cite{Brida:2016flw,DallaBrida:2016kgh,Bruno:2017gxd}
and mass anomalous dimension~\cite{Campos:2016eef,Campos:2016vxh,massnf3}.
Preliminary results indicate that
a precision $\sim 1\%$ for the running to low-energy scales is possible
even for values of the hadronic matching scale well below the one reached for $\NF=2$.
This is an essential ingredient in order to obtain matrix elements of
phenomenological interest with fully controlled uncertainties and target
precisions in the few percent ballpark.

\section*{Acknowledgements}
We are indebted to P.~Dimopoulos, M.~Guagnelli, J.~Heitger, G.~Herdo\'{\i}za, S.~Sint, and A.~Vladikas
for their r\^ole in earlier joint work of which this project is a spinoff.
The authors acknowledge support by Spanish MINECO
grants FPA2012-31686 and FPA2015-68541-P (MINECO/FEDER), and MINECO’s “Centro
de Excelencia Severo Ochoa” Programme under grant SEV-2012-0249.

\newpage
\begin{appendix}
\section{Perturbative improvement}
\label{app:cT}

The improvement coefficient $\icT$ for the tensor current can, by definition, be determined by requiring an $\Oa$ improved approach to the continuum of the renormalized correlation function at any given order in perturbation theory.
As discussed in the main text, the computation of $\icT$ to one loop has been carried out in~\cite{Sint:1997dj};
here we reproduce it, mainly as a crosscheck of our perturbative setup.

We introduce the following notation for the {\em renormalized} tensor correlator
${\kT}_{\rm\scriptscriptstyle ;R}$ in the chiral limit evaluated with SF boundary conditions
at $x_0=T/2$,
\begin{gather}
\hT(\theta,a/L)\equiv{\kT}_{\rm\scriptscriptstyle ;R}(T/2) \,.
\label{ratio}
\end{gather}
where the $\theta$ as well as the $a/L$ dependence have been made explicit.
The one-loop expansion reads
\begin{multline}
\hT=\kT^{(0)}(T/2) + \gbar^2  \{ \kT^{(1)}(T/2) + \icttil^{(1)}{\kT}_{\rm ;bi}^{(0)}(T/2) + am_0^{(1)}\frac{\partial \kT^{(0)}(T/2)}{\partial m_0} + \\
 \left ( \ZT^{(1)} + 2Z_{\xi}^{(1)} \right ) \kT^{(0)}(T/2) + \icT^{(1)}\tilde{\partial}_0\kV^{(0)}(T/2)  \} + \mathcal{O}(\gbar^4) \,,
\end{multline}
where $Z_{\xi}$ is the renormalization constant of the boundary fermionic fields, and $\icT$ is the coefficient we are interested in, providing the $\Oa$ improvement of the operator.
In order to determine $\icT^{(1)}$ we have adopted two different strategies.

The first one proceeds by imposing the condition
\begin{gather}
\frac{\hT(\theta,a/L)}{\hT(0,a/L)} = {\rm const}  + \mathcal{O}(a^2) \,.
\label{ratioA}
\end{gather}
With some trivial algebra, and observing that $\tilde{\partial}_0\kV^{(0)}(\theta=0)=0$, we end up with the relation 
\begin{gather}
\frac{\bar{k}_{\rm\scriptscriptstyle T}^{(1)}(\theta,a/L)}{\kT^{(0)}(\theta,a/L)} - \frac{\bar{k}_{\rm\scriptscriptstyle T}^{(1)}(0,a/L)}{\kT^{(0)}(0,a/L)} = -\icT^{(1)} \frac{\tilde{\partial}_0\kV^{(0)}(\theta,a/L)|_{x_0=T/2}}{\kT^{(0)}(\theta,a/L)} \,,
\label{ratioB}
\end{gather}
where $\bar{k}_{\rm\scriptscriptstyle T}$ is a shorthand notation for the correlator including the subtraction of the boundary and mass $\Oa$ terms. The divergent part of $\ZT^{(1)}$, as well as of $Z_{\xi}$, 
cancel out in the ratio, since they are independent of $\theta$ at one loop.
Following~\cite{Sint:1997jx}, in order to remove the constant term on the r.h.s. of Eq.~(\ref{ratioA}) ---
which is indeed proportional to the difference of the finite parts at two different values of $\theta$  --- we take a symmetric derivative in $L$, defined as
\begin{gather}
\tilde{\partial}_L f(L)=\frac{1}{2a}\left[f(L+a) - f(L-a)\right]\,,
\end{gather}
and apply it to both sides of Eq.~(\ref{ratioB}), obtaining 
\begin{gather}
R(\theta,a/L)=-\frac{\tilde{\partial}_LC(L)}{\tilde{\partial}_LA(L)}=\icT^{(1)} + \Oa\,,
\end{gather}
with $C(L)$ as the l.h.s of Eq.~(\ref{ratioB}), and $A(L)$ the r.h.s. without the
term with $\icT^{(1)}$.

As a second strategy to determine $\icT$ to one loop,
one can exploit the tree-level identities obtained in \cite{Sint:1997jx}, which relate $\kV^{(0)}, \kT^{(0)}, \fA^{(0)}$ and $\fP^{(0)}$, and impose
\begin{multline}
\bar{k}_{\rm\scriptscriptstyle T}^{(1)}+\frac{1}{3}\bar{f}_{\rm\scriptscriptstyle P}^{(1)}-\frac{2}{3}\bar{f}_{\rm\scriptscriptstyle A}^{(1)} + \ZT^{(1)}\kT^{(0)} + \frac{1}{3}\ZP^{(1)}\fP^{(0)} + \\
 \icT^{(1)}\tilde{\partial}_0\kV^{(0)}|_{x_0=T/2} - \frac{2}{3}c_A^{(1)}\tilde{\partial}_0\fP^{(0)}|_{x_0=T/2} ) = {\rm const } + \mathcal{O}(a^2).
\end{multline}
After some simple algebra we find 
\begin{gather}
F(\theta,a/L)\equiv -\frac{\tilde{\partial}_L C(L)}{\tilde{\partial}_LA(L)} + c_A^{(1)} = \icT^{(1)} + \Oa\,,
\end{gather}
where now 
\begin{multline}
C(L)=\bar{k}_{\rm\scriptscriptstyle T}^{(1)}(L/a) + \frac{1}{3}\bar{f}_{\rm\scriptscriptstyle P}^{(1)}(L/a) - \frac{2}{3}\bar{f}_{\rm\scriptscriptstyle A}^{(1)}(L/a) + \\
\frac{8}{3(4\pi^2)}\log(L/a)[\kT^{(0)}(L/a)-\fP^{(0)}(L/a)] \,,
\end{multline}
\begin{gather}
A(L)= \tilde{\partial}_0\fP^{(0)}(T/2) \,.
\end{gather}
Using the results for $\icA^{(1)}$ quoted in \cite{Sint:1997jx}, we reproduce within errors the value quoted in \cite{Sint:1997dj}, which reads
\begin{gather}
\icT^{(1)}=0.00896(1)\CF \,.
\end{gather}
The comparison between our determination and the one in \cite{Sint:1997dj} is displayed in Fig.~\ref{cTplot}.
In all cases, the continuum extrapolation has been performed using similar techniques
to the one employed for the finite part of renormalization constants (see App.~\ref{app:fitpt}).

\begin{center}
\begin{minipage}[t!]{1.0\textwidth}
\begin{center}
\includegraphics[width=100mm]{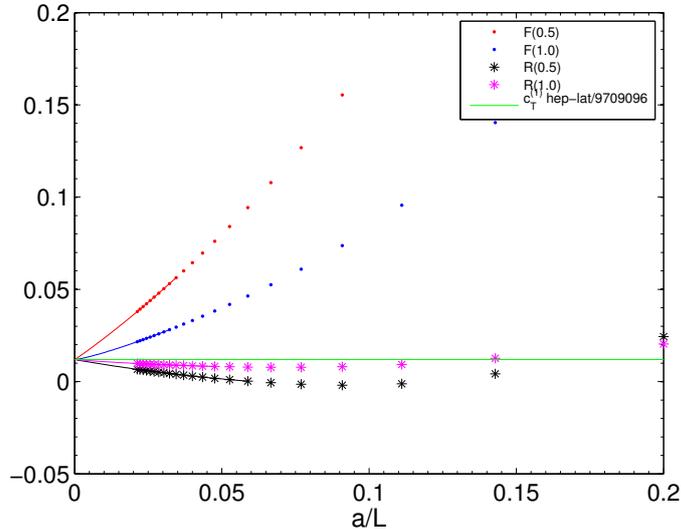}
\vspace{-2mm}
\captionof{figure}{Extraction of $\icT^{(1)}$, compared with the result in \cite{Sint:1997dj}.}
\label{cTplot}
\end{center}
\end{minipage}
\end{center}

\section{Continuum extrapolations in perturbation theory}
\label{app:fitpt}

In this appendix we summarize the techniques used to extrapolate
our perturbative computations to $a/L \to 0$, a necessary step in order
to obtain scheme-matching and improvement coefficients.
Our approach is essentially an application to the present context of the techniques discussed in 
Appendix~D of \cite{Bode:1999sm}, which have been applied in a number of cases,
see e.g. \cite{Palombi:2005zd}.

The typical outcome of a perturbative computation is a linear combination of one-loop Feynman diagrams, e.g. the one yielding the one-loop coefficient $Z^{(1)}$ of a renormalization constant, for $N$ values $\{l_1,\dots,l_N\}$ of the variable $l=L/a$. 
We consider the quantity to be a function of $l$ only. It is possible to identify all divergences appearing in the quantity of interest at one-loop, which in general means linear divergences related to the additive renormalization of the quark masses proportional to the one-loop critical mass $m_{\rm cr}^{(1)}$, and the logarithmic divergences proportional to the (one-loop) anomalous dimension. The latter is particularly relevant for the present analysis, since it allows to check the consistency of the fitting procedure and provides a natural criterion for the choice of the best fitting ansatz. In the following we consider finite quantities, since the leading divergence is subtracted, and the critical mass is appropriately tuned. Considering $F(l)$ as a generic one-loop interesting quantity, following \cite{Bode:1999sm} we conservatively assign the error
\begin{gather}
\delta F(l) = \epsilon(l)|F(l)| \, , 
\epsilon(l) = \left ( \frac{l}{2} \right )^3 \times 10^{-14} \, , 
\end{gather} 
since in this case the computation has been carried out in double precision. As expected, the asymptotic behaviour is (cf.~\req{Zasym})
\begin{gather}
F(l) = r_0 + \sum_{k=1}^n \frac{1}{l^n}(r_k + s_k {\rm ln}(l) ) + R_n(l)
\end{gather}
with a residue $R_n(l)$ that decreases faster than any of the terms in the sum as $l\to \infty$. In order to determine the coefficients $(r_k,s_k)$ we define as our likelihood function a $\chi^2$ given by 
\begin{gather}
\chi^2 = (F - f\xi)^T W (F-f\xi) \, ,
\end{gather}
where $F$ and $\xi$ are the $N-$column vectors $F=(F(l_1), \dots, F(l_N))^T$ and $(2n+1)-$column vector $\xi=(r_0,r_1,\dots,r_n,s_1,\dots,s_0)^T$, $f$ is the $N\times(2n+1)$ matrix 
\begin{gather}
f =
\begin{pmatrix}
1 		& l_1^{-1} & \dots  & l_1^{-n} & l_1^{-1} {\rm ln}(l_1) & \dots  & l_1^{-n} {\rm ln}(l_1) \\
1 		& l_2^{-1} & \dots  & l_2^{-n} & l_2^{-1} {\rm ln}(l_2) & \dots  & l_2^{-n} {\rm ln}(l_2) \\
\vdots  & \vdots   & \vdots & \vdots   & \vdots                 & \vdots & \vdots                 \\
1 		& l_N^{-1} & \dots  & l_N^{-n} & l_N^{-1} {\rm ln}(l_N) & \dots  & l_N^{-n} {\rm ln}(l_N) 
\end{pmatrix} \, ,
\end{gather}
and $W$ is in general a matrix with weights which, as suggested in \cite{Bode:1999sm}, is omitted 
from the actual $\chi^2$ used. The minimum condition for our likelihood function is given by
\begin{gather}
f\xi=PF\, ,
\label{eq:minchi2}
\end{gather}
where we are assuming that $2n+1<N$, and $P$ is the projector to the subspace spanned by the linearly independent column-vectors of $f$. A convenient and numerically stable way to solve \req{eq:minchi2} is the Singular Value Decomposition of $f$
\begin{gather}
f=USV^T \, ,
\label{eq:svd}
\end{gather}
where $U$ is an $N\times (2n+1)$ matrix such that 
\begin{gather}
U^TU=\mathbf{1} \, \quad \, UU^T=P \, ,
\end{gather}
$S$ is a diagonal and $V$ is an orthonormal $(2n+1)\times(2n+1)$ matrix. Inserting \req{eq:svd} into \req{eq:minchi2} one has 
\begin{gather}
\xi=VS^{-1}U^TF \, .
\end{gather}
Finally the uncertainty of the results $\xi_k$ is estimated to be
\begin{gather}
(\delta \xi_k)^2 = \sum_{l=1}^N [(VS^{-1}U^T)_{kl}]^2(\delta F_l)^2 \, , 
\end{gather}
with $\delta F_k=F(l_k)$. In order to avoid giving excessive weight to the coarsest lattices, we considered several possible fit ranges $[l_{\rm min},l_{\rm max}]$, where $l_{\rm max}=48$ and $l_{\rm min}$ is changed from $4$ to $20$. In order to account for a better description of the dependence on $l$ we explored different values of $n$ from $1$ to $4$. 

\begin{figure}[t!]
\begin{center}
\includegraphics[width=130mm]{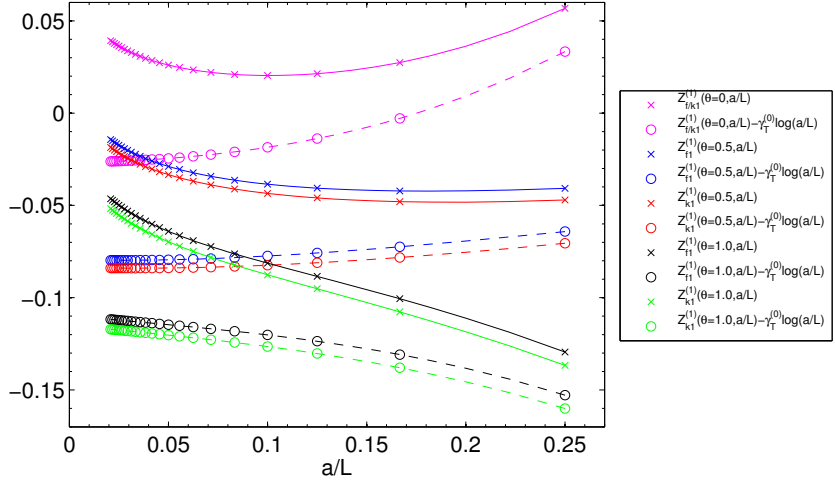}
\vspace{0mm}
\caption{One-loop renormalization constants for the three values of $\theta=0,0.5,1$. The $a/L$ dependence both before and after the subtraction of the leading logarithmic divergence is shown.}
\end{center}
\end{figure}

In particular, concerning the fit for the extraction of the finite parts, we chose as best ansatz the one reproducing the coefficient of the LO anomalous dimension $\gamma_T^{(0)}$. In particular for the Wilson action we find $s_0/\gamma_{\rm\scriptscriptstyle T}^{(0)}=0.998(5)$ for both $f_1$ and $k_1$ schemes using $n=3$ starting with $L/a=16$ as the smallest lattice. In the case with clover improvement of the action for the three values of $\theta=0$ for $n=3$ $L/a=14$ we have $\gamma_{\rm\scriptscriptstyle T}^{(0)}/s_0=1.001(3)$; for $\theta=0.5$, $n=3$, and $L/a=10$,
$\gamma_{\rm\scriptscriptstyle T}^{(0)}/s_0=1.000(6)$; and finally, for $\theta=1.0$, $n=3$, and $L/a=10$, $\gamma_T^{(0)}/s_0=1.000(3)$.

\end{appendix}
\appendixend

\newpage

\begin{landscape}
\begin{table}
\begin{center}
\begin{tabular}{ccccccc}
\toprule
$L/a$ &
[0.0,0,1,1l,*] &
[0.0,1/2,1,1l,*] &
[1.0,0,1,1l,0] &
[1.0,1/2,1,1l,0] &
[1.0,0,1,1l,1l] &
[1.0,1/2,1,1l,1l] \\
\midrule
4  & -4.024919 & -4.024919 & 1.973892 & 2.021956 & 2.500183 & 2.548247 \\
6  & -1.548675 & -1.548675 & 0.740155 & 0.831017 & 1.079337 & 1.170200 \\
8  & -0.826327 & -0.826327 & 0.400756 & 0.475909 & 0.651713 & 0.726867 \\
10 & -0.516240 & -0.516240 & 0.273405 & 0.331519 & 0.472838 & 0.530953 \\
12 & -0.353534 & -0.353534 & 0.209692 & 0.254876 & 0.375268 & 0.420452 \\
14 & -0.257339 & -0.257339 & 0.171319 & 0.207105 & 0.312917 & 0.348703 \\
16 & -0.195713 & -0.195713 & 0.145443 & 0.174349 & 0.269155 & 0.298061 \\
18 & -0.153857 & -0.153857 & 0.126680 & 0.150454 & 0.236532 & 0.260306 \\
20 & -0.124131 & -0.124131 & 0.112377 & 0.132244 & 0.211170 & 0.231037 \\
22 & -0.102261 & -0.102261 & 0.101072 & 0.117905 & 0.190834 & 0.207667 \\
24 & -0.085702 & -0.085702 & 0.091888 & 0.106322 & 0.174135 & 0.188570 \\
\toprule
\end{tabular}
\end{center}
\begin{center}
\begin{tabular}{ccccccc}
\toprule
$L/a$ &
[0.5,0,0,0,0] &
[0.5,1/2,0,0,0] &
[0.5,0,1,1l,0] &
[0.5,1/2,1,1l,0] &
[0.5,0,1,1l,1l] &
[0.5,1/2,1,1l,1l] \\
\midrule
4  & -1.467173 & -1.564753  & -1.120302 & -1.031861 & -0.990330 &  -0.901889 \\
6  & -1.318718 & -1.366570  & -0.587012 & -0.500733 & -0.501419 &  -0.415141 \\
8  & -1.097265 & -1.125110  & -0.351334 & -0.288400 & -0.287405 &  -0.224471 \\
10 & -0.919572 & -0.937671  & -0.225979 & -0.179971 & -0.174931 &  -0.128923 \\
12 & -0.785609 & -0.798283  & -0.153873 & -0.119221 & -0.111375 &  -0.076724 \\
14 & -0.683546 & -0.692903  & -0.109513 & -0.082621 & -0.073108 &  -0.046216 \\
16 & -0.603968 & -0.611155  & -0.080628 & -0.059210 & -0.048785 &  -0.027367 \\
18 & -0.540470 & -0.546161  & -0.060930 & -0.043495 & -0.032633 &  -0.015198 \\
20 & -0.488753 & -0.493370  & -0.046987 & -0.032532 & -0.021524 &  -0.007069 \\
22 & -0.445879 & -0.449699  & -0.036813 & -0.024641 & -0.013668 &  -0.001496 \\
24 & -0.409794 & -0.413007  & -0.029200 & -0.018813 & -0.018813 &   0.002400 \\
\bottomrule
\end{tabular}
\end{center}
\caption{Cutoff effects $\delta_k$ in tensor SSFs (see \req{eq:deltak})
for various schemes and amounts of $\Oa$ improvement.
The headers of the columns correspond to the values of the parameters $[\theta,\alpha,\icsw,\icttil,\icT]$
(``1l'' refers to the one-loop value of the coefficient).
For $\theta=0.0$ results at one loop
are independent of the value of $\alpha$, and the contribution from $\icT$ vanishes.}
\label{tabcutoff}
\end{table}
\end{landscape}

\clearpage

\begin{landscape}
\begin{table}[t!]
\centering
\begin{scriptsize}
\begin{tabular}{rrl@{\hspace{10mm}}llll@{\hspace{10mm}}llll}
\toprule
\multicolumn{3}{c}{} &
\multicolumn{4}{c}{Improved action} &
\multicolumn{4}{c}{Unimproved action} \\[1.0ex]
$\beta~~~$ & $L/a$ & $~~~\bar{g}^2(L)$ &
$~~~~~~\hopc$ & $\ZT\left(g_0^2,L/a\right)$ & $\ZT\left(g_0^2,2L/a\right)$ & $\SigmaT\left(u,L/a\right)$ &
$~~~~~~\hopc$ & $\ZT\left(g_0^2,L/a\right)$ & $\ZT\left(g_0^2,2L/a\right)$ & $\SigmaT\left(u,L/a\right)$ \\[1.0ex]
\midrule
10.7503 & 6 & 0.8873(5) & 0.130591(4) & 0.9781(7) & 0.9857(12) & 1.0078(14) & 0.134696(7) & 0.9571(8) & 0.9464(11) & 0.9888(14) \\ 
11.0000 & 8 & 0.8873(10) & 0.130439(3) & 0.9812(7) & 0.9923(12) & 1.0113(14) & 0.134548(6) & 0.9569(7) & 0.9522(12) & 0.9951(15) \\ 
11.3384 & 12 & 0.8873(30) & 0.130251(2) & 0.9878(11) & 1.0022(16) & 1.0146(20) & 0.134277(5) & 0.9605(11) & 0.9618(18) & 1.0014(22) \\ 
11.5736 & 16 & 0.8873(25) & 0.130125(2) & 0.9918(10) & 1.0061(23) & 1.0144(25) & 0.134068(6) & 0.9637(11) & 0.9686(20) & 1.0051(24) \\ 
\midrule
10.0500 & 6 & 0.9944(7) & 0.131073(5) & 0.9771(7) & 0.9868(14) & 1.0099(16) & 0.135659(8) & 0.9532(10) & 0.9428(12) & 0.9891(16) \\ 
10.3000 & 8 & 0.9944(13) & 0.130889(3) & 0.9820(11) & 0.9927(12) & 1.0109(17) & 0.135457(5) & 0.9535(8) & 0.9472(13) & 0.9934(16) \\ 
10.6086 & 12 & 0.9944(30) & 0.130692(2) & 0.9896(12) & 1.0047(18) & 1.0153(22) & 0.135160(4) & 0.9590(11) & 0.9624(20) & 1.0035(24) \\ 
10.8910 & 16 & 0.9944(28) & 0.130515(2) & 0.9936(11) & 1.0073(20) & 1.0138(23) & 0.134849(6) & 0.9641(13) & 0.9686(33) & 1.0047(37) \\ 
\midrule
9.5030 & 6 & 1.0989(8) & 0.131514(5) & 0.9766(9) & 0.9880(15) & 1.0117(18) & 0.136520(5) & 0.9516(10) & 0.9389(14) & 0.9867(18) \\ 
9.7500 & 8 & 1.0989(13) & 0.131312(3) & 0.9798(9) & 0.9964(16) & 1.0169(19) & 0.136310(3) & 0.9515(9) & 0.9475(13) & 0.9958(17) \\ 
10.0577 & 12 & 1.0989(40) & 0.131079(3) & 0.9874(12) & 1.0048(18) & 1.0176(22) & 0.135949(4) & 0.9574(13) & 0.9581(22) & 1.0007(27) \\ 
10.3419 & 16 & 1.0989(44) & 0.130876(2) & 0.9963(14) & 1.0090(19) & 1.0127(24) & 0.135572(4) & 0.9619(18) & 0.9676(22) & 1.0059(30) \\ 
\midrule
8.8997 & 6 & 1.2430(13) & 0.132072(9) & 0.9742(6) & 0.9908(12) & 1.0170(14) & 0.137706(5) & 0.9463(11) & 0.9363(14) & 0.9894(19) \\ 
9.1544 & 8 & 1.2430(14) & 0.131838(4) & 0.9806(8) & 0.9988(17) & 1.0186(19) & 0.137400(4) & 0.9487(10) & 0.9426(17) & 0.9936(21) \\ 
9.5202 & 12 & 1.2430(35) & 0.131503(3) & 0.9885(11) & 1.0062(23) & 1.0179(26) & 0.136855(2) & 0.9537(14) & 0.9558(16) & 1.0022(22) \\ 
9.7350 & 16 & 1.2430(34) & 0.131335(3) & 0.9971(21) & 1.0201(22) & 1.0231(31) & 0.136523(4) & 0.9564(14) & 0.9661(23) & 1.0101(28) \\ 
\midrule
8.6129 & 6 & 1.3293(12) & 0.132380(6) & 0.9732(9) & 0.9903(17) & 1.0176(20) & 0.138346(6) & 0.9455(12) & 0.9322(13) & 0.9859(19) \\ 
8.8500 & 8 & 1.3293(21) & 0.132140(5) & 0.9797(10) & 1.0036(18) & 1.0244(21) & 0.138057(4) & 0.9475(10) & 0.9397(18) & 0.9918(22) \\ 
9.1859 & 12 & 1.3293(60) & 0.131814(3) & 0.9914(15) & 1.0089(25) & 1.0177(30) & 0.137503(2) & 0.9534(15) & 0.9572(18) & 1.0040(25) \\ 
9.4381 & 16 & 1.3293(40) & 0.131589(2) & 0.9962(14) & 1.0207(30) & 1.0246(33) & 0.137061(4) & 0.9578(22) & 0.9645(23) & 1.0070(33) \\ 
\midrule
8.3124 & 6 & 1.4300(20) & 0.132734(10) & 0.9750(7) & 0.9908(14) & 1.0162(16) & 0.139128(11) & 0.9393(12) & 0.9299(15) & 0.9900(20) \\ 
8.5598 & 8 & 1.4300(21) & 0.132453(5) & 0.9800(9) & 1.0011(16) & 1.0215(19) & 0.138742(7) & 0.9445(11) & 0.9381(20) & 0.9932(24) \\ 
8.9003 & 12 & 1.4300(50) & 0.132095(3) & 0.9897(17) & 1.0188(26) & 1.0294(32) & 0.138120(8) & 0.9532(15) & 0.9574(25) & 1.0044(31) \\ 
9.1415 & 16 & 1.4300(58) & 0.131855(3) & 0.9976(12) & 1.0248(28) & 1.0273(31) & 0.137655(5) & 0.9592(16) & 0.9655(26) & 1.0066(32) \\ 
\midrule
7.9993 & 6 & 1.5553(15) & 0.133118(7) & 0.9726(7) & 0.9932(21) & 1.0212(23) & 0.140003(11) & 0.9385(13) & 0.9215(15) & 0.9819(21) \\ 
8.2500 & 8 & 1.5553(24) & 0.132821(5) & 0.9785(11) & 1.0073(22) & 1.0294(25) & 0.139588(8) & 0.9422(11) & 0.9359(20) & 0.9933(24) \\ 
8.5985 & 12 & 1.5533(70) & 0.132427(3) & 0.9927(17) & 1.0204(29) & 1.0279(34) & 0.138847(6) & 0.9532(16) & 0.9575(27) & 1.0045(33) \\ 
8.8323 & 16 & 1.5533(70) & 0.132169(3) & 0.9999(19) & 1.0305(35) & 1.0306(40) & 0.138339(7) & 0.9594(22) & 0.9671(34) & 1.0080(42) \\ 
  \bottomrule
\end{tabular}
\end{scriptsize}
\caption{
$\NF=0$ results for the renormalization constant $\ZP$ and the step scaling function $\SigmaT$
for the scheme $\alpha=0$.
}
\label{renormalization_constantnf0A}
\end{table}
\end{landscape}

\clearpage

\begin{landscape}
\begin{table}[t!]
\centering
\begin{scriptsize}
\begin{tabular}{rrl@{\hspace{10mm}}llll@{\hspace{10mm}}llll}
\toprule
\multicolumn{3}{c}{} &
\multicolumn{4}{c}{Improved action} &
\multicolumn{4}{c}{Unimproved action} \\[1.0ex]
$\beta~~~$ & $L/a$ & $~~~\bar{g}^2(L)$ &
$~~~~~~\hopc$ & $\ZT\left(g_0^2,L/a\right)$ & $\ZT\left(g_0^2,2L/a\right)$ & $\SigmaT\left(u,L/a\right)$ &
$~~~~~~\hopc$ & $\ZT\left(g_0^2,L/a\right)$ & $\ZT\left(g_0^2,2L/a\right)$ & $\SigmaT\left(u,L/a\right)$ \\[1.0ex]
\midrule
7.7170 & 6 & 1.6950(26) & 0.133517(8) & 0.9729(10) & 0.9977(7) & 1.0255(13) & 0.140954(12) & 0.9380(13) & 0.9199(18) & 0.9807(24) \\ 
7.9741 & 8 & 1.6950(28) & 0.133179(5) & 0.9787(9) & 1.0115(22) & 1.0335(24) & 0.140438(8) & 0.9402(12) & 0.9385(29) & 0.9982(33) \\ 
8.3218 & 12 & 1.6950(79) & 0.132756(4) & 0.9953(7) & 1.0268(23) & 1.0316(24) & 0.139589(6) & 0.9505(18) & 0.9616(28) & 1.0117(35) \\ 
8.5479 & 16 & 1.6950(90) & 0.132485(3) & 1.0014(19) & 1.0389(32) & 1.0374(38) & 0.139058(6) & 0.9579(20) & 0.9719(36) & 1.0146(43) \\ 
\midrule
7.4082 & 6 & 1.8811(22) & 0.133961(8) & 0.9702(10) & 0.9992(8) & 1.0299(13) & 0.142145(11) & 0.9346(14) & 0.9122(18) & 0.9760(24) \\ 
7.6547 & 8 & 1.8811(28) & 0.133632(6) & 0.9812(10) & 1.0175(22) & 1.0370(25) & 0.141572(9) & 0.9386(13) & 0.9347(19) & 0.9958(24) \\ 
7.9993 & 12 & 1.8811(38) & 0.133159(4) & 0.9980(7) & 1.0317(32) & 1.0338(33) & 0.140597(6) & 0.9498(18) & 0.9559(32) & 1.0064(39) \\ 
8.2415 & 16 & 1.8811(99) & 0.132847(3) & 1.0059(28) & 1.0445(27) & 1.0384(39) & 0.139900(6) & 0.9565(22) & 0.9776(38) & 1.0221(46) \\ 
\midrule
7.1214 & 6 & 2.1000(39) & 0.134423(9) & 0.9720(12) & 1.0039(9) & 1.0328(16) & 0.143416(11) & 0.9243(16) & 0.9067(21) & 0.9810(28) \\ 
7.3632 & 8 & 2.1000(45) & 0.134088(6) & 0.9833(12) & 1.0235(26) & 1.0409(29) & 0.142749(9) & 0.9312(14) & 0.9253(27) & 0.9937(33) \\ 
7.6985 & 12 & 2.1000(80) & 0.133599(4) & 0.9995(8) & 1.0427(25) & 1.0432(26) & 0.141657(6) & 0.9480(14) & 0.9564(22) & 1.0089(28) \\ 
7.9560 & 16 & 2.100(11) & 0.133229(3) & 1.0090(21) & 1.0564(27) & 1.0470(35) & 0.140817(7) & 0.9594(22) & 0.9749(35) & 1.0162(43) \\ 
\midrule
6.7807 & 6 & 2.4484(37) & 0.134994(11) & 0.9741(13) & 1.0160(10) & 1.0430(17) & 0.145286(11) & 0.9229(15) & 0.9003(21) & 0.9755(28) \\ 
7.0197 & 8 & 2.4484(45) & 0.134639(7) & 0.9866(13) & 1.0301(29) & 1.0441(32) & 0.144454(7) & 0.9318(15) & 0.9256(23) & 0.9933(29) \\ 
7.3551 & 12 & 2.4484(80) & 0.134141(5) & 1.0061(8) & 1.0618(30) & 1.0554(31) & 0.143113(6) & 0.9522(17) & 0.9572(38) & 1.0053(44) \\ 
7.6101 & 16 & 2.448(17) & 0.133729(4) & 1.0167(22) & 1.0808(32) & 1.0630(39) & 0.142107(6) & 0.9579(22) & 0.9851(39) & 1.0284(47) \\ 
\midrule
6.5512 & 6 & 2.770(7) & 0.135327(12) & 0.9798(14) & 1.0279(8) & 1.0491(17) & 0.146825(11) & 0.9208(18) & 0.8887(22) & 0.9651(30) \\ 
6.7860 & 8 & 2.770(7) & 0.135056(8) & 0.9910(13) & 1.0527(31) & 1.0623(34) & 0.145859(7) & 0.9311(16) & 0.9181(33) & 0.9860(39) \\ 
7.1190 & 12 & 2.770(11) & 0.134513(5) & 1.0097(10) & 1.0823(25) & 1.0719(27) & 0.144299(8) & 0.9489(21) & 0.9688(33) & 1.0210(41) \\ 
7.3686 & 16 & 2.770(14) & 0.134114(3) & 1.0215(27) & 1.1012(37) & 1.0780(46) & 0.143175(7) & 0.9663(31) & 1.0018(47) & 1.0367(59) \\ 
\midrule
6.3665 & 6 & 3.111(4) & 0.135488(6) & 0.9809(16) & 1.0384(30) & 1.0586(35) & 0.148317(10) & 0.9207(19) & 0.8802(19) & 0.9560(29) \\ 
6.6100 & 8 & 3.111(6) & 0.135339(3) & 0.9944(16) & 1.0711(37) & 1.0771(41) & 0.147112(7) & 0.9328(18) & 0.9189(27) & 0.9851(35) \\ 
6.9322 & 12 & 3.111(12) & 0.134855(3) & 1.0160(23) & 1.1093(35) & 1.0918(42) & 0.145371(7) & 0.9526(21) & 0.9740(35) & 1.0225(43) \\ 
7.1911 & 16 & 3.111(16) & 0.134411(3) & 1.0340(21) & 1.1222(42) & 1.0853(46) & 0.144060(8) & 0.9676(28) & 1.0092(45) & 1.0430(55) \\ 
 \midrule
6.2204 & 6 & 3.480(8) & 0.135470(15) & 0.9869(8) & 1.0678(27) & 1.0820(29) & 0.149685(15) & 0.9178(21) & 0.8709(23) & 0.9489(33) \\ 
6.4527 & 8 & 3.480(14) & 0.135543(9) & 1.0005(10) & 1.0909(46) & 1.0904(47) & 0.148391(9) & 0.9295(19) & 0.9140(44) & 0.9833(51) \\ 
6.7750 & 12 & 3.480(39) & 0.135121(5) & 1.0292(20) & 1.1281(41) & 1.0961(45) & 0.146408(7) & 0.9570(20) & 0.9793(49) & 1.0233(55) \\ 
7.0203 & 16 & 3.480(21) & 0.134707(4) & 1.0408(22) & 1.1420(45) & 1.0972(49) & 0.145025(8) & 0.9714(24) & 1.0264(51) & 1.0566(59) \\ 
\bottomrule
\end{tabular}
\end{scriptsize}
\addtocounter{table}{-1}
\caption{
(continued)
}
\end{table}
\end{landscape}

\clearpage

\begin{landscape}
\begin{table}[t!]
\centering
\begin{scriptsize}
\begin{tabular}{rrl@{\hspace{10mm}}llll@{\hspace{10mm}}llll}
\toprule
\multicolumn{3}{c}{} &
\multicolumn{4}{c}{Improved action} &
\multicolumn{4}{c}{Unimproved action} \\[1.0ex]
$\beta~~~$ & $L/a$ & $~~~\bar{g}^2(L)$ &
$~~~~~~\hopc$ & $\ZT\left(g_0^2,L/a\right)$ & $\ZT\left(g_0^2,2L/a\right)$ & $\SigmaT\left(u,L/a\right)$ &
$~~~~~~\hopc$ & $\ZT\left(g_0^2,L/a\right)$ & $\ZT\left(g_0^2,2L/a\right)$ & $\SigmaT\left(u,L/a\right)$ \\[1.0ex]
\midrule
10.7503 & 6 & 0.8873(5) & 0.130591(4) & 0.9687(6) & 0.9769(11) & 1.0085(13) & 0.134696(7) & 0.9497(8) & 0.9388(10) & 0.9885(13) \\ 
11.0000 & 8 & 0.8873(10) & 0.130439(3) & 0.9726(6) & 0.9835(11) & 1.0112(13) & 0.134548(6) & 0.9497(7) & 0.9446(11) & 0.9946(14) \\ 
11.3384 & 12 & 0.8873(30) & 0.130251(2) & 0.9795(10) & 0.9930(14) & 1.0138(18) & 0.134277(5) & 0.9529(10) & 0.9536(16) & 1.0007(20) \\ 
11.5736 & 16 & 0.8873(25) & 0.130125(2) & 0.9839(9) & 0.9974(20) & 1.0137(22) & 0.134068(6) & 0.9561(10) & 0.9603(18) & 1.0044(22) \\ 
\midrule
10.0500 & 6 & 0.9944(7) & 0.131073(5) & 0.9661(7) & 0.9761(11) & 1.0104(14) & 0.135659(8) & 0.9448(9) & 0.9339(11) & 0.9885(15) \\ 
10.3000 & 8 & 0.9944(13) & 0.130889(3) & 0.9716(9) & 0.9824(10) & 1.0111(14) & 0.135457(5) & 0.9450(8) & 0.9381(11) & 0.9927(14) \\ 
10.6086 & 12 & 0.9944(30) & 0.130692(2) & 0.9800(11) & 0.9942(16) & 1.0145(20) & 0.135160(4) & 0.9500(10) & 0.9521(18) & 1.0022(22) \\ 
10.8910 & 16 & 0.9944(28) & 0.130515(2) & 0.9845(9) & 0.9974(18) & 1.0131(20) & 0.134849(6) & 0.9554(11) & 0.9590(29) & 1.0038(32) \\ 
\midrule
9.5030 & 6 & 1.0989(8) & 0.131514(5) & 0.9642(8) & 0.9761(13) & 1.0123(16) & 0.136520(5) & 0.9419(9) & 0.9290(12) & 0.9863(16) \\ 
9.7500 & 8 & 1.0989(13) & 0.131312(3) & 0.9682(8) & 0.9842(13) & 1.0165(16) & 0.136310(3) & 0.9415(8) & 0.9369(11) & 0.9951(14) \\ 
10.0577 & 12 & 1.0989(40) & 0.131079(3) & 0.9766(10) & 0.9934(15) & 1.0172(19) & 0.135949(4) & 0.9471(12) & 0.9466(18) & 0.9995(23) \\ 
10.3419 & 16 & 1.0989(44) & 0.130876(2) & 0.9859(12) & 0.9975(16) & 1.0118(20) & 0.135572(4) & 0.9518(16) & 0.9568(19) & 1.0053(26) \\ 
\midrule
8.8997 & 6 & 1.2430(13) & 0.132072(9) & 0.9598(6) & 0.9759(10) & 1.0168(12) & 0.137706(5) & 0.9351(10) & 0.9243(12) & 0.9885(17) \\ 
9.1544 & 8 & 1.2430(14) & 0.131838(4) & 0.9673(7) & 0.9840(14) & 1.0173(16) & 0.137400(4) & 0.9374(9) & 0.9305(15) & 0.9926(19) \\ 
9.5202 & 12 & 1.2430(35) & 0.131503(3) & 0.9762(9) & 0.9926(19) & 1.0168(22) & 0.136855(2) & 0.9421(12) & 0.9433(13) & 1.0013(19) \\ 
9.7350 & 16 & 1.2430(34) & 0.131335(3) & 0.9849(18) & 1.0057(18) & 1.0211(26) & 0.136523(4) & 0.9453(13) & 0.9530(20) & 1.0081(25) \\ 
\midrule
8.6129 & 6 & 1.3293(12) & 0.132380(6) & 0.9577(8) & 0.9742(15) & 1.0172(18) & 0.138346(6) & 0.9332(11) & 0.9191(12) & 0.9849(17) \\ 
8.8500 & 8 & 1.3293(21) & 0.132140(5) & 0.9652(8) & 0.9871(15) & 1.0227(18) & 0.138057(4) & 0.9349(9) & 0.9262(15) & 0.9907(19) \\ 
9.1859 & 12 & 1.3293(60) & 0.131814(3) & 0.9776(13) & 0.9934(20) & 1.0162(25) & 0.137503(2) & 0.9403(13) & 0.9428(15) & 1.0027(21) \\ 
9.4381 & 16 & 1.3293(40) & 0.131589(2) & 0.9832(12) & 1.0055(25) & 1.0227(28) & 0.137061(4) & 0.9456(19) & 0.9504(20) & 1.0051(29) \\ 
\midrule
8.3124 & 6 & 1.4300(20) & 0.132734(10) & 0.9579(6) & 0.9731(11) & 1.0159(13) & 0.139128(11) & 0.9263(11) & 0.9153(13) & 0.9881(18) \\ 
8.5598 & 8 & 1.4300(21) & 0.132453(5) & 0.9642(8) & 0.9833(13) & 1.0198(16) & 0.138742(7) & 0.9312(10) & 0.9233(17) & 0.9915(21) \\ 
8.9003 & 12 & 1.4300(50) & 0.132095(3) & 0.9748(14) & 1.0001(21) & 1.0260(26) & 0.138120(8) & 0.9396(13) & 0.9411(20) & 1.0016(25) \\ 
9.1415 & 16 & 1.4300(58) & 0.131855(3) & 0.9835(10) & 1.0085(24) & 1.0254(27) & 0.137655(5) & 0.9452(14) & 0.9497(22) & 1.0048(28) \\ 
\midrule
7.9993 & 6 & 1.5553(15) & 0.133118(7) & 0.9537(6) & 0.9725(17) & 1.0197(19) & 0.140003(11) & 0.9239(11) & 0.9066(13) & 0.9813(18) \\ 
8.2500 & 8 & 1.5553(24) & 0.132821(5) & 0.9614(9) & 0.9873(18) & 1.0269(21) & 0.139588(8) & 0.9273(10) & 0.9197(17) & 0.9918(21) \\ 
8.5985 & 12 & 1.5533(70) & 0.132427(3) & 0.9765(14) & 1.0006(24) & 1.0247(29) & 0.138847(6) & 0.9376(14) & 0.9403(24) & 1.0029(30) \\ 
8.8323 & 16 & 1.5533(70) & 0.132169(3) & 0.9837(16) & 1.0102(27) & 1.0269(32) & 0.138339(7) & 0.9441(18) & 0.9499(28) & 1.0061(35) \\ 
\bottomrule
\end{tabular}
\end{scriptsize}
\caption{
$\NF=0$ results for the renormalization constant $\ZP$ and the step scaling function $\SigmaT$
for the scheme $\alpha=1/2$.
}
\label{renormalization_constantnf0B}
\end{table}
\end{landscape}

\clearpage

\begin{landscape}
\begin{table}[t!]
\centering
\begin{scriptsize}
\begin{tabular}{rrl@{\hspace{10mm}}llll@{\hspace{10mm}}llll}
\toprule
\multicolumn{3}{c}{} &
\multicolumn{4}{c}{Improved action} &
\multicolumn{4}{c}{Unimproved action} \\[1.0ex]
$\beta~~~$ & $L/a$ & $~~~\bar{g}^2(L)$ &
$~~~~~~\hopc$ & $\ZT\left(g_0^2,L/a\right)$ & $\ZT\left(g_0^2,2L/a\right)$ & $\SigmaT\left(u,L/a\right)$ &
$~~~~~~\hopc$ & $\ZT\left(g_0^2,L/a\right)$ & $\ZT\left(g_0^2,2L/a\right)$ & $\SigmaT\left(u,L/a\right)$ \\[1.0ex]
\midrule
7.7170 & 6 & 1.6950(26) & 0.133517(8) & 0.9522(9) & 0.9747(6) & 1.0236(12) & 0.140954(12) & 0.9215(12) & 0.9026(15) & 0.9795(21) \\ 
7.9741 & 8 & 1.6950(28) & 0.133179(5) & 0.9599(7) & 0.9887(18) & 1.0300(20) & 0.140438(8) & 0.9234(11) & 0.9204(24) & 0.9968(29) \\ 
8.3218 & 12 & 1.6950(79) & 0.132756(4) & 0.9769(5) & 1.0042(19) & 1.0279(20) & 0.139589(6) & 0.9333(15) & 0.9408(22) & 1.0080(29) \\ 
8.5479 & 16 & 1.6950(90) & 0.132485(3) & 0.9839(16) & 1.0160(26) & 1.0326(31) & 0.139058(6) & 0.9412(17) & 0.9522(31) & 1.0117(38) \\ 
\midrule
7.4082 & 6 & 1.8811(22) & 0.133961(8) & 0.9472(9) & 0.9730(6) & 1.0272(12) & 0.142145(11) & 0.9162(12) & 0.8933(15) & 0.9750(21) \\ 
7.6547 & 8 & 1.8811(28) & 0.133632(6) & 0.9597(8) & 0.9912(18) & 1.0328(21) & 0.141572(9) & 0.9197(11) & 0.9129(16) & 0.9926(21) \\ 
7.9993 & 12 & 1.8811(38) & 0.133159(4) & 0.9771(6) & 1.0066(27) & 1.0302(28) & 0.140597(6) & 0.9306(15) & 0.9337(26) & 1.0033(32) \\ 
8.2415 & 16 & 1.8811(99) & 0.132847(3) & 0.9865(24) & 1.0178(22) & 1.0317(34) & 0.139900(6) & 0.9380(18) & 0.9547(32) & 1.0178(39) \\ 
\midrule
7.1214 & 6 & 2.1000(39) & 0.134423(9) & 0.9454(10) & 0.9731(7) & 1.0293(13) & 0.143416(11) & 0.9044(14) & 0.8854(17) & 0.9790(24) \\ 
7.3632 & 8 & 2.1000(45) & 0.134088(6) & 0.9585(9) & 0.9914(19) & 1.0343(22) & 0.142749(9) & 0.9104(12) & 0.9021(22) & 0.9909(27) \\ 
7.6985 & 12 & 2.1000(80) & 0.133599(4) & 0.9764(6) & 1.0128(20) & 1.0373(21) & 0.141657(6) & 0.9265(11) & 0.9304(17) & 1.0042(22) \\ 
7.9560 & 16 & 2.100(11) & 0.133229(3) & 0.9862(17) & 1.0250(20) & 1.0393(27) & 0.140817(7) & 0.9380(19) & 0.9471(27) & 1.0097(35) \\ 
\midrule
6.7807 & 6 & 2.4484(37) & 0.134994(11) & 0.9431(11) & 0.9768(8) & 1.0357(15) & 0.145286(11) & 0.8989(13) & 0.8745(17) & 0.9729(24) \\ 
7.0197 & 8 & 2.4484(45) & 0.134639(7) & 0.9571(10) & 0.9933(23) & 1.0378(26) & 0.144454(7) & 0.9066(13) & 0.8959(19) & 0.9882(25) \\ 
7.3551 & 12 & 2.4484(80) & 0.134141(5) & 0.9777(7) & 1.0229(23) & 1.0462(25) & 0.143113(6) & 0.9260(14) & 0.9250(29) & 0.9989(35) \\ 
7.6101 & 16 & 2.448(17) & 0.133729(4) & 0.9905(18) & 1.0406(25) & 1.0506(32) & 0.142107(6) & 0.9325(18) & 0.9520(29) & 1.0209(37) \\ 
\midrule
6.5512 & 6 & 2.770(7) & 0.135327(12) & 0.9431(11) & 0.9807(6) & 1.0399(14) & 0.146825(11) & 0.8932(16) & 0.8591(17) & 0.9618(26) \\ 
6.7860 & 8 & 2.770(7) & 0.135056(8) & 0.9572(10) & 1.0057(24) & 1.0507(27) & 0.145859(7) & 0.9026(13) & 0.8859(26) & 0.9815(32) \\ 
7.1190 & 12 & 2.770(11) & 0.134513(5) & 0.9782(8) & 1.0326(18) & 1.0556(20) & 0.144299(8) & 0.9195(17) & 0.9287(25) & 1.0100(33) \\ 
7.3686 & 16 & 2.770(14) & 0.134114(3) & 0.9910(21) & 1.0505(28) & 1.0600(36) & 0.143175(7) & 0.9365(24) & 0.9595(36) & 1.0246(47) \\ 
\midrule
6.3665 & 6 & 3.111(4) & 0.135488(6) & 0.9399(13) & 0.9825(21) & 1.0453(27) & 0.148317(10) & 0.8889(16) & 0.8452(15) & 0.9508(24) \\ 
6.6100 & 8 & 3.111(6) & 0.135339(3) & 0.9572(13) & 1.0133(28) & 1.0586(33) & 0.147112(7) & 0.8999(15) & 0.8802(20) & 0.9781(28) \\ 
6.9322 & 12 & 3.111(12) & 0.134855(3) & 0.9803(18) & 1.0474(25) & 1.0684(32) & 0.145371(7) & 0.9189(16) & 0.9258(26) & 1.0075(33) \\ 
7.1911 & 16 & 3.111(16) & 0.134411(3) & 0.9988(16) & 1.0633(30) & 1.0646(35) & 0.144060(8) & 0.9349(22) & 0.9601(31) & 1.0270(41) \\ 
\midrule
6.2204 & 6 & 3.480(8) & 0.135470(15) & 0.9405(6) & 0.9952(19) & 1.0582(21) & 0.149685(15) & 0.8833(17) & 0.8316(19) & 0.9415(28) \\ 
6.4527 & 8 & 3.480(14) & 0.135543(9) & 0.9575(8) & 1.0198(31) & 1.0651(34) & 0.148391(9) & 0.8933(15) & 0.8701(34) & 0.9740(41) \\ 
6.7750 & 12 & 3.480(39) & 0.135121(5) & 0.9871(15) & 1.0568(29) & 1.0706(34) & 0.146408(7) & 0.9199(16) & 0.9247(35) & 1.0052(42) \\ 
7.0203 & 16 & 3.480(21) & 0.134707(4) & 1.0003(16) & 1.0705(32) & 1.0702(36) & 0.145025(8) & 0.9349(19) & 0.9657(36) & 1.0329(44) \\ 
\bottomrule
\end{tabular}
\end{scriptsize}
\addtocounter{table}{-1}
\caption{
(continued)
}
\end{table}
\end{landscape}


\clearpage

\begin{figure}[h!]
\begin{center}
\includegraphics[width=70mm]{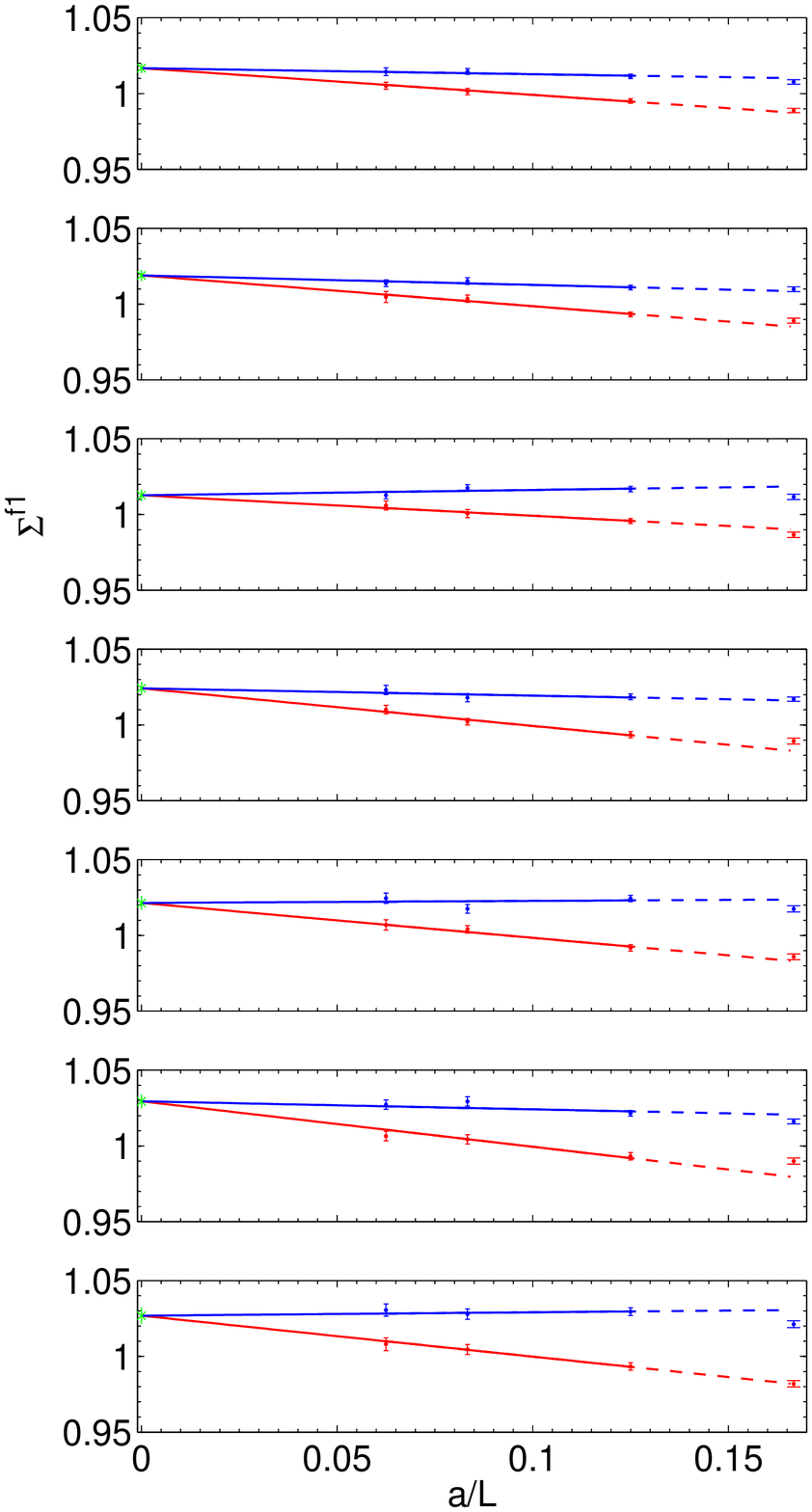}
\hspace{-3mm}
\includegraphics[width=70mm]{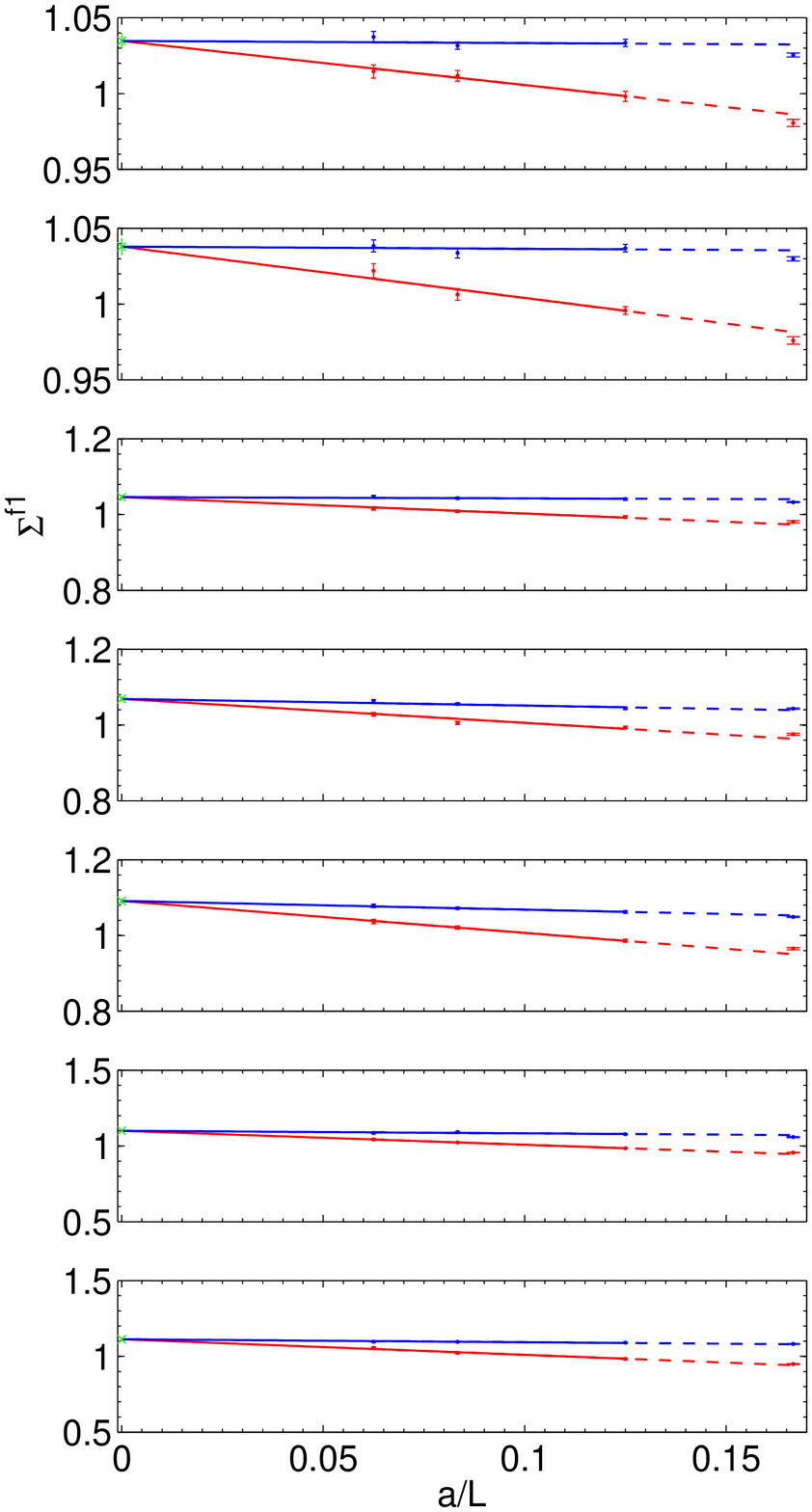}
\vspace{0mm}
\end{center}
\caption{Continuum limit extrapolations of the $\NF=0$ SSF for the renormalization scheme $\alpha=0$.
Blue (red) points correspond to results with the $\Oa$ improved (unimproved) action, respectively.}
\label{clnf0f1}
\end{figure}

\begin{figure}[h!]
\begin{center}
\includegraphics[width=70mm]{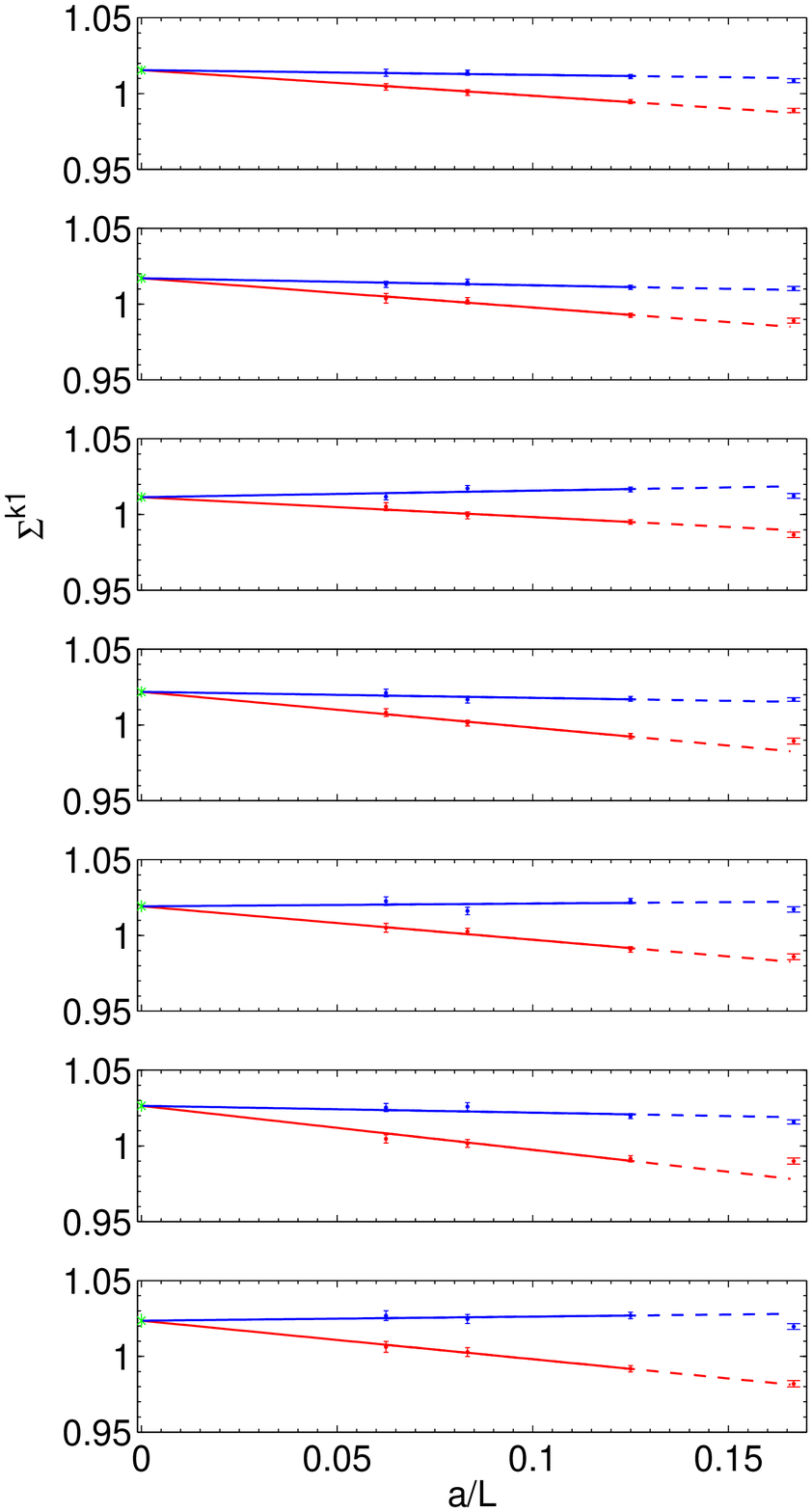}
\hspace{-3mm}
\includegraphics[width=70mm]{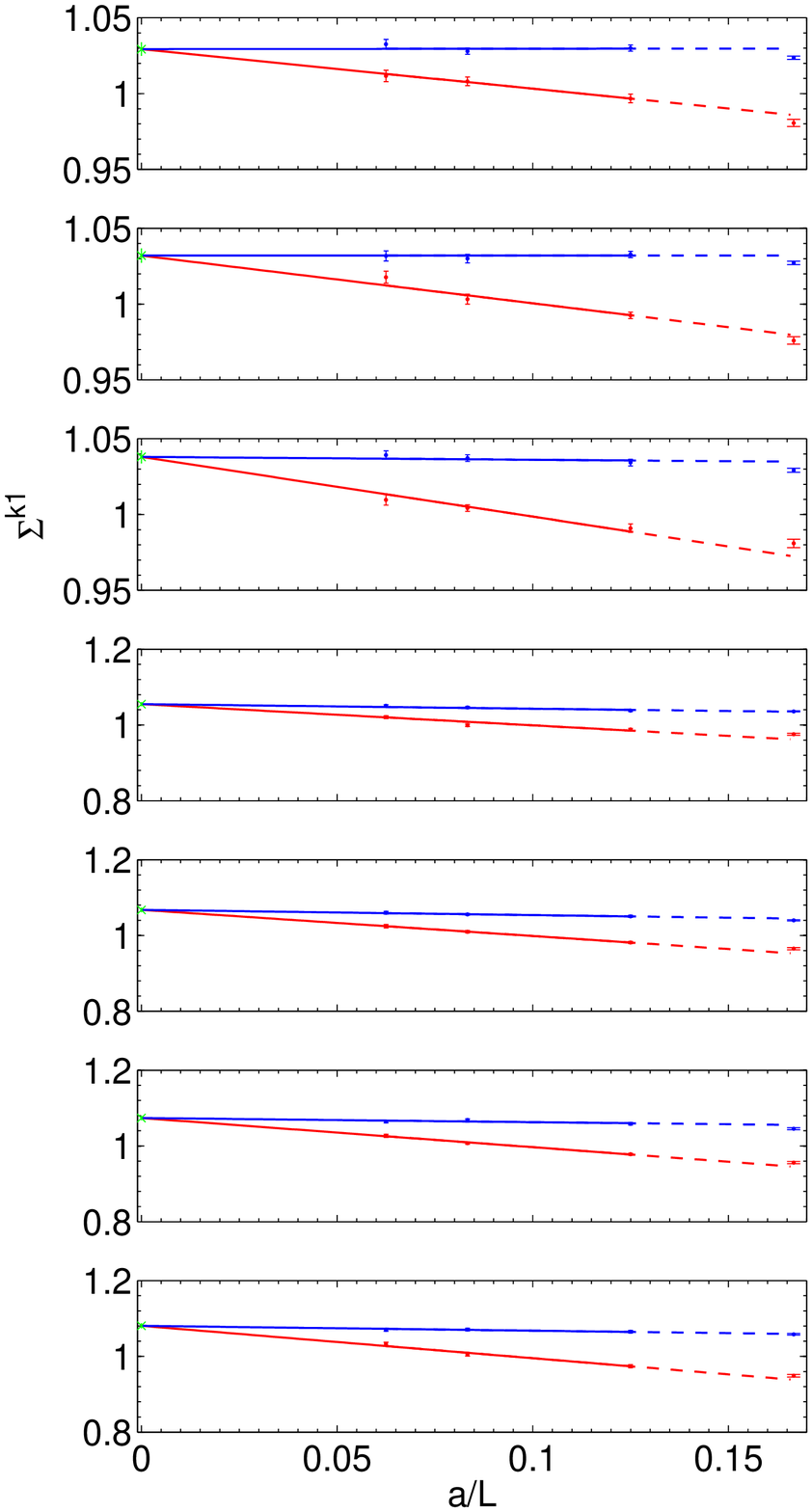}
\vspace{0mm}
\end{center}
\caption{Continuum limit extrapolations of the $\NF=0$ SSF for the renormalization scheme $\alpha=1/2$.
Blue (red) points correspond to results with the $\Oa$ improved (unimproved) action, respectively.}
\label{clnf0k1}
\end{figure}

\begin{table}[t!]
\noindent
\begin{center}
\begin{tabular}{c@{\hspace{10mm}}cc@{\hspace{10mm}}cc}
\toprule
& \multicolumn{2}{c}{$\alpha=0$} & \multicolumn{2}{c}{$\alpha=1/2$} \\
$u$ & $\sigmaT(u)$ & $\chi^2/{\rm dof}$ & $\sigmaT(u)$ & $\chi^2/{\rm dof}$   \\
\midrule
0.8873 & 1.0168(31) & 0.23 & 1.0155(27) & 0.20 \\
0.9944 & 1.0190(34) & 0.46 & 1.0171(30) & 0.41 \\
1.0989 & 1.0127(34) & 0.69 & 1.0115(30) & 1.18 \\
1.2430 & 1.0242(38) & 0.61 & 1.0219(33) & 0.54 \\
1.3293 & 1.0215(42) & 1.49 & 1.0192(36) & 1.83 \\
1.4300 & 1.0295(42) & 1.48 & 1.0265(36) & 1.52 \\
1.5553 & 1.0268(51) & 0.20 & 1.0235(43) & 0.20 \\
1.6950 & 1.0347(50) & 0.64 & 1.0294(42) & 0.60 \\
1.8811 & 1.0380(53) & 1.01 & 1.0320(45) & 1.03 \\
2.1000 & 1.0461(50) & 0.58 & 1.0381(40) & 1.08 \\
2.4484 & 1.0688(57) & 3.41 & 1.0550(45) & 3.65 \\
2.7700 & 1.0912(63) & 0.06 & 1.0677(50) & 0.05 \\
3.1110 & 1.1001(67) & 1.00 & 1.0738(51) & 0.86 \\
3.4800 & 1.1128(76) & 1.00 & 1.0806(57) & 1.09 \\
\bottomrule
\end{tabular}
\end{center}
\caption{Continuum-extrapolated values for the SSFs for $\NF=0$.}
\label{tabcontlimnf0}
\end{table}

\begin{table}[t!]
\begin{center}
\noindent\begin{tabular}{ccccccc}
\toprule
& fit & $p_1$ & $p_2$  & $p_3$ & $p_4$ & $\chi^2/{\rm dof}$ \\
\midrule
\multirow{5}*{$\alpha=0$}
& A & 0.011705 & 0.00611(32)		& --- 				& ---	 				& 1.16 \\ 
& B & 0.011705 & 0.0042(12) 		& 0.00072(45)		& ---	 				& 1.04 \\ 
& C & 0.011705 & 0.005449 			& 0.00028(11)		& ---	 				& 1.04 \\ 
& D & 0.011705 & 0.005449 			&-0.00005(66)		&0.00011(22)	 	& 1.11 \\ 
& E & 0.011705 & -0.0006(37)		&0.0051(32)			&-0.00089(64)	 	& 0.96 \\ 
\midrule
\multirow{5}*{$\alpha=1/2$}
& A & 0.011705 & 0.00370(25) 		& --- 				& ---	 				& 0.88 \\ 
& B & 0.011705 & 0.0035(10) 		&0.000072(36)		& ---				& 0.95 \\ 
& C & 0.011705 & 0.005043 			&-0.000455(88)		& ---	 				& 1.05 \\ 
& D & 0.011705 & 0.005043 			&-0.00098(54)		& 0.00017(17)	 	& 1.06 \\ 
& E & 0.011705 & -0.0003(31)		& 0.0034(26)		& -0.00068(52)	 	& 0.88 \\ 
\toprule
\end{tabular}
\caption{Fits to the continuum $\NF=0$ SSFs for various choices of polynomial ansatz, cf. \req{eq:ssfpoly}.}
\label{fitssfnf0}
\end{center}
\end{table}

\begin{landscape}
\begin{table}
\begin{center}
\noindent\begin{tabular}{lllllll}
\toprule
k & $u_{k}$ 		    &
$[U(\mu_{\rm\scriptscriptstyle had},2^{k+1}\mu_{\rm\scriptscriptstyle had})]^{-1}$  &
$\orgi{c}^{1/2}(\mu_{\rm\scriptscriptstyle had})$   &
$\orgi{c}^{2/2}(\mu_{\rm\scriptscriptstyle had})$   &
$\mathbf{\orgi{c}^{2/3}(\ci{\mu}_{\rm\scriptscriptstyle\bf had})}$   &
$\orgi{c}^{3/3}(\mu_{\rm\scriptscriptstyle had})$   \\
\midrule
0 & 3.480        & 0.8916(45) & 1.0655(53)   & 0.9099(45) & 0.9133(46) & 0.8201(41) \\
1 & 2.455(18)    & 0.8376(51) & 1.0377(64)   & 0.9256(59) & 0.9272(59) & 0.8768(59) \\
2 & 1.918(15)    & 0.8031(54) & 1.0218(70)   & 0.9332(66) & 0.9342(66) & 0.9021(65) \\
3 & 1.584(13)    & 0.7783(57) & 1.0113(76)   & 0.9378(72) & 0.9384(72) & 0.9160(72) \\
4 & 1.353(13)    & 0.7592(60) & 1.0039(82)   & 0.9408(78) & 0.9412(78) & 0.9246(78) \\
5 & 1.184(12)    & 0.7436(63) & 0.9983(87)   & 0.9429(84) & 0.9433(84) & 0.9304(84) \\
6 & 1.053(12)    & 0.7306(66) & 0.9939(93)   & 0.9446(90) & 0.9448(90) & 0.9346(90) \\
7 & 0.950(11)    & 0.7195(68) & 0.9905(98)   & 0.9459(95) & 0.9461(95) & 0.9377(95) \\
8 & 0.865(10)    & 0.7097(70) & 0.9876(102)  & 0.9469(99) & 0.9471(99) & 0.9401(99) \\
\bottomrule
\end{tabular}
\end{center}
\caption{Non-perturbative $\NF=0$ running in the scheme $\alpha=0$. (Our best result $\orgi{c}^{2/3}(\mu_{\rm\scriptscriptstyle had})$
is stressed.)}
\label{tab:run_nf0_f1}
\end{table}

\begin{table}
\begin{center}
\noindent\begin{tabular}{lllllll}
\toprule
k &  $u_{k}$ 		    &
$[U(\mu_{\rm\scriptscriptstyle had},2^{k+1}\mu_{\rm\scriptscriptstyle had})]^{-1}$  &
$\orgi{c}^{1/2}(\mu_{\rm\scriptscriptstyle had})$   &
$\orgi{c}^{2/2}(\mu_{\rm\scriptscriptstyle had})$   &
$\mathbf{\orgi{c}^{2/3}(\ci{\mu}_{\rm\scriptscriptstyle\bf had})}$   &
$\orgi{c}^{3/3}(\mu_{\rm\scriptscriptstyle had})$   \\
\midrule
0 & 3.480        & 0.9207(36) & 1.1003(44) & 0.9522(38) & 0.9556(38) & 0.8732(35) \\
1 & 2.455(18)    & 0.8762(41) & 1.0855(52) & 0.9776(49) & 0.9792(49) & 0.9344(49) \\
2 & 1.918(15)    & 0.8459(44) & 1.0761(57) & 0.9904(54) & 0.9914(54) & 0.9628(54) \\
3 & 1.584(13)    & 0.8231(47) & 1.0695(63) & 0.9981(60) & 0.9987(60) & 0.9787(60) \\
4 & 1.353(13)    & 0.8051(50) & 1.0646(69) & 1.0031(66) & 1.0036(66) & 0.9888(66) \\
5 & 1.184(12)    & 0.7902(54) & 1.0608(74) & 1.0068(72) & 1.0071(72) & 0.9956(72) \\
6 & 1.053(12)    & 0.7775(57) & 1.0577(80) & 1.0095(78) & 1.0098(78) & 1.0006(78) \\
7 & 0.950(11)    & 0.7665(59) & 1.0552(85) & 1.0116(83) & 1.0119(83) & 1.0043(83) \\
8 & 0.865(10)    & 0.7568(62) & 1.0531(89) & 1.0133(87) & 1.0135(87) & 1.0072(87) \\
\bottomrule
\end{tabular}
\end{center}
\caption{Non-perturbative $\NF=0$ running in the scheme $\alpha=1/2$. (Our best result $\orgi{c}^{2/3}(\mu_{\rm\scriptscriptstyle had})$
is stressed.)}
\label{tab:run_nf0_k1}
\end{table}
\end{landscape}

\begin{table}
\centering
\begin{tabular}{cc@{\hspace{10mm}}ll@{\hspace{10mm}}ll}
\toprule
\multicolumn{2}{c}{} &
\multicolumn{2}{c}{$\icsw={\rm NP}$} &
\multicolumn{2}{c}{$\icsw=0$} \\
$\beta$ &
$\frac{L}{a}$ &
$\qquad\hopc$ &
$\,\,\quad\ZT$ &
$\qquad\hopc$ &
$\,\,\quad\ZT$ \\
\midrule
6.0219 & 8 & 0.135043(17) & 1.0401(21) & 0.153371(10) & 0.9407(19) \\ 
6.1628 & 10 & 0.135643(11) & 1.0606(13) & 0.152012(7) & 0.9617(16) \\ 
6.2885 & 12 & 0.135739(13) & 1.0738(15) & 0.150752(10) & 0.9792(24) \\ 
6.4956 & 16 & 0.135577(7) & 1.0950(35) & 0.148876(13) & 1.0022(35) \\ 
\bottomrule
\end{tabular}
\caption{
Renormalization constants $\ZT(g_0^2,L/a)$ at $L=1/\mu_{\rm\scriptscriptstyle had}$ for $\NF=0$, scheme $\alpha=0$.
}
\label{renormalization_matchingnf0_f1}
\end{table}

\begin{table}
\centering
\begin{tabular}{cc@{\hspace{10mm}}ll@{\hspace{10mm}}ll}
\toprule
\multicolumn{2}{c}{} &
\multicolumn{2}{c}{$\icsw={\rm NP}$} &
\multicolumn{2}{c}{$\icsw=0$} \\
$\beta$ &
$\frac{L}{a}$ &
$\qquad\hopc$ &
$\,\,\quad\ZT$ &
$\qquad\hopc$ &
$\,\,\quad\ZT$ \\
\midrule
6.0219 & 8 & 0.135043(17) & 0.9715(15) & 0.153371(10) & 0.8853(15) \\ 
6.1628 & 10 & 0.135643(11) & 0.9909(9) & 0.152012(7) & 0.9033(13) \\ 
6.2885 & 12 & 0.135739(13) & 1.0044(11) & 0.150752(10) & 0.9178(18) \\ 
6.4956 & 16 & 0.135577(7) & 1.0236(24) & 0.148876(13) & 0.9399(27) \\ 
\bottomrule
\end{tabular}
\caption{
Renormalization constants $\ZT(g_0^2,L/a)$ at $L=1/\mu_{\rm\scriptscriptstyle had}$ for $\NF=0$, scheme $\alpha=1/2$.
}
\label{renormalization_matchingnf0_k1}
\end{table}

\begin{table}[t!]
\begin{center}
\noindent\begin{tabular}{c@{\hspace{10mm}}ll@{\hspace{10mm}}ll}
\toprule
& \multicolumn{2}{c}{$\icsw={\rm NP}$} & \multicolumn{2}{c}{$\icsw=0$} \\
$\beta$ &
\multicolumn{1}{c}{$\hat{Z}_{\rm\scriptscriptstyle T}^{\alpha=0}$} &
\multicolumn{1}{c}{$\hat{Z}_{\rm\scriptscriptstyle T}^{\alpha=1/2}$} &
\multicolumn{1}{c}{$\hat{Z}_{\rm\scriptscriptstyle T}^{\alpha=0}$} &
\multicolumn{1}{c}{$\hat{Z}_{\rm\scriptscriptstyle T}^{\alpha=1/2}$} \\
\midrule
6.0129 & 0.984(10) & 0.983(8) & 0.890(9)  & 0.896(8) \\
6.1628 & 1.003(10) & 1.003(8) & 0.910(9)  & 0.914(8) \\
6.2885 & 1.016(10) & 1.016(8) & 0.926(10) & 0.929(8) \\
6.4956 & 1.036(11) & 1.036(9) & 0.948(10) & 0.951(8) \\
\bottomrule
\end{tabular}
\end{center}
\caption{RGI renormalization factors $\hat{Z}_{\rm\scriptscriptstyle T}$ for $\NF=0$.}
\label{RGInf0}
\end{table}

\clearpage

\begin{landscape}
\begin{table}[t!]
\begin{scriptsize}
\noindent\begin{tabular}{lllr@{\hspace{15mm}}lll@{\hspace{15mm}}lll}
\toprule
\multicolumn{4}{c}{} & \multicolumn{3}{c}{$\alpha=0$} & \multicolumn{3}{c}{$\alpha=1/2$} \\
$\gbar^2_{\rm\scriptscriptstyle SF}(L)$ & $\beta$ & $\hopc$ & $L/a$ & $\ZT(g_0^2,L/a)$ & $\ZT(g_0^2,2L/a)$ & $\SigmaT(g_0^2,L/a)$ & $\ZT(g_0^2,L/a)$ & $\ZT(g_0^2,2L/a)$ & $\SigmaT(g_0^2,L/a)$ \\
\midrule
\multirow{3}*{0.9793}
& 9.50000 & 0.131532 & 6 & 0.97814(88) & 0.9895(14)   & 1.0116(17) & 0.96579(74) & 0.9778(11) & 1.0124(14) \\ 
& 9.73410 & 0.131305 & 8 & 0.98014(87) & 0.9914(20)   & 1.0115(22) & 0.96924(76) & 0.9806(17) & 1.0118(19) \\ 
& 10.05755 & 0.131069 & 12 & 0.98792(92) & 1.0062(25) & 1.0185(27) & 0.97789(79) & 0.9947(23) & 1.0172(25) \\ 
\midrule
\multirow{3}*{1.1814}
& 8.50000 & 0.132509 & 6 & 0.97574(94) & 0.9902(39)   & 1.0148(42) & 0.95998(79) & 0.9755(33) & 1.0161(35) \\ 
& 8.72230 & 0.132291 & 8 & 0.9819(17) & 1.0032(15)    & 1.0217(24) & 0.9674(15) & 0.9881(13) & 1.0214(21) \\ 
& 8.99366 & 0.131975 & 12 & 0.9902(11) & 1.0135(37)   & 1.0236(39) & 0.97665(91) & 0.9987(32) & 1.0226(35) \\ 
\midrule
\multirow{2}*{1.5078}
& 7.54200 & 0.133705 & 6 & 0.97715(94) & 1.0019(30)   & 1.0253(32) & 0.95543(79) & 0.9792(24) & 1.0249(26) \\ 
& 7.72060 & 0.133497 & 8 & 0.9830(22) & 1.0131(39)    & 1.0306(46) & 0.9631(18) & 0.9909(33) & 1.0289(39) \\ 
\multirow{2}*{1.5031}
& 7.50000 & 0.133815 & 6 & 0.97702(84) & 0.9931(35)   & 1.0164(37) & 0.95515(71) & 0.9721(29) & 1.0177(31) \\ 
& 8.02599 & 0.133063 & 12 & 0.9925(23) & 1.01982(58)  & 1.0275(25) & 0.9743(21) & 0.9988(29) & 1.0252(37) \\ 
\midrule
\multirow{3}*{2.0142}
& 6.60850 & 0.135260 & 6 & 0.9808(13) & 1.0158(22)    & 1.0357(26) & 0.9484(10) & 0.9808(17) & 1.0342(21) \\ 
& 6.82170 & 0.134891 & 8 & 0.9879(22) & 1.0311(26)    & 1.0438(35) & 0.9592(18) & 0.9966(22) & 1.0390(30) \\ 
& 7.09300 & 0.134432 & 12 & 1.0042(23) & 1.0433(13)   & 1.0389(27) & 0.9770(18) & 1.0112(10) & 1.0351(22) \\ 
\midrule
\multirow{3}*{2.4792}
& 6.13300 & 0.136110 & 6 & 0.9890(21) & 1.0407(70)    & 1.0523(74) & 0.9460(17) & 0.9916(56) & 1.0483(62) \\ 
& 6.32290 & 0.135767 & 8 & 0.9952(14) & 1.0435(84)    & 1.0485(86) & 0.9568(11) & 0.9981(66) & 1.0431(70) \\ 
& 6.63164 & 0.135227 & 12 & 1.0123(32) & 1.0756(13)   & 1.0625(36) & 0.9787(24) & 1.03073(99) & 1.0531(28) \\ 
\midrule
\multirow{3}*{3.3340}
& 5.62150 & 0.136665 & 6 & 1.0024(38) & 1.080(15)     & 1.077(16)  & 0.9405(29) & 0.9993(92) & 1.062(10) \\ 
& 5.80970 & 0.136608 & 8 & 1.0171(39) & 1.129(12)     & 1.110(13)  & 0.9608(27) & 1.0402(95) & 1.083(10) \\ 
& 6.11816 & 0.136139 & 12 & 1.0335(56) & 1.126(12)    & 1.090(13)  & 0.9844(41) & 1.0479(67) & 1.0645(81) \\ 
\bottomrule\end{tabular}
\caption{
$\NF=2$ results for the renormalization constant $\ZT$ and the step scaling function $\SigmaT$.
}
\label{tabZTnf2}
\end{scriptsize}
\end{table}
\end{landscape}

\newpage

\begin{center}
\begin{minipage}[t!]{1\textwidth}
\begin{center}
\includegraphics[width=70mm]{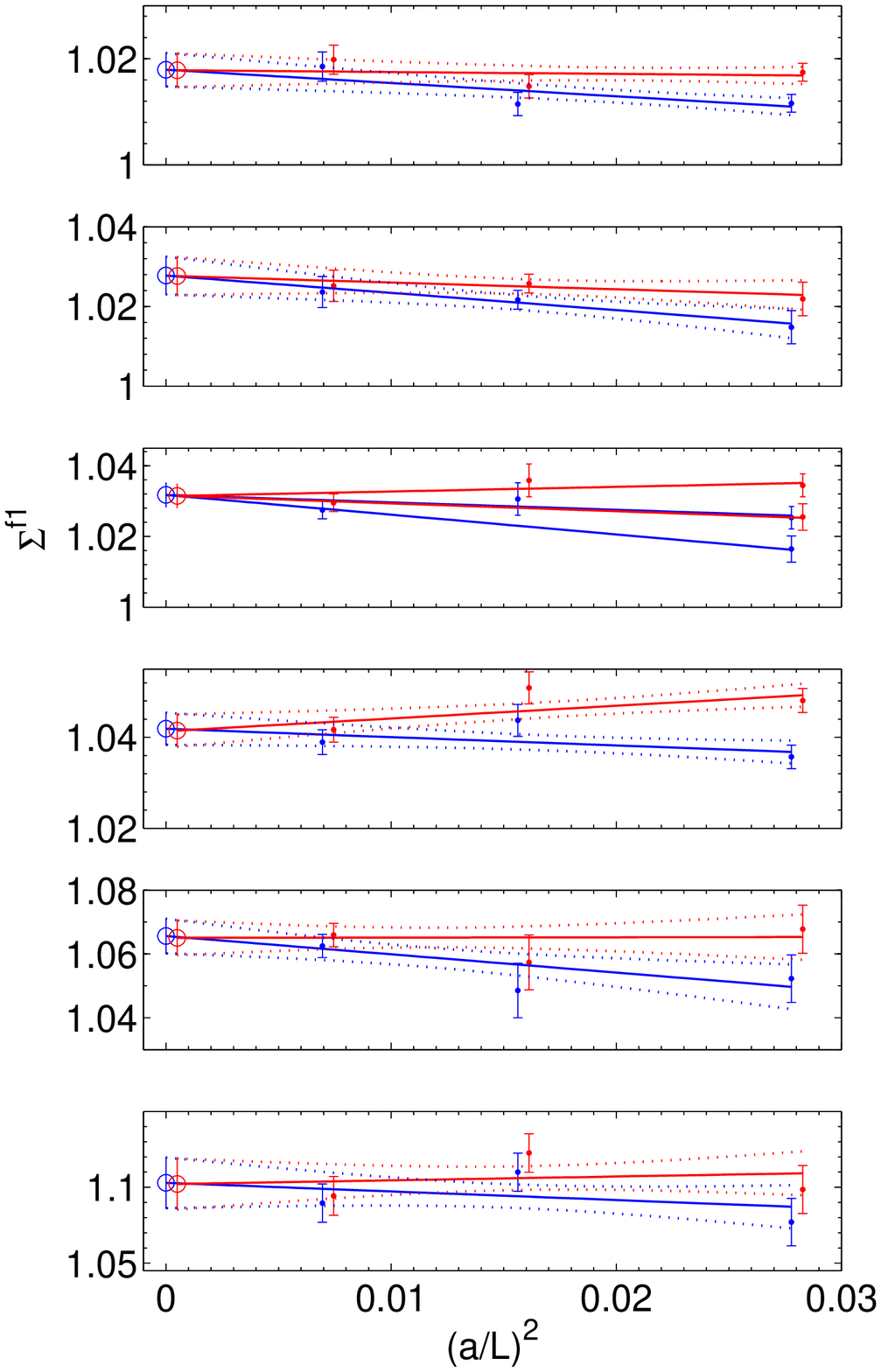}
\hspace{-4mm}
\includegraphics[width=70mm]{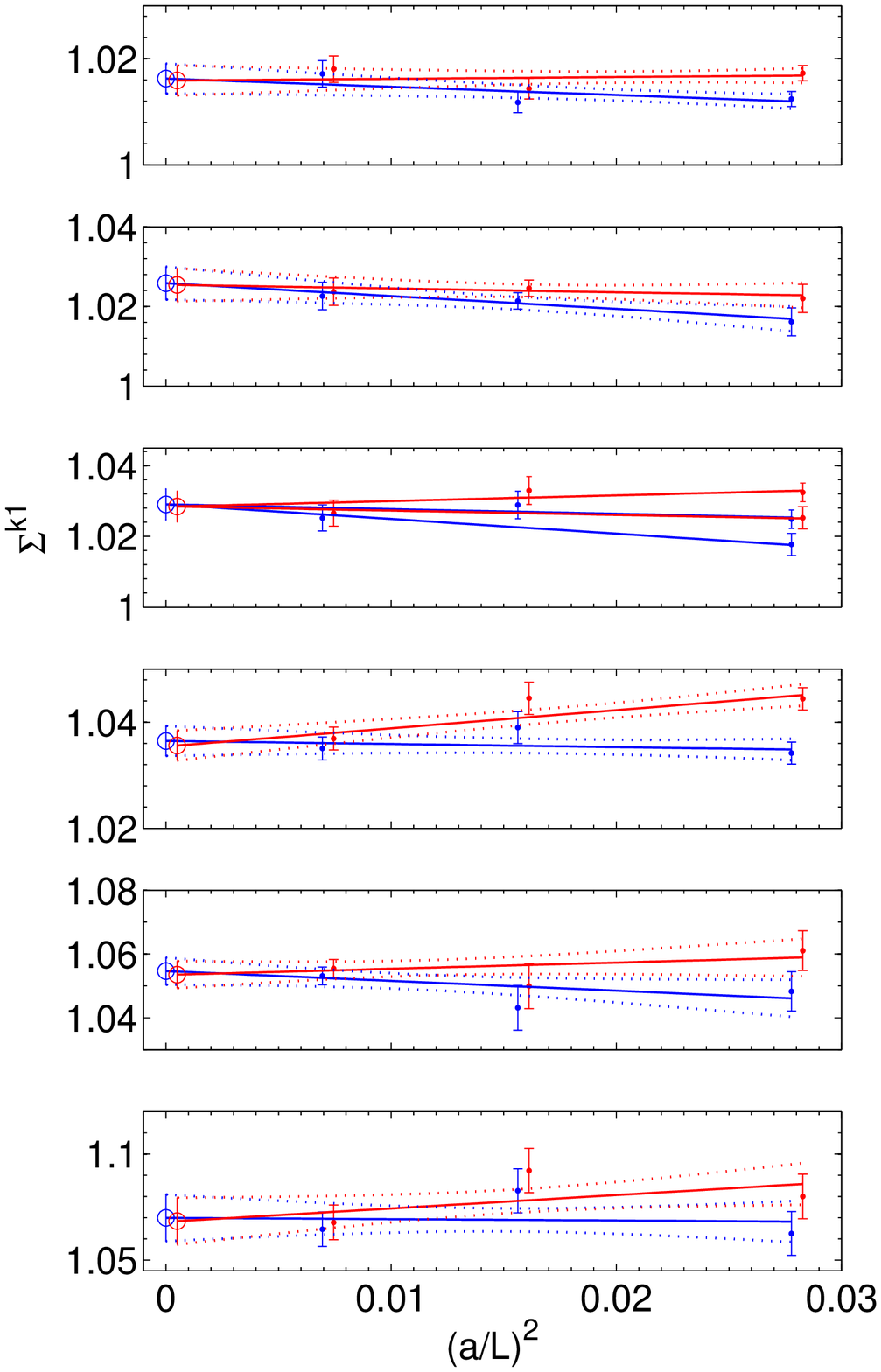}
\vspace{10mm}
\end{center}
\label{contlimnf0k1}
\captionof{figure}{Continuum extrapolations of SSFs for $\NF=2$ in the schemes $\alpha=0$ (left) and $\alpha=1/2$ (right).
Blue points are the data in Table~\ref{tabZTnf2}; red points result from subtracting the one-loop value of cutoff effects.}
\end{minipage}
\end{center}

\newpage

\begin{table}
\begin{center}
\noindent\begin{tabular}{c@{\hspace{10mm}}ccc@{\hspace{10mm}}ccc}
\toprule
& \multicolumn{3}{c}{$\alpha=0$} & \multicolumn{3}{c}{$\alpha=1/2$} \\
$u$ & $\sigmaT$ & $\rho(u)$ & $\chi^2/{\rm dof}$ & $\sigmaT$ & $\rho(u)$ & $\chi^2/{\rm dof}$ \\
\midrule
0.9793 & 1.0179(32)  & -0.25(14) & 2.22 & 1.0163(28) & -0.15(13) &  1.87 \\
1.1814 & 1.0278(48)  & -0.43(27) & 0.22 & 1.0258(41) & -0.32(23) & 0.21 \\
\multirow{2}{*}{1.5078} & \multirow{2}{*}{1.0317(34)} & -0.21(11) & \multirow{2}{*}{0.28} & \multirow{2}{*}{1.0291(44)} & -0.13(12) & \multirow{2}{*}{0.35} \\
				     & 				                 & -0.55(20) & 		    	       &                             & -0.41(19) & \\
2.0142 & 1.0419(35)  & -0.18(18) & 2.37 & 1.0365(28) & -0.06(14) & 1.61 \\
2.4792 & 1.0656(54)  & -0.57(39) & 1.08 & 1.0546(42) & -0.31(32) & 1.09 \\
3.3340 & 1.103(17)  & -0.57(96) & 2.59 & 1.070(11) & -0.06(63) & 2.42 \\
\bottomrule
\end{tabular}
\caption{$\NF=2$ continuum-extrapolated values of $\sigmaT$ without subtraction of perturbative cutoff effects. The two lines for $u=1.5078$ correspond to the use of the one- and two-loop
value of $\ict$, respectively.}
\label{tabcontlimnf2NOimp}
\end{center}
\end{table}

\begin{table}
\begin{center}
\noindent\begin{tabular}{c@{\hspace{10mm}}ccc@{\hspace{10mm}}ccc}
\toprule
& \multicolumn{3}{c}{$\alpha=0$} & \multicolumn{3}{c}{$\alpha=1/2$} \\
$u$ & $\sigmaT(u)$ & $\rho(u)$ & $\chi^2/{\rm dof}$ & $\sigmaT(u)$ & $\rho(u)$ & $\chi^2/{\rm dof}$ \\
\midrule
0.9793 & 1.0178(32) & -0.04(14) & 2.05 & 1.0159(28) & 0.03(13)  & 1.78\\
1.1814 & 1.0276(48) & -0.17(27) & 0.27 & 1.0254(41) & -0.09(24) & 0.24 \\
\multirow{2}{*}{1.5078} & \multirow{2}{*}{1.0315(34)}& 0.13(12)  & \multirow{2}{*}{0.33} & \multirow{2}{*}{1.0285(44)} &  0.16(12) & \multirow{2}{*}{0.376} \\
				    & 								 &  -0.22(20)& 				  	      &					            & -0.12(19) &  \\
2.0142 & 1.0415(35) & 0.28(18) & 2.63 & 1.0356(28) & 0.34(14) & 1.73\\
2.4792 & 1.0651(55) & 0.01(39) & 0.99 & 1.0535(43) & 0.19(32) & 1.04\\
3.3340 & 1.102(17) & 0.25(97) & 2.70 & 1.068(11) & 0.63(64) & 2.49 \\
\bottomrule
\end{tabular}
\caption{$\NF=2$ continuum-extrapolated values of $\sigmaT$ with subtraction of perturbative cutoff effects. The two lines for $u=1.5078$ correspond to the use of the one- and two-loop
value of $\ict$, respectively.}
\label{tabcontlimnf2SIimp}
\end{center}
\end{table}

\begin{table}[t!]
\begin{center}
\noindent\begin{tabular}{ccccccc}
\toprule
& fit & $p_1$ & $p_2$  & $p_3$ & $p_4$ & $\chi^2/{\rm dof}$ \\
\midrule
\multirow{5}*{$\alpha=0$}
& A & 0.011705 & 0.0055(5)	& - 				& -	 			& 0.73 \\ 
& B & 0.011705 & 0.0059(22) 	& -0.00018(94)	& -	 			& 0.90 \\ 
& C & 0.011705 & 0.005070 		& 0.00015(23)		& -	 			& 0.75 \\ 
& D & 0.011705 & 0.005070		& 0.00016(10)	    & -0.0000(4) 	& 0.93 \\ 
& E & 0.011705 & 0.0116(62)	& -0.0055(55)	    & 0.0012(12)  & 0.87 \\ 
\midrule
\multirow{5}*{$\alpha=1/2$}
& A & 0.011705 & 0.00351(42) 	& - 				& -	 	   		& 1.00 \\ 
& B & 0.011705 & 0.0054(18)	& -0.00080(74)	& -	 			& 0.96 \\ 
& C & 0.011705 & 0.004713		& -0.00053(17)	& -	 			& 0.80 \\ 
& D & 0.011705 & 0.004713	 	& -0.00034(83)	& -0.00007(31)	& 0.98 \\ 
& E & 0.011705 & 0.0076(55)	& -0.0027(46)	    & 0.00040(95) & 1.22 \\ 
\toprule
\end{tabular}
\caption{Fits to the continuum $\NF=2$ SSFs for various choices of polynomial ansatz, cf. \req{eq:ssfpoly}.}
\label{fitssfnf2}
\end{center}
\end{table}

\begin{landscape}

\begin{table}
\begin{center}
\noindent\begin{tabular}{lllllll}
\toprule
k & $u_k$ 		    &
$[U(\mu_{\rm\scriptscriptstyle had},2^{k+1}\mu_{\rm\scriptscriptstyle had})]^{-1}$  &
$\orgi{c}^{1/2}(\mu_{\rm\scriptscriptstyle had})$   &
$\orgi{c}^{2/2}(\mu_{\rm\scriptscriptstyle had})$   &
$\mathbf{\orgi{c}^{2/3}(\ci{\mu}_{\rm\scriptscriptstyle\bf had})}$   &
$\orgi{c}^{3/3}(\mu_{\rm\scriptscriptstyle had})$   \\
\midrule
-1 & 4.610     & 1         & 1.1818    & 0.9483    & 0.9495    & 0.7871      \\
0  & 3.032(16) & 0.922(7)  & 1.144(9)  & 0.985(8)  & 0.986(8)  & 0.905(7)  \\
1  & 2.341(21) & 0.872(8)  & 1.117(11) & 0.993(10) & 0.993(10) & 0.943(10)  \\
2 & 1.918(20) & 0.837(9)  & 1.098(12) & 0.996(11) & 0.996(11) & 0.962(11) \\
3 & 1.628(17) & 0.809(9)  & 1.084(13) & 0.998(12) & 0.998(12) & 0.973(12) \\
4 & 1.414(14) & 0.787(9)  & 1.074(13) & 0.999(12) & 0.999(12) & 0.980(12) \\
5 & 1.251(12) & 0.769(9)  & 1.067(14) & 1.000(13) & 1.000(13) & 0.985(13) \\
6 & 1.121(11) & 0.754(10) & 1.060(14) & 1.001(14) & 1.001(14) & 0.988(13) \\
7 & 1.017(10) & 0.741(10) & 1.055(15) & 1.001(14) & 1.001(14) & 0.991(14) \\
\bottomrule
\end{tabular}
\end{center}
\caption{Non-perturbative $\NF=2$ running in the scheme $\alpha=0$. (Our best result $\orgi{c}^{2/3}(\mu_{\rm\scriptscriptstyle had})$
is stressed.)}
\label{tab:run_nf2_f1}
\end{table}

\begin{table}
\begin{center}
\noindent\begin{tabular}{lllllll}
\toprule
k & $u_k$ 		    &
$[U(\mu_{\rm\scriptscriptstyle had},2^{k+1}\mu_{\rm\scriptscriptstyle had})]^{-1}$  &
$\orgi{c}^{1/2}(\mu_{\rm\scriptscriptstyle had})$   &
$\orgi{c}^{2/2}(\mu_{\rm\scriptscriptstyle had})$   &
$\mathbf{\orgi{c}^{2/3}(\ci{\mu}_{\rm\scriptscriptstyle\bf had})}$   &
$\orgi{c}^{3/3}(\mu_{\rm\scriptscriptstyle had})$   \\
\midrule
-1 & 4.610     & 1        & 1.1818    & 0.9650    & 0.9661    & 0.8241      \\
0 & 3.032(16) & 0.941(5) & 1.167(7)  & 1.017(6)  & 1.018(6)  & 0.947(6)  \\
1 & 2.341(21) & 0.899(7) & 1.151(9)  & 1.033(8)  & 1.033(8)  & 0.989(8)  \\
2 & 1.918(20) & 0.867(7) & 1.138(10) & 1.041(9)  & 1.041(9)  & 1.010(9)  \\
3 & 1.628(17) & 0.842(7) & 1.128(10) & 1.045(10) & 1.045(10) & 1.023(10)  \\
4 & 1.414(14) & 0.821(8) & 1.121(11) & 1.048(10) & 1.048(10) & 1.031(10) \\
5 & 1.251(12) & 0.804(8) & 1.115(12) & 1.051(11) & 1.051(11) & 1.037(11) \\
6 & 1.121(11) & 0.789(8) & 1.110(12) & 1.052(12) & 1.052(12) & 1.041(12) \\
7 & 1.017(10) & 0.776(9) & 1.106(13) & 1.053(12) & 1.053(12) & 1.044(12) \\
\bottomrule
\end{tabular}
\end{center}
\caption{Non-perturbative $\NF=2$ running in the scheme $\alpha=1/2$. (Our best result $\orgi{c}^{2/3}(\mu_{\rm\scriptscriptstyle had})$
is stressed.)}
\label{tab:run_nf2_k1}
\end{table}

\end{landscape}

\begin{table}[t!]
\begin{center}
\noindent\begin{tabular}{cccccc}
\toprule
$\beta$ & $\hopc$ & $L/a$ & $\bar{g}^2_{\rm\scriptscriptstyle SF}(L)$ & $\ZT^{\alpha=0}$ & $\ZT^{\alpha=1/2}$ \\
\midrule
\multirow{2}*{5.20} & \multirow{2}*{0.13600}
&   4  & 3.65(3) 		& 1.0256(13)  	& 0.9263(10)\\ 
& & 6  & 4.61(4)  		& 1.0678(17)  	& 0.9608(11)\\
\midrule
\multirow{3}*{5.29} & \multirow{3}*{0.13641}
&   4  & 3.394(17) 		& 1.0133(12)  	& 0.9251(9)\\ 
& & 6  & 4.297(37) 		& 1.0487(20)  	& 0.9556(14)\\ 
& & 8  & 5.65(9) 		& 1.0958(22)  	& 0.9886(15)\\ 
\midrule
\multirow{3}*{5.40} & \multirow{3}*{0.13669}
&   4  & 3.188(24) 		& 1.0054(11)  & 0.9270(9)\\ 
& & 6  & 3.864(34) 		& 1.0306(16)  & 0.9500(12)\\ 
& & 8  & 4.747(63) 		& 1.0671(17)  & 0.9781(12)\\ 
\bottomrule
\end{tabular}
\end{center}
\caption{Renormalization constants $\ZT(g_0^2,L/a)$ at $L=1/\mu_{\rm\scriptscriptstyle had}$ for $\NF=2$.}
\label{renormalization_matchingnf2}
\end{table}

\begin{figure}[h!]
\begin{center}
\includegraphics[width=70mm]{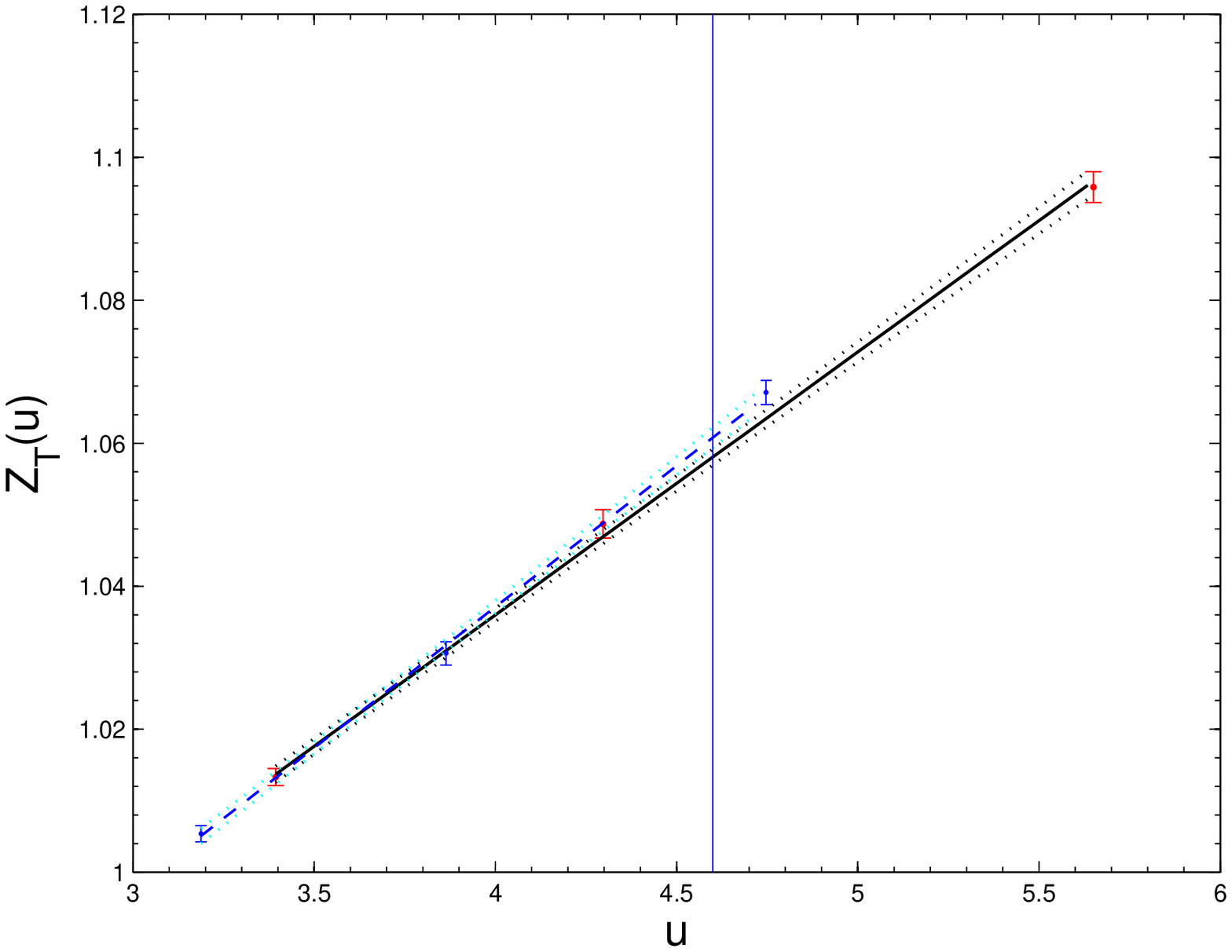}
\hspace{-3mm}
\includegraphics[width=70mm]{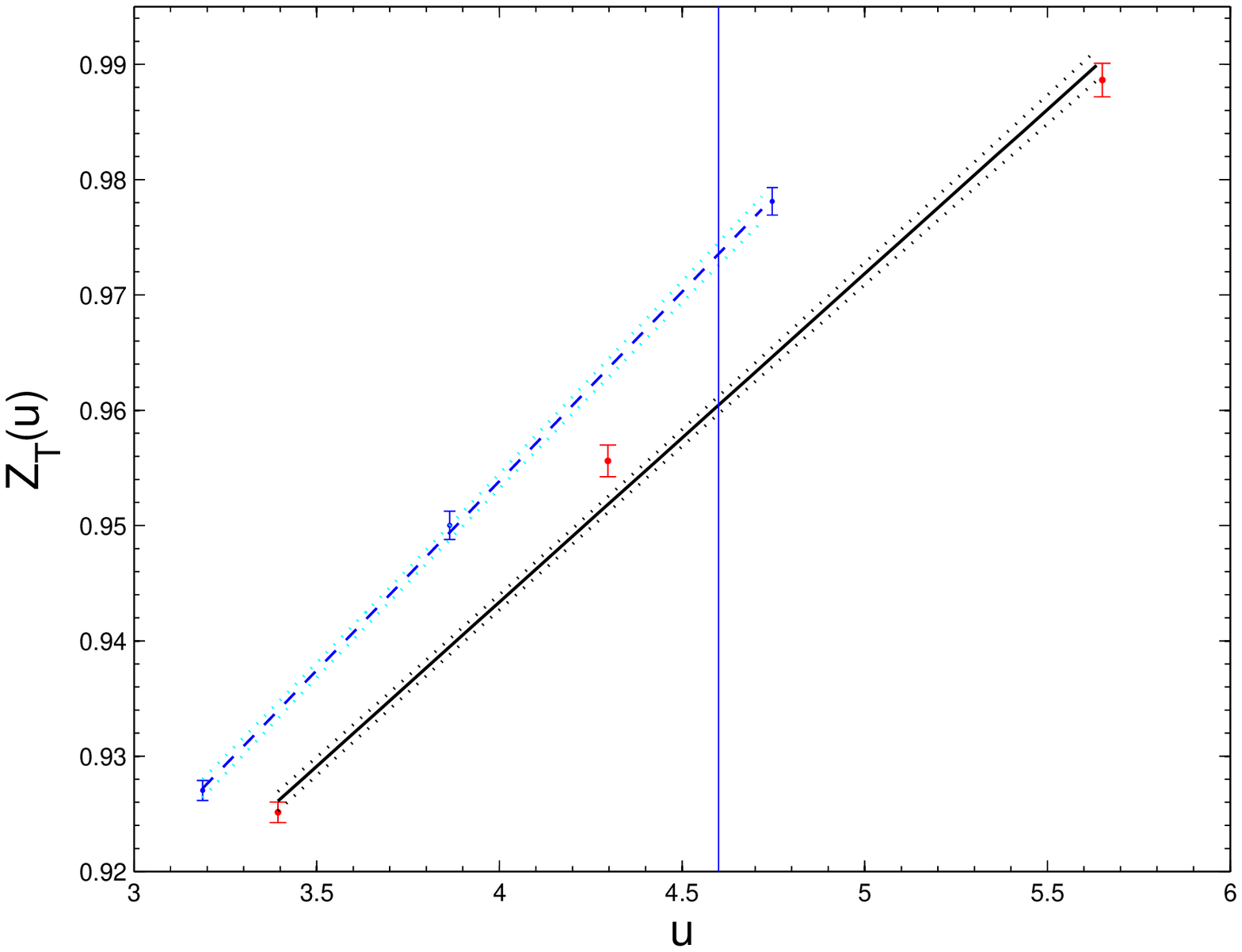}
\vspace{-7mm}
\end{center}
\caption{$\NF=2$, interpolation to $u_{\rm had}$.}
\label{interpolation_hadronic}
\end{figure}

\begin{table}[h!]
\begin{center}
\noindent\begin{tabular}{c@{\hspace{10mm}}ll}
\toprule
$\beta$ &
\multicolumn{1}{c}{$\hat{Z}_{\rm\scriptscriptstyle T}^{\alpha=0}$} &
\multicolumn{1}{c}{$\hat{Z}_{\rm\scriptscriptstyle T}^{\alpha=1/2}$} \\
\midrule
5.20 & 1.069(15) &  1.012(12) \\
5.29 & 1.060(15) &  1.012(12) \\
5.40 & 1.062(15) &  1.026(12) \\
\bottomrule
\end{tabular}
\end{center}
\caption{RGI renormalization factors $\hat{Z}_{\rm\scriptscriptstyle T}$ for $\NF=2$.}
\label{RGInf2}
\end{table}

\bibliographystyle{JHEPjus}
\bibliography{biblio}
\end{document}